\documentclass[useAMS,usenatbib]{mn2e}
\pdfoutput=1
\usepackage{times}  
\usepackage{listings}
\usepackage{graphics}
\usepackage{subfigure}
\usepackage{graphicx}
\usepackage{amssymb}
\usepackage{amsmath}
\usepackage[draft]{hyperref}
\usepackage{aas_macros}
\usepackage{verbatim}
\usepackage{textgreek}
\usepackage{color}
\usepackage{url}
\usepackage{multirow}
\usepackage[normalem]{ulem}
\usepackage[dvipsnames]{xcolor}

\long\def\symbolfootnote[#1]#2{\begingroup%
\def\thefootnote{\fnsymbol{footnote}}\footnote[#1]{#2}\endgroup} 

\newcommand{\euclid}{\emph{Euclid}}

\newcommand{\galacticus}{\textsc{Galacticus}}

\newcommand{\cloudy}{\textsc{Cloudy}}

\newcommand{\halpha}{{\rm H\alpha}}
\newcommand{\Halpha}{{\rm H\alpha}}
\newcommand{\nii}{\left [{\rm N_{II}} \right ]}
\newcommand{\oii}{\left [{\rm O_{II}} \right ]}

\newcommand{\Mpc}{{\,\rm Mpc}}
\newcommand{\hMpc}{\,h^{-1}{\rm Mpc}}
\newcommand{\hGpc}{\,h^{-1}{\rm Gpc}}

\newcommand{\ergPerSecondPerCM}{{\rm erg}\,{\rm s}^{-1}{\rm cm}^{-2}}
\newcommand{\ergPerSecond}{{\rm erg}\,{\rm s}^{-1}}
\newcommand{\Msol}{{\rm M_{\odot}}}
\newcommand{\hMsol}{h^{-1}{\rm M_{\odot}}}

\input{epsf} \title{Linear bias forecasts for emission line cosmological surveys}

\author[Merson {\it et al.}]
       {\parbox[h]{\textwidth}{Alexander~Merson$^{1,2}$\thanks{E-mail:
       alex.i.merson@gmail.com}, Alex Smith$^{3,4}$, Andrew~Benson$^5$, Yun~Wang$^{2}$, Carlton Baugh$^4$}     
  \vspace*{10pt}\\
  \noindent$^1$Jet Propulsion Laboratory, California Institute of Technology, 4800 Oak Grove Drive, Pasadena, CA 91109, USA\\
  $^2$IPAC, Mail Code 314-6, California Institute of Technology, 1200 East California Boulevard, Pasadena, CA 91125, USA\\
  $^3$IRFU, CEA, Universit\'{e} Paris-Saclay, F-91191 Gif-sur-Yvette, France\\
$^4$Institute for Computational Cosmology, Department of Physics, University of Durham, South Road, Durham DH1 3LE, UK\\
$^5$Carnegie Observatories, 813 Santa Barbara Street, Pasadena, CA 91101, USA}
\date{}

\pagerange{\pageref{firstpage}--\pageref{lastpage}}
\pubyear{\the\year}

\begin{document}

\maketitle
\title{Linear bias forecasts}
\label{firstpage}

\begin{abstract}
We forecast the linear bias for $\halpha$-emitting galaxies at high redshift. To simulate a Euclid-like and a WFIRST-like survey, we place galaxies into a large-volume dark matter halo lightcone by sampling a library of luminosity-dependent halo occupation distributions (HODs), which is constructed using a physically motivated galaxy formation model. We calibrate the dust attenuation in the lightcones such that they are able to reproduce the $\halpha$ luminosity function or the $\halpha$ cumulative number counts. The angle-averaged galaxy correlation function is computed for each survey in redshift slices of width $\Delta z=0.2$. In each redshift bin the linear bias can be fitted with a single, scale-independent value that increases with increasing redshift.  Fitting for the evolution of linear bias with redshift, we find that our Euclid-like and  WFIRST-like surveys are both consistent within error with the relation $b(z)= 0.7z+0.7$. Our bias forecasts are consistent with bias measurements from the HiZELS survey. We find that the Euclid-like and WFIRST-like surveys yield linear biases that are broadly consistent within error, most likely due to the HOD for the WFIRST-like survey having a steeper power-law slope towards larger halo masses.
\end{abstract}



\section{Introduction}
\label{sec:intro}

Probing the nature of the driving force behind the observed accelerated expansion of the Universe continues to be one of the major goals of modern cosmology. Most of the observational evidence gathered to date is consistent with the theory that the expansion is the result of a mysterious phenomenon known as \emph{dark energy}, although existing observations lack the statistical precision to allow alternative theories to be differentiated and conclusively ruled out. Up and coming cosmological missions aim to make very precise measurements of different cosmological probes of the expansion history of the Universe and the growth rate of cosmic large-scale structure in order to distinguish between the competing theories \citep[e.g.][]{Albrecht06,Guzzo08,Wang08a,Wang08b}.  

For missions such as the ESA-led {\euclid} mission \citep{Laureijs11} and the NASA-led \emph{Wide Field Infrared Survey Telescope} mission \citep[WFIRST,][]{Dressler12, Green12, Spergel15}, dark energy will be probed using a spectroscopic galaxy redshift survey, which will measure the galaxy clustering and redshift-space distortions of the galaxy distribution, and a photometric survey, which will be used to measure weak gravitational lensing shear of the galaxies. Specifically for the galaxy clustering measurements, these missions will measure the \emph{baryon acoustic oscillations} (BAO) in the galaxy population, which can be used as a standard ruler to probe the expansion history of the Universe \citep{Blake03,Seo03}, and redshift-space distortions, which are sensitive to the growth rate of the large-scale structure \citep{Kaiser87,Song09}.

The spectroscopic redshift surveys of the Euclid and WFIRST missions will target many tens of millions of emission line galaxies (ELGs), which will be identified using near-IR grism spectroscopy. The primary targets for the spectroscopic galaxy redshift surveys are $\halpha$-emitting galaxies between redshifts $0.9\lesssim z \lesssim 2$. The Euclid wide-area survey will cover 15,000 square degrees to an $\halpha$ line flux limit of $2\times 10^{-16}\,\ergPerSecondPerCM$ and is expected to detect $\Halpha$-emitting galaxies in the redshift range $0.9\lesssim z\lesssim 1.8$. In a similar fashion, the WFIRST High Latitude Survey (HLS) will cover $\sim2200$ square degrees to a fainter $\halpha$ line flux limit of $1\times 10^{-16}\,\ergPerSecondPerCM$ and is expected to detect $\Halpha$-emitting galaxies in a redshift range of $1\lesssim z\lesssim 2$. Due to their observing strategy, both missions are expected to detect additional emission lines belonging to galaxies outside of these redshift ranges. For instance, both Euclid and WFIRST expect to detect [OIII] emission from galaxies at $z\gtrsim 2$. In this work  we focus on the redshift surveys of $\Halpha$-emitting galaxies. The Euclid and WFIRST dark energy missions are designed to be highly complementary: the large area and shallower depth of Euclid is ideal for optimising statistical precision in cosmological measurements, whilst the smaller area and greater depth of WFIRST is ideal for understanding the systematics that are expected to dominate over measurement uncertainties.

One of the systematics that must be taken into account is \emph{galaxy bias}, which describes how galaxies trace the underlying dark matter distribution. Galaxy formation does not occur uniformly in space, but occurs primarily in the peaks of the matter density field. Galaxies are therefore biased tracers of the density field, sampling only the over-dense regions (e.g. \citealt{Kaiser84,Bardeen86}; see \citealt{Desjacques18} for a recent review). The bias, $b$, of a population of galaxies is defined according to,
\begin{equation}
\xi_{\rm gal}(r) = b^2(r)\xi_{\rm DM}(r),
\label{eq:bias}
\end{equation}
where $\xi_{\rm gal}(r)$ is the galaxy correlation function and $\xi_{\rm DM}(r)$ is the dark matter correlation function, expressed as a function of spatial separation $r$. Observational evidence indicates that different galaxy populations, identified for example by luminosity, stellar mass or morphological type, display different clustering amplitudes \citep[e.g.][]{Davis76, Dressler80, Guzzo97, Norberg01, Norberg02, Zehavi02, Zehavi11, Guo13, Skibba14, Kim15, McCracken15, Hatfield16, Favole17, Law-Smith17, Cochrane17, Cochrane18a, Durkalec18}. This has been further confirmed with simulations of galaxy formation \citep[e.g.][]{Kauffmann97, Benson00b, Orsi10, Contreras13}. As such, different galaxy populations are expected to display different bias values. This can be understood in terms of the halo model, which predicts that different galaxy populations commonly reside in dark matter haloes of different mass. The hierarchical nature of structure formation means that the bias of galaxies will additionally vary with redshift \citep[e.g.][]{Fry96, Tegmark98, Hui08,Basilakos12,Mirbabayi15}. In the non-linear regime the bias is typically scale-dependent, but in the linear regime, at scales $r\gtrsim 50\hMpc$, the bias approaches a constant value \citep{Mann98, Peacock00, Seljak00, Benson00a, Blanton00, Berlind02, Verde02}. We refer to the constant bias on large-scales as linear bias, $b(r)\sim b_{\rm lin}$.

To date, many of the clustering analyses of emission line galaxies at high redshift have been carried out using data from the High-z Emission Line Survey \citep[HiZELS,][]{Geach08,Sobral09,Sobral13} -- a ground-based, narrow band survey capable of detecting emission line galaxies up to $z\sim 9$. Using a sample of $\sim700$ $\halpha$-emitters at $z=0.84$, \citet{Sobral10} observed that the clustering strength of $\halpha$-emitters is strongly dependent on the $\halpha$ luminosity. \citet{Geach12} examined the clustering of 370 $\halpha$-emitters at $z=2.23$ to measure a bias of $2.4^{+0.1}_{-0.2}$ and estimate that the dark matter haloes hosting $\halpha$-emitters have a typical mass of $\log_{10}\left (M_h/ h^{-1}\Msol\right )=11.7\pm0.1$. \citet{Geach12} additionally construct a parametric model for the \emph{halo occupation distribution} (HOD) of their $\halpha$-emitters and attempt to fit this model to their data. Due to the size of their galaxy sample, \citet{Geach12} were unable to place strong constraints on the HOD parameters. However, despite this, they were able to constrain the effective halo mass of the $\halpha$-emitters to be $\log_{10}\left(M_{\rm eff}/\hMsol \right )=12.1^{+0.1}_{-0.2}$. Recently \citet{Cochrane17} built upon the previous HiZELS results by carrying out an extensive clustering analysis of $\sim3000$ $\halpha$-emitters at $z=0.8$, $\sim 450$ emitters at $z=1.47$ and $\sim 730$ emitters at $z=2.23$, split into  both differential and cumulative luminosity bins. Using Markov Chain Monte Carlo sampling, \citet{Cochrane17} fit a parametrised HOD to their clustering results to estimate the bias and halo mass of their $\halpha$-emitters. They find that the bias of the emitters increases both with redshift and with luminosity, and that the emitters are hosted by haloes with mass typically $\sim 10^{12}\hMsol$. 

In this work we use simulated galaxy catalogues to forecast the linear bias as a function of redshift for $\halpha$-emitting galaxies in a Euclid-like redshift survey and a WFIRST-like redshift survey. Forecasts of the galaxy bias are important for two reasons. First, understanding the bias of the $\halpha$-emitting galaxies will be crucial for the Euclid and WFIRST missions if we are to correctly infer the dark matter clustering from the galaxy clustering measurements, and thereby accurately estimate cosmological parameters \citep{Gaztanaga12,Salvador19}. \citet{Clerkin15} provide a comparison of several bias evolution prescriptions and the impact that choice of prescription has on cosmological parameter constraints. Second, the \emph{figure-of-merit} of a cosmological mission, which indicates the ability of a mission to successfully measure dark energy, is sensitive to the understanding of galaxy bias. As such forecasting of the bias is vital for helping optimise the observational strategy of a cosmological mission. For example, \citet{Orsi10} used galaxy mock catalogues built using the \textsc{Galform} semi-analytical galaxy formation model to compare the effective volumes that could be probed by an $\halpha$-selected galaxy survey and an H-band selected galaxy survey. They determine that for an $\halpha$-selected galaxy survey to probe an effective volume comparable to that of an ${\rm H_{AB}}=22$ slit-based survey, the $\halpha$-selected survey would need to probe down to a flux depth of $10^{-16}\ergPerSecondPerCM$. \citet{Orsi10} additionally provide some forecasts for the luminosity dependence of the bias of $\halpha$-emitters between $z=0$ and $z=2$, though due to the limited volume of the underlying N-body simulation they are unable to estimate the bias well into the linear regime ($r\gtrsim 50\hMpc$). 

The structure of the paper is as follows: in Section~\ref{sec:mocks} we give details of how we construct our lightcone galaxy catalogues; in Section~\ref{sec:calibration} we discuss how we calibrate our lightcone catalogues to match the luminosity function and cumulative number counts of $\halpha$-emitters; in Section~\ref{sec:linear_bias} we present our linear bias forecasts; in Section~\ref{sec:discussion} we compare the linear bias values for the Euclid-like and WFIRST-like surveys; and in Section~\ref{sec:conclusions} we summarise and conclude.

\section{Mock Catalogue Construction}
\label{sec:mocks}

For this work we use a pair of mock catalogues built using the methodology of \citet{Smith17}, who populate the dark matter haloes of the \textit{Millennium XXL} simulation \citep[MXXL,][]{Angulo12} using a set of luminosity-dependent HODs. In this section we provide further details on the construction of these mock catalogues.

\begin{figure}
  \centering
  \includegraphics[width=0.48\textwidth]{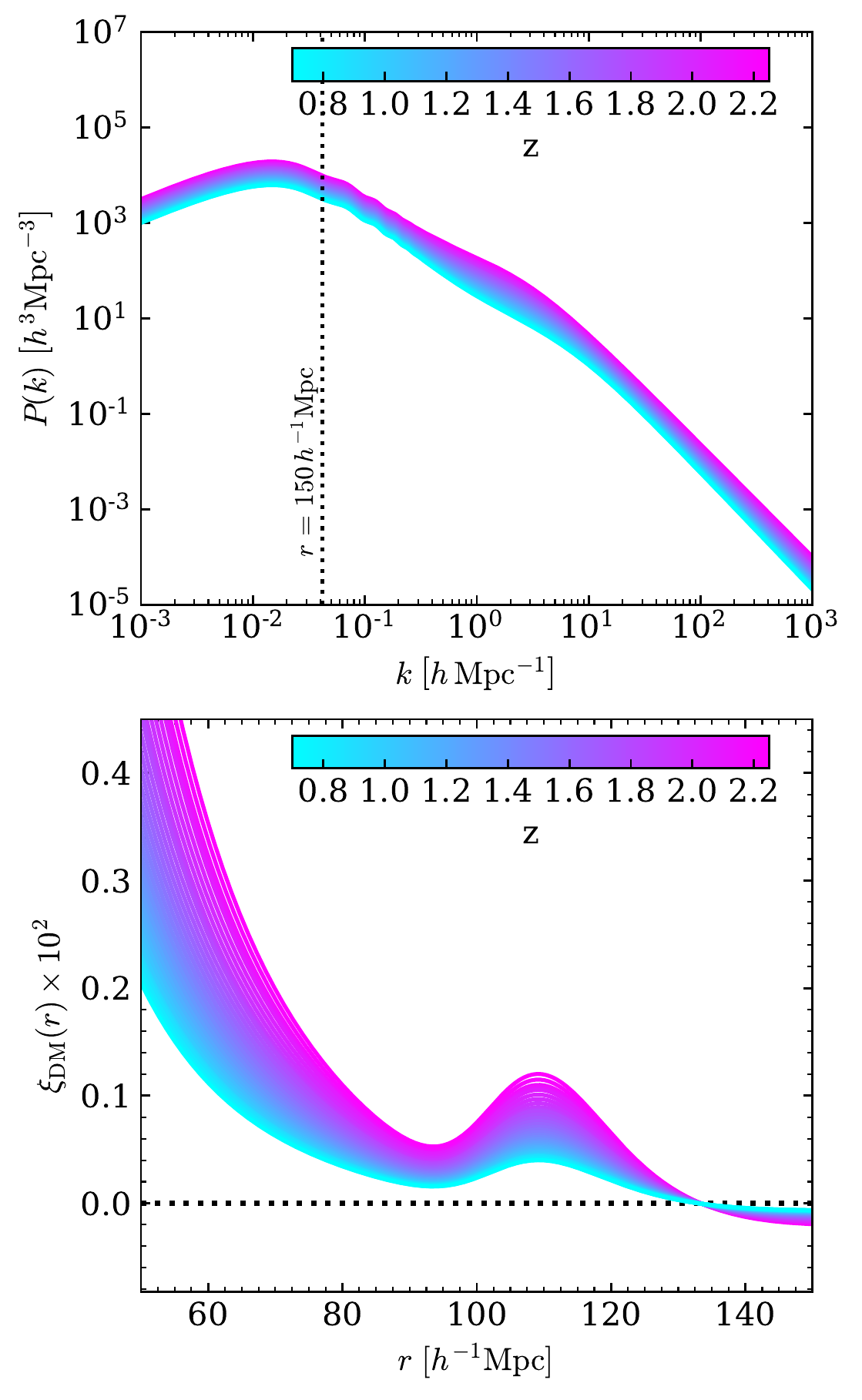}
  \caption{Non-linear matter power spectra (top panel) and dark matter two-point correlation function (bottom panel) for an MXXL cosmology for redshifts in range $0.8\lesssim z\lesssim 2$, as indicated by the inset colour bar. All power spectra and correlation functions were computed using the \textsc{CLASS} and \textsc{HaloFit} functionality in the \texttt{NBodyKit} python package \protect{\citep{Hand18}}. In the upper panel a vertical dotted line indicates the wavenumber corresponding to a co-moving distance of $150\hMpc$. In the lower panel the BAO peak is clearly visible at approximately $110\hMpc$.}
  \label{fig:theoryPk}
\end{figure}

\subsection{The Millennium XXL Simulation}
\label{sec:mxxl}

In this work we use the \emph{Millennium XXL} simulation \citep[MXXL,][]{Angulo12}, which uses $6720^3=303,464,448,000$ particles of mass  $6.17\times10^{9}\hMsol$  to follow the non-linear, hierarchical growth of dark matter structure within a cubic volume of $3\,\hGpc$. The cosmology adopted in the MXXL simulation is a  ${\rm \Lambda}$CDM cosmology with parameters identical to the cosmological parameters adopted for the \emph{Millennium Simulation} \citep{Springel05}. These parameters are: a baryon matter density $\Omega_{{\rm b}} = 0.045$; a total matter density $\Omega_{{\rm m}} = \Omega_{{\rm b}} + \Omega_{{\rm CDM}} = 0.25$, where $\Omega_{{\rm CDM}}$ is the density in cold dark matter; a dark energy density $\Omega_{{\rm \Lambda}} = 0.75$; a Hubble constant $H_0 = 100h \,{\rm km\,s}^{-1}\Mpc^{-1}$, where $h = 0.73$; a primordial scalar spectral index $n_{\rm s}=1$ and a fluctuation amplitude $\sigma_{8}=0.9$. This cosmological parameter set is consistent with the first year results from the Wilkinson Microwave Anisotropy Probe \citep{Spergel03}. We note that  this cosmology is discrepant with the latest cosmological parameters measured from the Planck mission \citep{Planck18e}, but argue that this should not be of major concern since our ignorance in the modelling of galaxy formation will dominate over any discrepancies in cosmology. In Fig.~\ref{fig:theoryPk} we show, for reference, the non-linear matter power spectrum and corresponding real-space correlation function as computed for the MXXL cosmology using the open source \texttt{Nbodykit}\footnote{\url{https://github.com/bccp/nbodykit}} python package \citep{Hand18}. The BAO peak is clearly visible and, given the paremeters of the simulation, is located at approximately $110\hMpc$.

Particle positions and velocities are stored in 63 snapshots at fixed epochs spaced approximately logarithmically in expansion factor between $z=20$ and $z=0$, as was done for the Millennium Simulation \citep[see ][]{Lemson06}. Halo merger trees have also been constructed using the \textsc{Subfind} structure finding algorithm \citep{Springel01}. \citet{Smith17} additionally constructed friends-of-friends (FOF) merger trees using the most massive subhalo in each FOF group. Further details regarding the MXXL simulation can be found in \citet{Angulo12}.

\citet{Smith17} constructed a lightcone catalogue of dark matter haloes from the MXXL simulation using the methodology described in \citet{Merson13}. An observer is first placed in the simulation box at a randomly chosen position and the simulation box is replicated about the observer so as to generate a sufficiently large volume reaching out to the user-specified redshift. Using the positions of haloes and their associated descendants, one can determine the pair of snapshots between which any particular merger tree intercepted the past lightcone of the observer (if it intercepts the lightcone at all). Once the pair of snapshots has been identified, a binary search algorithm is used to interpolate along the trajectory of the halo to compute the exact location at which the halo crosses the lightcone. Interpolation of the halo trajectories between pairs of snapshots is done using cubic interpolation, with the initial position and velocity of a halo and its corresponding descendant used to set the boundary conditions. The lightcone that \citet{Smith17} built is all-sky and extends out to redshift $z=2.2$.

\subsection{Populating the MXXL dark matter haloes}
\label{sec:populating_the_mxxl_haloes}

Ideally we would like to populate the haloes of the MXXL simulation with galaxies by running a physically motivated galaxy formation model directly on the MXXL halo merger trees. Unfortunately the mass resolution of the MXXL simulation is too poor, with a minimum resolved halo mass of $1.22\times 10^{11}\hMsol$, such that running a semi-analytical galaxy formation model on the MXXL trees directly would yield an unrealistic galaxy population \citep[e.g. see][]{Angulo14}. We note, however, that this minimum halo mass is just under an order of magnitude smaller than the typical halo mass for $\halpha$-emitters estimated by \citet{Geach12}.

To populate the MXXL haloes we therefore use the pipeline presented in \citet{Smith17}. This pipeline follows the methodology of \citet{Skibba06}, whereby haloes are populated with galaxies of different luminosities by random sampling of probability distribution functions created from sets of luminosity-dependent halo occupation distributions (HODs) that evolve with redshift. 

\citet{Smith17} originally developed this pipeline for the purpose of constructing an $r$-band selected galaxy catalogue to mimic the Dark Energy Spectroscopic Instrument Bright Galaxy Survey \citep[DESI BGS,][]{DESI16a}. For broad-band photometrically-selected samples such as this, the HOD is well understood and can be parametrised easily, \citep[e.g.][]{Zehavi11,Zhai17}. However, the HOD of $\halpha$-emitting galaxies is much less well understood, although some parametrisation has been attempted based upon clustering results from HiZELS \citep{Geach12, Cochrane17}. Some attempts have also been made to fit the HOD parameters for simulated star-forming galaxies and simulated [OII]-emitters \citep{Contreras13,Gonzalez-Perez18,Cochrane18b}.

Rather than adopting any parametrisation for the HODs of $\halpha$-emitting galaxies, we again consider using a semi-analytical galaxy formation model, but in this case using a model to generate a library of luminosity-dependent HODs that we can interpolate over as a function of luminosity and redshift. We choose to use the open source galaxy formation model \galacticus{} \citep{Benson12}.

\subsubsection{The \galacticus{} galaxy formation model}
\label{sec:galacticus}

The {\galacticus}\footnote{Here we use version 0.9.4 of {\galacticus}, which is publicly available from: \url{bitbucket.org/galacticusdev/galacticus}. The Mercurial hash I.D. for the particular revision used is: 4787d94cd86e.} semi-analytical galaxy formation model is designed to follow the formation and evolution of a galaxy population within a merging hierarchical distribution of dark matter haloes. The astrophysics governing the baryonic processes occurring within the dark matter haloes is described using sets of coupled ordinary differential equations (ODEs). These processes include the rate of radiative gas cooling, the quiescent star formation rate, the chemical enrichment of the stellar and gaseous components, as well as the regulation of feedback processes from supernovae and active galactic nuclei. By calling the ODE solver within \galacticus{} at various epochs, one can compute the star formation histories of a population of galaxies from high redshift to the present day. Given a stellar initial mass function (IMF), these histories can be convolved with a single stellar population synthesis model to generate a spectral energy distribution for each galaxy, with which we can compute photometric luminosities for a specified set of filter transmission curves. By default \galacticus{} adopts the Flexible Stellar Population Synthesis code of \citet{Conroy10}, with a \citet{Chabrier03} IMF.

Emission line luminosites are computed for the \galacticus{} galaxies by interpolating over tabulated libraries generated from the \cloudy{} photo-ionisation code \citep{Ferland13}. These libraries store emission line luminosities as a function of (i) the ionizing continuua luminosities for various species (\textsc{Hi}, He\textsc{i} and \textsc{Oii}), (ii) the hydrogen gas density, (iii) the metallicity of the interstellar medium and (iv) the volume filling factors of \textsc{Hii} regions. All of these properties can be computed for the galaxies from \galacticus{} and so by interpolating over the \cloudy{} tables we can obtain emission line luminosities that are consistent with the other galaxy properties. Further details regarding the calculation of the emission line luminosities are provided in \citet{Merson18}. 

Typically the \galacticus{} model parameters are calibrated to reproduce numerous observational statistics of the galaxy population, in particular those statistics of the local Universe that are the most tightly constrained. The version of \galacticus{} that we use here is the same as the version used in \citet{Merson18}, which employs the standard parameter set.  This version of \galacticus{} has been calibrated to reproduce a range of galaxy statistics, particularly those in the local Universe. Emphasis has been placed on reproducing the $z=0$ galaxy stellar mass function from the Sloan Digital Sky Survey, as measured by \citet{Li09}. Full details of the calibration procedure for \galacticus{} can be found in \citet{Benson14} and \citet{Knebe15}.

\begin{figure*}
  \centering
  \includegraphics[width=0.99\textwidth]{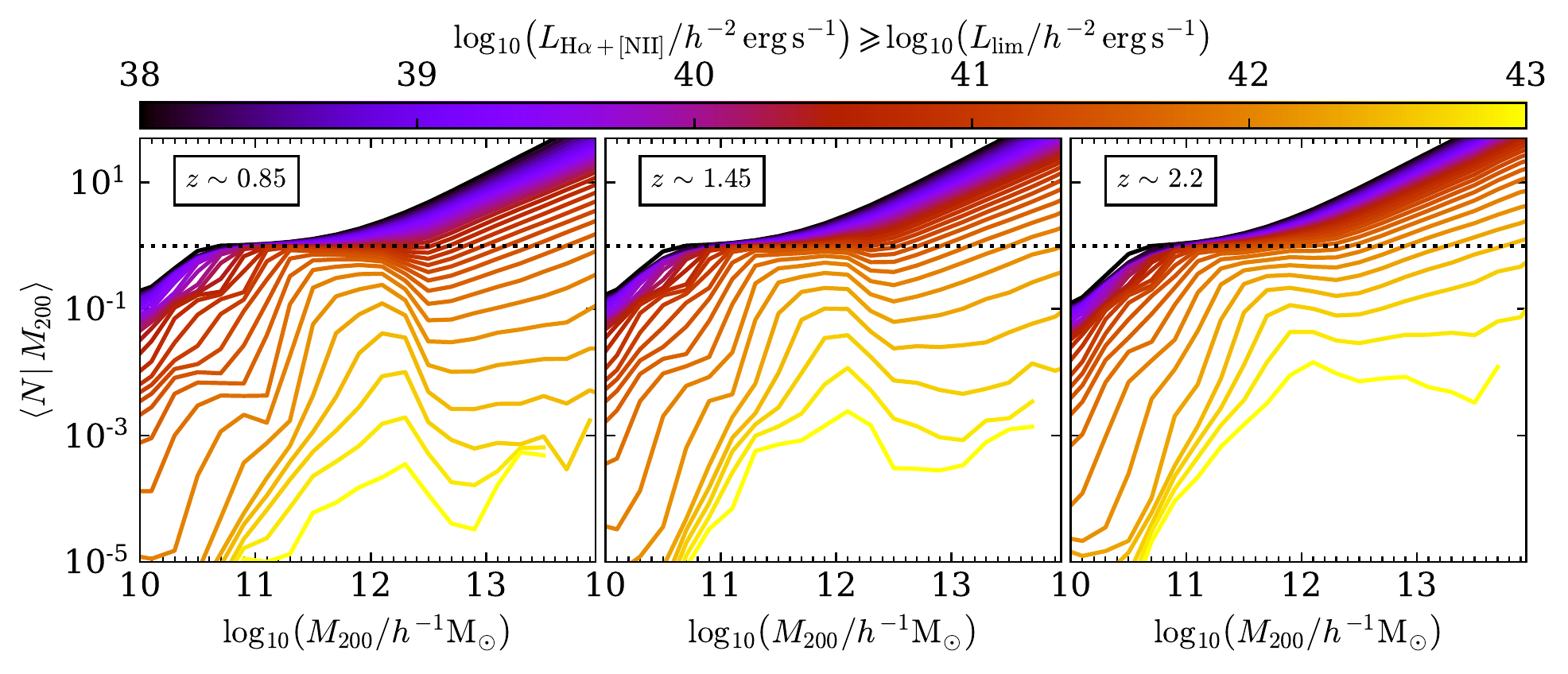}
  \caption{Halo Occupation Distributions (HODs) as measured from running \galacticus{} on three snapshots of the Millennium Simulation: $z=0.85$ (left-hand panel), $z=1.45$ (centre panel) and $z=2.2$ (right-hand panel). The various colours of the lines indicate the blended $\halpha+\nii$ luminosity limit used to select the galaxies, as indicated in the colour bar. The luminosities have not been attenuated due to dust. The halo mass assumed is the mass within an over-density with average density corresponding to 200 times the mean density of the Universe.}
  \label{fig:hizelsHalphaHOD_luminositySelection}
\end{figure*}

\subsubsection{Generating a library of luminosity-dependent HODs}
\label{sec:generating_hods}
As we have previously mentioned, the poor mass resolution of the MXXL simulation means that we are unable to run \galacticus{} directly on the MXXL merger trees. Instead we run the model on the halo merger trees of the Millennium Simulation, which we remind the reader has a cosmology identical to that of the MXXL simulation. Populations of galaxies are output at 31 snapshots with redshifts in the range $0.7<z<2.2$ (i.e. completely spanning the redshift ranges of Euclid and WFIRST). For each snapshot we split the galaxies into 31 cumulative luminosity-limited samples, with blended luminosity limits evenly separated in log-space between $L_{\halpha+[NII]}=10^{38} h^{-2}\ergPerSecond{}$ and  $L_{\halpha+[NII]}=10^{43} h^{-2}\ergPerSecond{}$. We select galaxies by their blended $\halpha{}+\nii$ luminosity as the resolution of the grisms proposed for Euclid will be too poor to deblend the $\halpha$ line and the $\nii$ doublet. See \citet{Faisst18} and \citet{Martens19} for discussions on the impact that $\halpha$ and $\nii$ line blending can have on Euclid and WFIRST cosmological measurements. To introduce $\nii$ contamination into the $\halpha$ luminosities from \galacticus{}, we follow the methodology adopted by \citet{Merson18}. The blended luminosities have no dust attenuation applied. For each luminosity-limited sample we construct the HOD by computing the mean number of central and satellite galaxies in haloes binned by halo mass, $M_{200}$, which we define as the mass within an over-density with an average density corresponding to 200 times the mean density of the Universe. Each HOD is measured in 26 mass bins evenly separated in log-space between $M_{200}=10^{9.7}\hMsol$ and $M_{200}=10^{14.7}\hMsol$. At this stage we now have a tabulated library of luminosity-dependent HODs over which we can interpolate as a function of halo mass, luminosity and redshift. Note that in the library we store three sets of HODs: the HODs for the central galaxies only, the HOD for the satellite galaxies only, and the combined HOD.

In Fig.~\ref{fig:hizelsHalphaHOD_luminositySelection} we show the combined HODs for our 31 blended luminosity-limited samples as measured from running \galacticus{} on three of the redshift snapshots of the Millennium Simulation. The shapes of these HODs are consistent with the shapes of emission line luminosity-selected HODs presented elsewhere in the literature \citep[e.g.][]{Geach12, Contreras13, Gonzalez-Perez18,Cochrane18b}. For the faintest luminosity limits considered, the HODs show a smooth step-like function, similar to the HODs for galaxy samples selected using broad-band photometry. For masses below $10^{11}\,h^{-1}\Msol$ there is evidence for a secondary step feature, particularly towards lower redshift. However, we note that at such low masses we are approaching the resolution limit of the Millennium Simulation and so the extent to which this secondary step is real is uncertain. 

Towards brighter luminosity limits the amplitude of the HOD decreases with the step-like shape of the HOD disappearing as the number of galaxies with bright $\halpha$ luminosity rapidly declines, particularly for very massive haloes. For the brightest luminosity limits considered, the distribution of central galaxies is much more peaked, with a maximum value at a halo mass of approximately $10^{12}\,h^{-1}\Msol$. This is understandable as $\halpha$ emission occurs in regions of ongoing star-formation and so $\halpha$ luminosity is correlated with galaxy star-formation rate, which, due to feedback from active galactic nuclei, does not increase monotonically with increasing halo mass. The quenching of satellite galaxies in massive haloes also leads to a decline in the number of satellite galaxies with bright $\halpha$ emission. In haloes with mass above $10^{12}\hMsol$, for $L_{\halpha+\nii}\gtrsim 10^{41}\ergPerSecond$ the satellite fraction is below approximately 5 per cent. Indeed the occupation number for satellite galaxies still appears to follow a power law relation, but with a slope that decreases with increasing luminosity. The impact of these quenching mechanisms is that for bright luminosities the HOD turns over for haloes with mass $M_{200}\gtrsim 10^{12}\,h^{-1}\Msol$. We can see from  Fig.~\ref{fig:hizelsHalphaHOD_luminositySelection} that the turn-over is more significant towards lower redshifts. In addition we can see that the amplitude of the HOD decreases with increasing luminosity limit, suggesting that galaxies that are bright in $\halpha$-emission are becoming increasingly rare. This change in the amplitude of the HOD, along with the change in the strength of the turnover, is consistent with the observed decline in the global star-formation rate density towards the present day \citep[e.g.][]{Madau98,Arnouts05,Ly07,Shim09,Brammer11,Madau14}. 

\begin{figure*}
  \centering
  \includegraphics[width=0.99\textwidth]{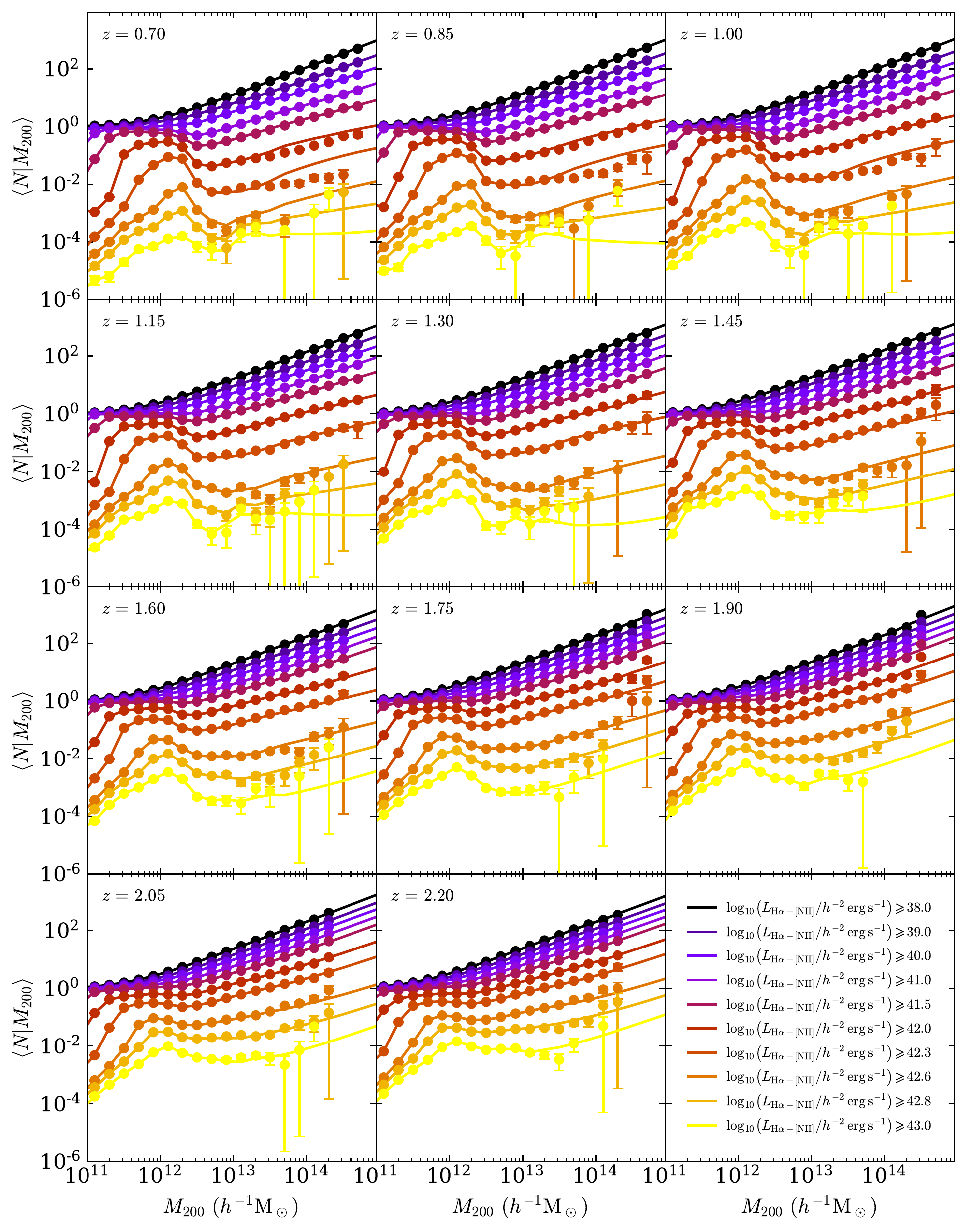}
  \caption{Tabulated halo occupation distributions (HODs) from \galacticus{} for 10 luminosity-selected samples of $\halpha$-emitting galaxies shown for various redshifts between $z=0.7$ and $z=2.2$. Solid lines show the functional fits to the HODs. The various colours of the lines indicate the blended $\halpha+\nii$ luminosity limit used to select the galaxies, as indicated in the lower right-hand panel. Luminosities have not had any dust attenuation applied. The halo mass assumed is the mass within an over-density with average density corresponding to 200 times the mean density of the Universe.}
  \label{fig:halphaHOD_fits}
\end{figure*}


\begin{table*}
\centering
\caption{Specifications for the lightcone catalogues considered in this work. Note that the Euclid-like lightcone is constructed by applying a completeness limit to the MXXL-15K. The WFIRST-like lightcone is derived from the MXXL-2K lightcone in a similar manner. The sky areas for the MXXL-15K and the Euclid-like lightcones have been rounded to the nearest integer -- in reality these lightcones each have an area of $(3/8)\times4\pi$ steradians. }

\begin{tabular}{|c|c|c|c|c|c|c|}
\hline
Catalogue&Flux Limit&Right&Declination&Area&Redshift&Completeness\\
Name&$\left [{\rm erg}\,{\rm s}^{-1}\,{\rm cm}^{-2}\right ]$&Ascension&&$\left [{\rm deg}^2\right ]$&Range&\\
\hline\hline
MXXL-15K&$2\times10^{-16}$&$0^{\circ}\leqslant\alpha\leqslant270^{\circ}$&$0^{\circ}\leqslant \delta\leqslant 90^{\circ}$&15470&$0.7\leqslant z<2.2$&1.0\\
Euclid-like&$2\times10^{-16}$&$0^{\circ}\leqslant\alpha\leqslant270^{\circ}$&$0^{\circ}\leqslant\delta\leqslant 90^{\circ}$&15470&$0.9\leqslant z<1.9$&0.45\\
MXXL-2K&$1\times10^{-16}$&$50^{\circ}\leqslant\alpha\leqslant100^{\circ}$&$-20^{\circ}\leqslant\delta\leqslant 20^{\circ}$&2000&$0.7\leqslant z<2.2$&1.0\\
WFIRST-like&$1\times10^{-16}$&$50^{\circ}\leqslant\alpha\leqslant100^{\circ}$&$-20^{\circ}\leqslant\delta\leqslant 20^{\circ}$&2000&$1.0\leqslant z<2.0$&0.7\\
\hline
\end{tabular}
\label{tab:surveys}
\end{table*}

\subsubsection{Placing galaxies into the MXXL haloes}
\label{sec:generating_galaxies}

Having generated a library of HODs we now randomly sample the HODs to populate the haloes in the MXXL halo lightcone. 
Note that by doing so, we lose any environmental-dependence (i.e., assembly bias). This does not matter for our present purposes, since we are only interested in measuring the linear bias on large scales. 

In order to sample the HODs we will need to interpolate over the library as a function of halo mass, luminosity and redshift. At low masses we can interpolate linearly in log-space between the tabulated values. However, for the most massive haloes in the MXXL lightcone, approaching $10^{15}\hMsol$, we will need to extrapolate beyond the halo mass range of the tabulated HODs. The limited volume of the Millennium Simulation means that our library of tabulated HODs from \galacticus{} extends only to $M_{200}=10^{14.7}\hMsol$. In order to extrapolate to the higher halo masses we therefore decide to fit smoothed functional forms to the tabulated occupation distributions for the central galaxies and the satellite galaxies. Details of this fitting procedure are provided in Appendix~\ref{sec:hod_component_fits}. 

The functional fits to the tabulated HODs are shown in Fig.~\ref{fig:halphaHOD_fits}. In the majority of cases the fits provide a reasonably good description of the tabulated HODs. In some cases, particularly at lower redshifts, the noise in the tabulated HODs for the brightest luminosity limits leads to overlap between the HODs of different luminosity limits. In these instances fits to the HODs will cross such that interpolation between the HODs as a function of luminosity would yield a constant HOD with increasing luminosity, which can be a problem in the random sampling of the HODs. This issue can be resolved by applying small vertical shifts of less than 0.3 dex to the amplitude of the HODs of the brightest luminosity limits. Given the rarity of bright $\halpha$-emitting galaxies in massive haloes at lower redshift, applying these offsets leads to only negligible differences in the number density of galaxies in our lightcone catalogues \citep{SmithThesis}.

Galaxies are then placed into each halo using a methodology presented in \citet{Smith17} and \citet{SmithThesis}. The first step in the process is to select a minimum luminosity, $L_{\rm min}$, for the lightcone catalogue. Here we adopt $L_{\rm min}=10^{40}h^{-2}\ergPerSecond{}$, which, at the lowest redshift in the mock, is about an order of magnitude fainter than the luminosity limit for a WFIRST-like selection. Following this we identify all of the haloes that could host a central galaxy with $L_{\halpha+\nii}\geqslant L_{\rm min}$ by interpolating over the HOD with the corresponding luminosity threshold. We assign a central galaxy to each of these haloes for which $x_1<\left <N_{\rm cen}\left (>L_{\rm min}|M_{200},z\right )\right >$, where $\left <N_{\rm cen}\left (>L_{\rm min}|M_{200},z\right )\right >$ is the mean number of central galaxies brighter than $L_{\rm lim}$ that could be found in a halo of mass $M_{200}$ at redshift $z$, and $x_1$ is a random number drawn from a uniform distribution $0\leqslant x_1\leqslant 1$. Central galaxies are placed at the centre-of-mass of the haloes, with a velocity equal to the velocity of their host halo. The luminosity of the galaxy, $L_{\rm cen}$, is determined by interpolating over the library of HODs to find the luminosity that satisfies,
\begin{equation}
    \frac{\left <N_{\rm cen}\left (>L_{\rm cen}|M_{200},z\right )\right >}{\left <N_{\rm cen}\left (>L_{\rm min}|M_{200},z\right )\right >}=x_2,
    \label{eq:random_sample_cen}
\end{equation}
where $x_2$ is a second random number, again drawn from a  $0\leqslant x_2\leqslant 1$ uniform distribution.

The number of satellite galaxies assigned to a halo is determined by drawing from a Poisson distribution with a mean equal to $\left <N_{\rm sat}\left (>L_{\rm min}|M_{200},z\right )\right >$, where $\left <N_{\rm sat}\left (>L_{\rm min}|M_{200},z\right )\right >$ is the mean number of satellite galaxies brighter than $L_{\rm lim}$ that could be found in a halo of mass $M_{200}$ at redshift $z$. Satellite galaxies are placed randomly within the halo following a \citet{Navarro97} profile and are given random velocities relative to the velocity of the halo. These velocities are drawn from an isotropic Maxwell-Boltzmann distribution with a line-of-sight velocity dispersion, $\sigma^2\left (M_{200}\right )$, given by,
\begin{equation}
    \sigma^2\left (M_{200}\right ) = \frac{G M_{200}}{2R_{200}},
    \label{eq:satellite_velocity_dispersion}
\end{equation}
where $R_{200}$ is the radius of a sphere, centred on the halo centre-of-mass, within which the enclosed density is 200 times the mean density of the Universe. The luminosities of the satellite galaxies are assigned using random number generation in the same manner as was done for assigning the luminosities of the central galaxies.

\subsection{Catalogue specifications}
\label{sec:specifcations}

We employ the methodology laid out above to construct two lightcone catalogues: one covering an area of $15\,470\,{\rm deg}^2$ (MXXL-15K), and another covering a smaller area of $2\,000\,{\rm deg}^2$ (MXXL-2K). Note that in reality the MXXL-15K lightcone covers $3/8^{\rm ths}$ of the sky. The specifications for these two lightcones are shown in Table~\ref{tab:surveys}. With these two lightcones we are able to simulate a Euclid-like survey and a WFIRST-like survey (see $\S$~\ref{sec:catalogue_preparation}).

At this point we note that due to the wide-angle area of the lightcones and their redshift extent, as well as the finite volume of the MXXL simulation, we expect these lightcones to contain repeated structures. Given the cosmology of the simulation, \citet{Smith17} determined that an observer placed at the very centre of the box would not observe any repeated structures for $z\lesssim0.5$. Therefore we expect to observe repeated structures, although we do not expect this to significantly affect our clustering measurements. However, we should expect this feature to lead to under-estimates in the uncertainties on our clustering measurements, and therefore the uncertainties on the bias estimates.
Note that those effects should be negligible, since the mass density variance is very small on scales larger than $3\hGpc$.

\section{Mock Catalogue Calibration}
\label{sec:calibration}

\begin{figure*}
  \centering
  \includegraphics[width=0.98\textwidth]{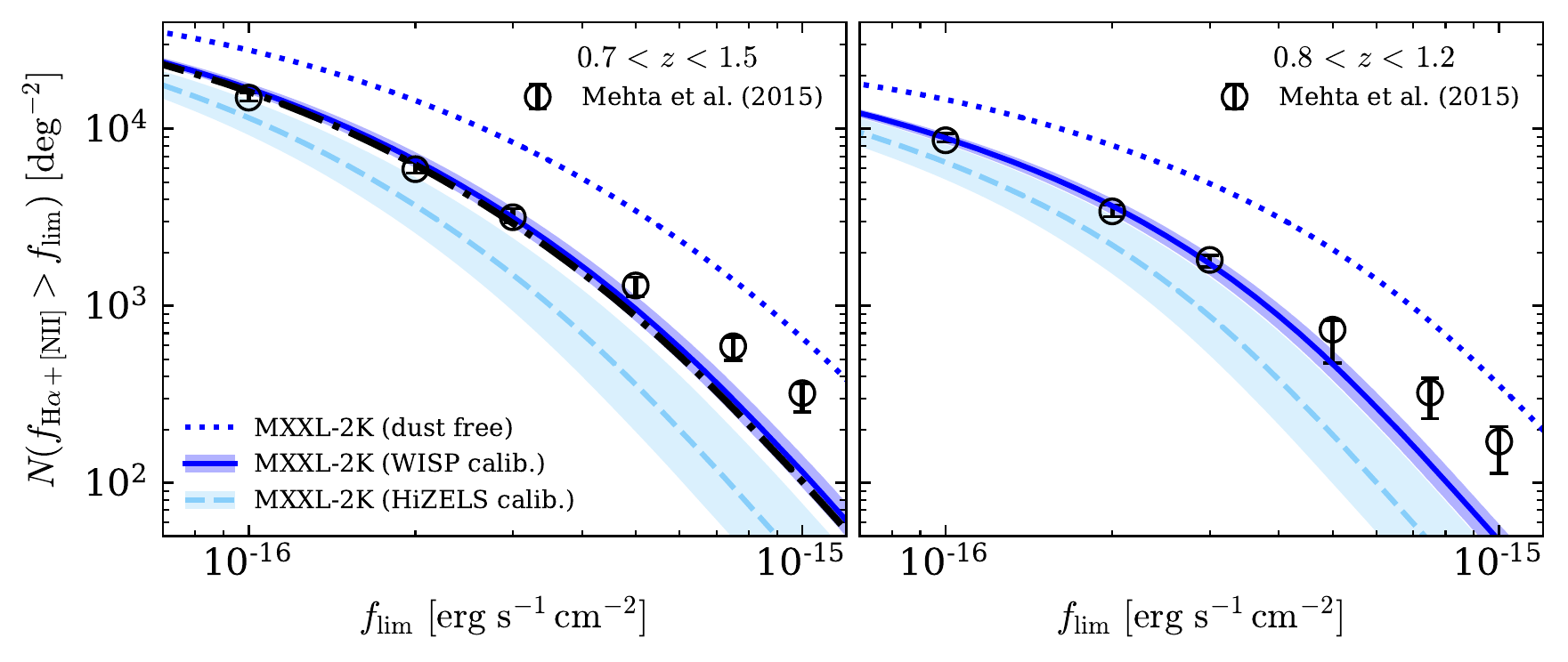}
  \caption{Cumulative number counts measured from the MXXL-2K lightcone over the redshift range $0.7<z<1.5$ (left-hand panel) and the redshift range $0.8<z<1.2$ (right-hand panel). Black data points show the WISP number counts measured by \protect\citet{Mehta15}. In each panel the blue dotted line shows the MXXL-2K number counts assuming no dust attenuation. The blue solid lines show the number counts for the MXXL galaxies when assuming a dust attenuation, $A_{\halpha}$, that leads to the best match to the WISP counts in that redshift range. In the left-hand panel the black dot-dashed line shows the number counts obtained when assuming the value for $A_{\halpha}$ that leads to the best match to the counts over the redshift range $0.8<z<1.2$. The light blue dashed lines show the number counts for the MXXL galaxies in the corresponding redshift range assuming a HiZELS-calibrated dust attenuation. Shaded regions show the spread in the counts given the corresponding uncertainty in $A_{\halpha}$ (see text for details). All fluxes correspond to blended $\halpha+\nii$ fluxes.}
  \label{fig:lightconeCounts}
\end{figure*}

\begin{figure}
  \centering
  \includegraphics[width=0.45\textwidth]{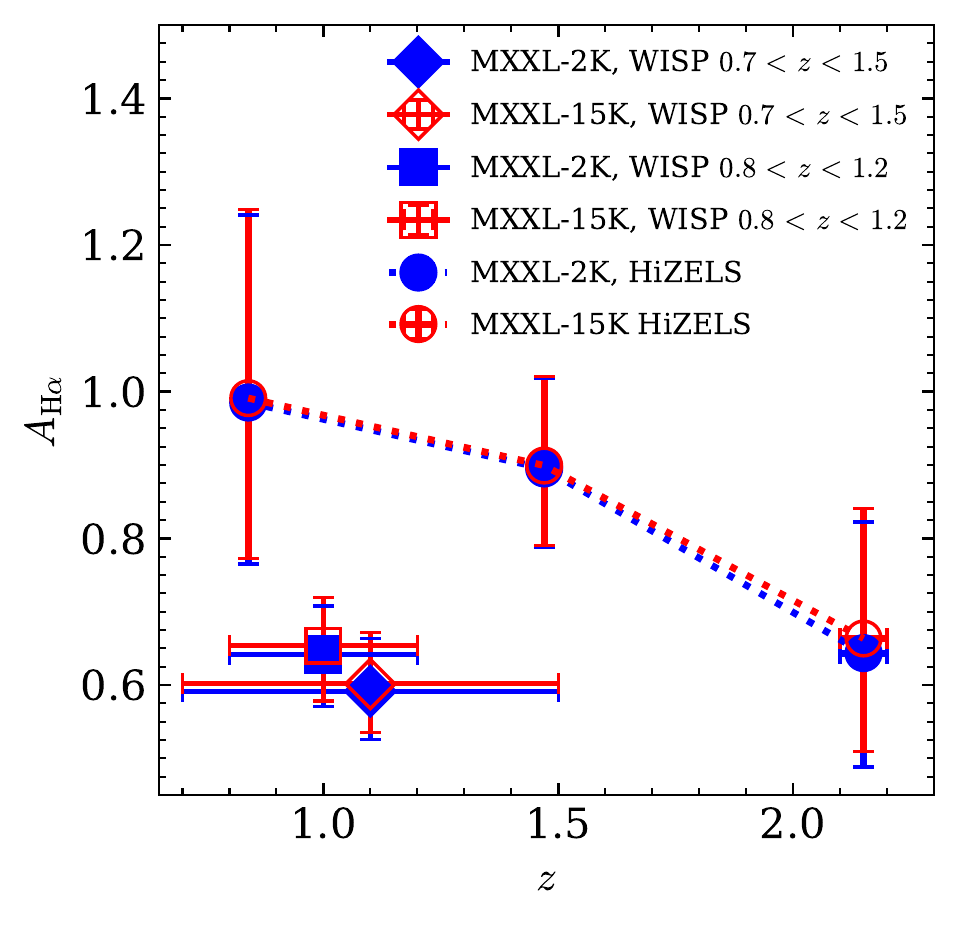}
  \caption{Dust attenuation values, $A_{\rm \halpha}$, recovered from our calibration procedure. Filled symbols correspond to the MXXL-2K lightcone, whilst empty symbols correspond to the MXXL-15K lightcone. Circles show the attenuation values required for a lightcone to reproduce the HiZELS luminosity functions, squares show the attenunation values requred to reproduce the WISP number counts in the redshift range $0.8<z<1.2$ and diamonds show the attenunation values requred to reproduce the WISP number counts in the redshift range $0.7<z<1.5$. Uncertainties in the attenuation correspond to the range of  $A_{\rm \halpha}$ values whose $\chi^2$ value is within one standard deviation of the minimum $\chi^2$ value. Error bars in the redshift direction indicate the width of the redshift range used to compute the counts or luminosity function.}
  \label{fig:dustAttenuation}
\end{figure}

Before we can compute the galaxy bias it is necessary to calibrate the lightcone catalogues so that they reproduce, as closely as possible, existing observations of $\halpha$-emitting galaxies. To calibrate the mock catalogue we apply dust attenuation to the galaxy luminosities, using $\chi^2$ minimisation to determine the dust attenuation, $A_{\rm \halpha}$, that leads to the best agreement with a particular observational dataset. The dust attenuation $A_{\rm \halpha}$ relates the un-attenuated luminosity, $L^0_{\rm \halpha}$, and attenuated luminosity, $L^{\rm att}_{\rm \halpha}$, according to,
\begin{equation}
    \log_{10}\left (L^{\rm att}_{\rm \halpha}\right ) = \log_{10}\left ( L^0_{\rm \halpha}\right ) - 0.4 A_{\rm \halpha}.
\label{eq:attenuation}
\end{equation}

Here we choose to calibrate the catalogues to match the luminosity functions and cumulative number counts of $\halpha$-emitters. Being able to reproduce the luminosity function helps ensure that we have the correct number density, whilst reproducing the cumulative number counts confirms that we have the correct total number of galaxies \citep[e.g.][]{Pozzetti16,Valentino17,Merson18}. 

\subsection{Observational datasets}
\label{sec:observations}

For calibration of the MXXL-15K and MXXL-2K lightcones we adopt (i) the cumulative number counts of $\halpha$-emitters from the Wide Field Camera 3 (WFC3) Infrared Spectroscopic Parallels Survey \citep[WISP,][]{Atek10, Atek11} as measured by \citet{Mehta15}, and (ii) the luminosity function of $\halpha$-emitters from HiZELS as measured by \citet{Sobral13}.

The WISP survey is a pure-parallel near-infrared grism spectroscopic survey that was carried out using the G141 ($1.2\,-\,1.7\mu{\rm m}$, $R\sim130$) and G102 ($0.8\,-\,1.2 \mu{\rm m}$, $R\sim 210$) grisms on the Hubble Space Telescope WFC3. Given the wavelength ranges of these grisms, WISP is able to directly detect the $\halpha$ emission line in galaxies out to $z\lesssim1.5$. Note that the Euclid and WFIRST observing strategies will be comparable to the WISP observing strategy. The survey targeted emission line galaxies in several hundred high-latitude fields. Estimates for the stellar mass of a subset of the WISP galaxies analysed by \citet{Dominguez13}, suggest these galaxies have stellar masses in the approximate range $10^{7}\hMsol\lesssim M_{\star}\lesssim 10^{11}\hMsol$, assuming a \citet{Chabrier03} initial mass function. For calibration we use the most recent published cumulative number counts measured by \citet{Mehta15} who in their analysis used galaxies from 52 separate fields covering an area of $182\, {\rm arcmin}^2$ with a quality flag $<16$ down to a flux limit of $(3-5)\times 10^{-17}\ergPerSecondPerCM$. Specifically, the counts that we use correspond to the cumulative counts of $\halpha$-emitters without any correction for $\nii$ contamination, i.e. counting the galaxies according to their $\halpha+\nii$ blended flux (see Table~5 of \citealt{Mehta15}). We consider the blended flux counts in the redshift ranges $0.7 < z < 1.5$ and $0.8 < z < 1.2$.

In contrast to the WISP survey, HiZELS is a ground-based panoramic survey of emission line galaxies covering several square degrees that was carried out on the United Kingdom Infrared Telescope, the Subaru Telescope and the Very Large Telescope. Custom-made ${\rm NB}_J$, ${\rm NB}_H$, ${\rm NB}_K$ and ${\rm NB921}$ narrow band filters were employed to target emission line galaxies out to $z\sim 9$. The specifications of the narrow band filters allowed $\halpha$-emitters to be detected in thin redshift slices centred approximately at $z=0.4$, $0.84$, $1.47$ and $2.23$. Further details regarding the survey design can be found in \citet{Geach08}, \citet{Sobral09}, \citet{Sobral12} and \citet{Sobral13}. Estimates by \citet{Cochrane18a} of the stellar masses of the subset of HiZELS galaxies used in their clustering analysis suggest that these galaxies have stellar masses in the approximate range $10^{8}\hMsol\lesssim M_{\star}\lesssim 10^{11}\hMsol$, assuming a \citet{Chabrier03} initial mass function. For our calibration we use the completeness corrected luminosity functions of $\halpha$-emitters presented in Table~4 of \citet{Sobral13}. For their analysis, \citet{Sobral13} selected $\halpha$-emitters using a combination of colour-colour selections and photometric redshifts when available.

\begin{figure*}
  \centering
  \includegraphics[width=0.98\textwidth]{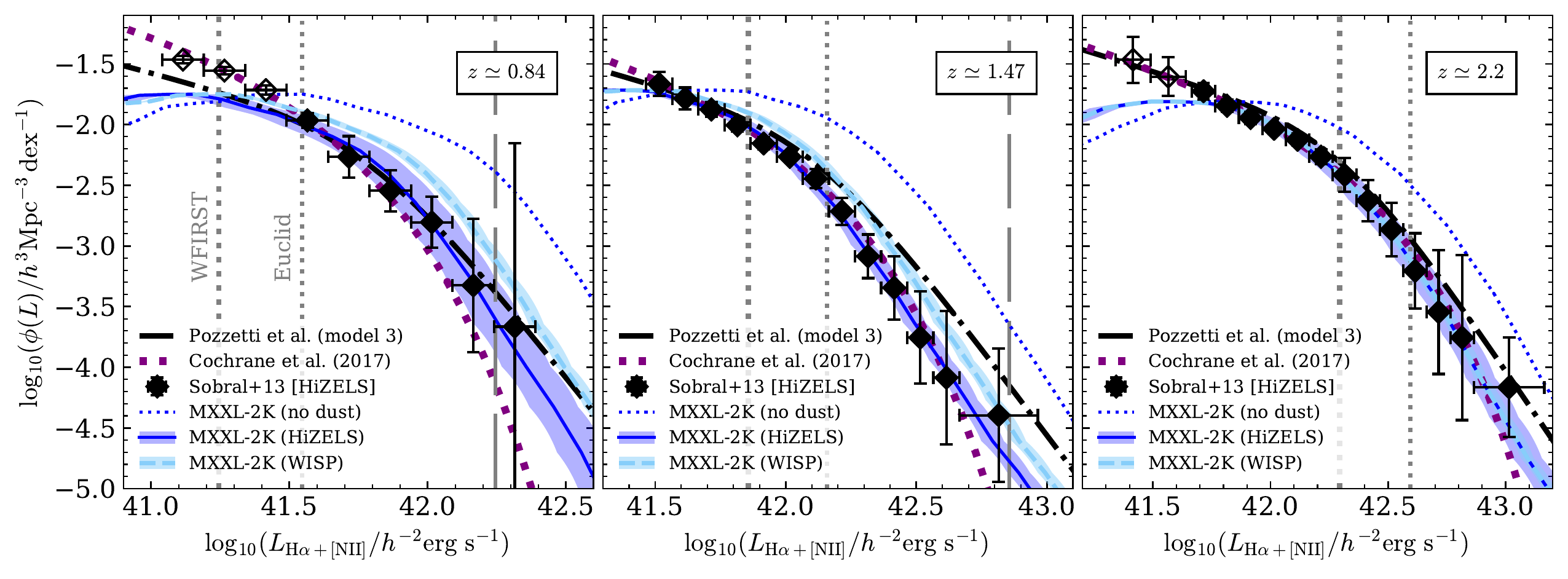}
  \caption{Luminosity functions for $\halpha$-emitting galaxies as measured from the MXXL-2K lightcone and compared to the HiZELS luminosity functions \protect\citep{Sobral13} measured in three redshift slices centred at: $z\simeq0.84$ (left-hand panel), $z\simeq1.47$ (centre panel) and $z\simeq2.2$ (right-hand panel). In each panel the blue dotted line shows the luminosity functions of the MXXL galaxies computed using the dust-free luminosities. The solid blue line shows the luminosity function for the MXXL galaxies computed assuming the dust attenuation, $A_{\halpha}$, that leads to the best match to the HiZELS luminosity function at that particular redshift. The symbols show the HiZELS observations from \protect\citet{Sobral13}, with the empty symbols corresponding to the luminosity bins that were excluded from the $\chi^2$ minimisation procedure due to incompleteness of the MXXL-2K lightcone. The light blue dashed line shows the luminosity function obtained assuming the dust attenuation obtained from calibration to the $0.8<z<1.2$ WISP number counts. Shaded regions indicate the range of allowable luminosity functions given the uncertainty on the dust attenuation  $A_{\halpha}$ (see text for details). The black dot-dashed line shows the luminosity function predicted by the third of the three empirical models presented by \protect\citet{Pozzetti16}. The thick dotted purple line shows the \protect\citet{Schechter76} fits from \protect\citet{Cochrane17}. All luminosities correspond to blended $\halpha+\nii$ luminosities. The HiZELS and \protect\citet{Pozzetti16} luminosities have been shifted by one magnitude faintwards to re-introduce dust attenuation and boosted by an additional factor of $4/3$ to account for $\nii$ contamination. Vertical dotted lines indicate the luminosities that correspond at each redshift to the Euclid and WFIRST flux limits. Vertical dashed lines indicate the luminosities corresponding to a flux of $1\times 10^{15}\ergPerSecondPerCM$.}
  \label{fig:lightconeLF}
\end{figure*}

\subsection{Number counts calibration}
\label{sec:calibration_counts}
As mentioned in the previous subsection, for calibrating the number counts of the lightcone we use the WISP cumulative number counts from \citet{Mehta15}, who measured the number counts in two broad redshift bins, $0.7<z<1.5$ and $0.8<z<1.2$. For each of these redshift bins we compute the number counts using the dust-free luminosities from the lightcone catalogue and use $\chi^2$ minimisation to determine the optimum value for $A_{\rm \halpha}$ needed to shift these counts into agreement with the observations. Note that this approach assumes that $A_{\rm \halpha}$ has a single fixed value for each redshift bin. We note that \citet{Dominguez13} observe a dust attenuation that varies with observed $\halpha$ luminosity, though their attenuation estimates are consistent with a fixed value attenuation that is independent of luminosity (see also \citealt{Merson18}).

The results of our calibration procedure are shown in Fig.~\ref{fig:lightconeCounts} where we compare the number counts for the MXXL-2K lightcone to the observed cumulative number counts from \citet{Mehta15} in the redshift ranges $0.7<z<1.5$ (left-hand panel) and $0.8<z<1.2$ (right-hand panel). In each panel the dotted blue line shows the number counts assuming the dust-free luminosities from the lightcone, and the solid blue line shows the number counts for the value of $A_{\rm \halpha}$ that leads to the best match to the observed counts in that particular redshift bin, i.e. the value that minimises the $\chi^2$ statistic. The blue shaded region shows the range of allowed number counts given the uncertainty on $A_{\rm \halpha}$, which we define as the range of $A_{\rm \halpha}$ values whose $\chi^2$ value is within one standard deviation of the minimum $\chi^2$ value. We see that for both redshift bins the calibrated counts from the lightcone are in excellent agreement with the observed counts, particularly at faint flux limits. Towards brighter flux limits, however, the lightcone under-predicts the number of galaxies. Since the number of bright $\halpha$-emitters is only a very small fraction of the samples for Euclid and WFIRST, we expect this deficit to have negligible effect on our results. Thus, for a Euclid-like flux limit of $2\times 10^{-16} \ergPerSecondPerCM{}$, or a WFIRST-like flux limit of $1\times 10^{-16} \ergPerSecondPerCM{}$, the lightcone shows a deficit of bright galaxies but predicts the correct overall number of galaxies. We do not show the counts for the MXXL-15K lightcone but confirm that they are in equally good agreement with the WISP observations as the MXXL-2K lightcone.

The WISP-calibrated values for $A_{\rm \halpha}$ are shown in Fig.~\ref{fig:dustAttenuation}, where the diamond symbols indicate the calibration results for the $0.7<z<1.5$ redshift bin and square symbols indicate the calibration results for the $0.8<z<1.2$ redshift bin. Filled symbols correspond to the results for the MXXL-2K lightcone and empty symbols correspond to the results for the MXXL-15K lightcone. For both redshift bins the calibration results for the MXXL-2K lightcone are in excellent agreement with the results for the MXXL-15K lightcone. Furthermore, for both lightcones the value for $A_{\rm \halpha}$ obtained by calibrating to the $0.7<z<1.5$ counts is consistent within error with the value obtained by calibrating to the $0.8<z<1.2$ counts. Therefore, adopting the calibration result for either redshift bin has negligible impact on the counts. This can be seen in the left-hand panel of  Fig.~\ref{fig:lightconeCounts} where the black dot-dashed line shows the counts over the redshift range $0.7<z<1.5$ computed assuming the value of $A_{\rm \halpha}$ obtained by calibrating to the counts in the $0.8<z<1.2$ bin. Going forward, we therefore adopt the results from the calibration to the counts in the  $0.8<z<1.2$ bin, which we note in the next section are marginally closer in value to the dust attenuation values obtained from calibrating the lightcone luminosity function.
The calibration using the luminosity function, which yields the HiZELS attenuation factors, is discussed in the next subsection.

\subsection{Luminosity function calibration}
\label{sec:calibration_lf}

For calibrating the lightcone luminosity function we use the HiZELS luminosity functions, as presented by \citet{Sobral13}, measured at $z=0.84$, $z=1.47$ and $z=2.23$. For each redshift, we compute the dust-free luminosity function from the lightcone using a thin redshift slice centred on the appropriate HiZELS redshift. Note that since $z\simeq2.23$ is beyond the upper redshift extent of our lightcone, in this case we use a thin redshift slice with $z=2.2$ as the upper bound. As with the cumulative counts, we use $\chi^2$ minimisation to determine the optimum value for $A_{\rm \halpha}$ needed to shift these luminosity functions into agreement with the observational estimates. Note that prior to our calibration we have adjusted the HiZELS luminosity functions in two ways. First, \citet{Sobral13} originally corrected for dust attenuation by shifting the $\halpha$ luminosity bright-wards by $A_{\rm \halpha}=1\,{\rm mag}$ (0.4 dex). Therefore, to ensure that we are comparing to the dust-attenuated HiZELS luminosity function we have removed this shift and made the luminosities fainter by 0.4 dex. Second, the HiZELS luminosity functions assumed $\halpha$ luminosities that had been corrected to remove $\nii$ contamination and so for our calibration we re-introduce the $\nii$ contamination. We do this by boosting the HiZELs luminosity function bright-wards by a factor of $4/3$. This factor was chosen to match the median correction of $\nii/\left (\halpha+\nii\right )\approx0.25$ reported by \citet{Sobral13}. Note that this correction factor shows some variation with redshift (with the correction being slightly smaller at higher redshift), though this variation is negligible in size.

The calibrated luminosity functions for the MXXL-2K lightcone are shown in Fig.~\ref{fig:lightconeLF}. In each panel the blue dotted line shows the luminosity function computed using the dust-free luminosities and the symbols correspond to the observed HiZELS luminosity functions, with the open symbols corresponding to luminosity bins that were excluded in our calibration procedure. These bins were left out following a preliminary calibration that indicated that the MXXL-2K lightcone is incomplete at these faint luminosities and so including these luminosity bins would unfairly bias our calibration results. The two vertical dotted lines indicate the luminosities at each redshift that  correspond to the Euclid and WFIRST flux limits. For reference we also show the \citet{Schechter76} functional fits presented by \citet{Cochrane17} and the luminosity functions predicted by the third of the three empirical models presented by \citet{Pozzetti16}, both which we have similarly corrected by a factor of $4/3$ in order to introduce $\nii$ contamination and applied an attenuation of $A_{\rm \halpha}=1\,{\rm mag}$.

In each panel the solid blue lines show the MXXL-2K luminosity function for the value of $A_{\rm \halpha}$ that minimises the $\chi^2$ statistic and leads to the best match to the HiZELS luminosity function at that particular redshift. The blue shaded region shows the range of allowed luminosity functions given the uncertainty on $A_{\rm \halpha}$, which again is defined as the range of $A_{\rm \halpha}$ values whose $\chi^2$ value is within one standard deviation of the minimum $\chi^2$ value. We can see that at each redshift the MXXL-2K lightcone is able to reproduce very well the HiZELS luminosity function brightwards of the Euclid flux limit. For $z\simeq1.47$ and $z\simeq2.2$ the lightcone provides a good match to the HiZELS luminosity function down to the WFIRST flux limit. At $z\simeq0.84$ we see the impact of incompleteness in the lightcone luminosity function as it falls below the observations for luminosities faintwards of the Euclid flux limit. We do not expect this to have a significant impact our clustering analysis however as we note that $z\simeq0.84$ is the below the lower redshift limit of WFIRST. We do not show the luminosity functions for the MXXL-15K lightcone but can confirm that they are in equally good agreement to the HiZELS observations as the MXXL-2K lightcone.

The dust attenuation values for our luminosity function calibration are also shown with circular symbols in Fig.~\ref{fig:dustAttenuation}. As with the number counts calibration, the attenuations obtained for the MXXL-15K and the MXXL-2K lightcones are in excellent agreement. The values for $A_{\rm \halpha}$ from our luminosity function calibration are much larger than the values obtained from the WISP number counts calibration, particularly at low redshift. However, we note that at $z\simeq0.84$ and $z\simeq2.2$ the $A_{\halpha}$ values that we have obtained are consistent with the $A_{\halpha}=1\,{\rm mag}$ that was originally assumed by \citet{Sobral13}. Using the median $\oii/\halpha$ line ratio as a dust attenuation indicator (see \citealt{Sobral12} for details), \citet{Sobral13} estimated that their $z\simeq1.47$ galaxy sample should have an attenuation of $A_{\halpha}=0.8\,{\rm mag}$, which, as seen in Fig.~\ref{fig:dustAttenuation}, is consistent within error with our $z\simeq 1.47$ calibration result. Additionally, \citet{Garn10a} determined that $A_{\halpha}\approx 1.2$ at $z\simeq0.84$, which is also consistent within the errors with our calibration result at $z\simeq0.84$. At $z\simeq2.2$ our luminosity function calibrated attenuation is consistent with the attenuation obtained from our calibration to the WISP number counts.

We note that \citet{Cochrane17} presented updated \citet{Schechter76} functional fits to the measured HiZELS luminosity functions, which at $z=0.84$ differ from the functional fits presented in \citet{Sobral13}. It is clear from  Fig.~\ref{fig:lightconeLF}, where the \citet{Cochrane17} fits are shown as thick dotted lines, that calibrating the MXXL lightcones to reproduce the \citet{Cochrane17} fits  would require a dust attenuation that is equal to or stronger than the attenuation required to reproduce the \citet{Sobral13} measurements. Adopting the \citet{Cochrane17} fits, instead of the \citet{Sobral13} measurements, may therefore widen the discrepancy between the dust attenuation needed to reproduce the WISP number counts and the HiZELS luminosity functions.

To see how much of an impact the discrepancy between the WISP-calibrated attenuation and the HiZELS-calibrated attenuation has on the luminosity function, we show in Fig.~\ref{fig:lightconeLF} the luminosity function obtained when assuming the dust attenuation $A_{\rm \halpha}$ calibrated to reproduce the WISP counts in the redshift range $0.8<z<1.2$. Using the WISP-calibrated attenuation leads to the lightcone over-predicting the luminosity function, particularly for bright luminosities at low redshift. At $z\simeq2.2$ the luminosity function assuming the HiZELS-calibrated attenuation and the luminosity function assuming the WISP-calibrated attenuation are virtually identical. For $z\simeq0.84$ and  $z\simeq1.47$ the difference between the HiZELS-calibrated luminosity function and the WISP-calibrated luminosity function is negligible for luminosities corresponding to the Euclid and WFIRST flux limits. We additionally show in Fig.~\ref{fig:lightconeCounts} the cumulative number counts obtained when assuming the HiZELS-calibrated dust attenuation. As expected, the HiZELS-calibrated counts are lower than the WISP-calibrated counts, but the discrepancy decreases towards fainter flux limits, such that at the Euclid and WFIRST flux limits the discrepancy is less than a factor of two. 

\subsection{Discussion of calibration results}
\label{sec:calibration_discussion}

Why the calibration to the WISP number counts and the calibration to the HiZELS luminosity functions lead to different values for the dust attenuation is not immediately clear. One possibility is that the WISP number counts and the HiZELS luminosity functions are probing different luminosity ranges. As shown in Fig.~\ref{fig:lightconeCounts}, the WISP cumulative number counts from \citet{Mehta15} have been measured at flux limits spanning $1\times 10^{-16}\ergPerSecondPerCM$ to  $1\times 10^{-15}\ergPerSecondPerCM$. Note that the fainter limit corresponds to the WFIRST flux limit. These two flux limits have been plotted as vertical dotted and dashed lines in the panels of Fig.~\ref{fig:lightconeLF}. We can see that at $z\simeq 0.84$ and $z\simeq1.47$ this flux range covers most of the luminosity range measured by the \citet{Sobral13} luminosity functions, with only the faintest few luminosity bins outside of the flux range. This would suggest that the small difference in luminosity ranges probed by the cumulative counts and the luminosity functions is less likely to be the cause of the discrepancy. At $z\simeq2.2$ the flux range is restricted to brighter luminosities, though at this redshift the WISP-calibrated attenuations and HiZELS-calibrated attenuations are consistent.

The most likely cause of the discrepancy is the different observing strategies and selection functions employed by the WISP and HiZELS surveys. For example, narrow band surveys, such as HiZELS, rely on continuum detection to identify emission lines which means that these surveys typically miss the lowest mass galaxies that spectroscopic grism surveys, such as WISP, are more sensitive to \citep{Mehta15}. As such, for a given $\halpha$ flux limit the galaxies detected by WISP will typically have a lower stellar mass compared to those detected by HiZELS (J. Colbert, private communication). Therefore, given the relations observed between stellar mass and dust attenuation $A_{\halpha}$ \citep{Garn10b,Dominguez13}, we expect the WISP dataset to typically have a weaker dust attenuation, as we can see in Fig.~\ref{fig:dustAttenuation}. Additionally, the thin-shell volumes probed by narrow band surveys mean that volume densities estimated by such surveys are sensitive to cosmic variance introduced by the presence of large-scale structure \citep{Mehta15}. As such, cosmic variance may also be contributing to the discrepancy in dust attenuation that our calibration procedure has returned.

Using the HiZELS-calibrated attenuation results in Fig.~\ref{fig:dustAttenuation} we can construct a toy model of how the attenuation evolves with redshift, such that we can apply a HiZELS-calibrated attenuation to any galaxy in the mock catalogue by interpolating over these results as a function of redshift. Finally, we note that in our toy model we assume that $A_{\rm \halpha}$ depends only on redshift and not additionally on the $\halpha$ luminosity. However, for our purposes we argue that adding a luminosity-depenence is not necessary given that we are able to reproduce the HiZELS luminosity functions very well down to the luminosity limits of relevance for Euclid and WFIRST.

In summary, given the difference in the dust attenuation required to reproduce the WISP number counts and the HiZELS luminosity functions, we choose to carry out our bias forecasts using both a WISP-calibrated version of the lightcones (adopting the values for calibration to the counts in the $0.8<z<1.2$ bin) and a HiZELS-calibrated version of the lightcones. 

\subsection{Galaxy number densities}
\label{sec:number_densities}

\begin{figure}
  \centering
  \includegraphics[width=0.48\textwidth]{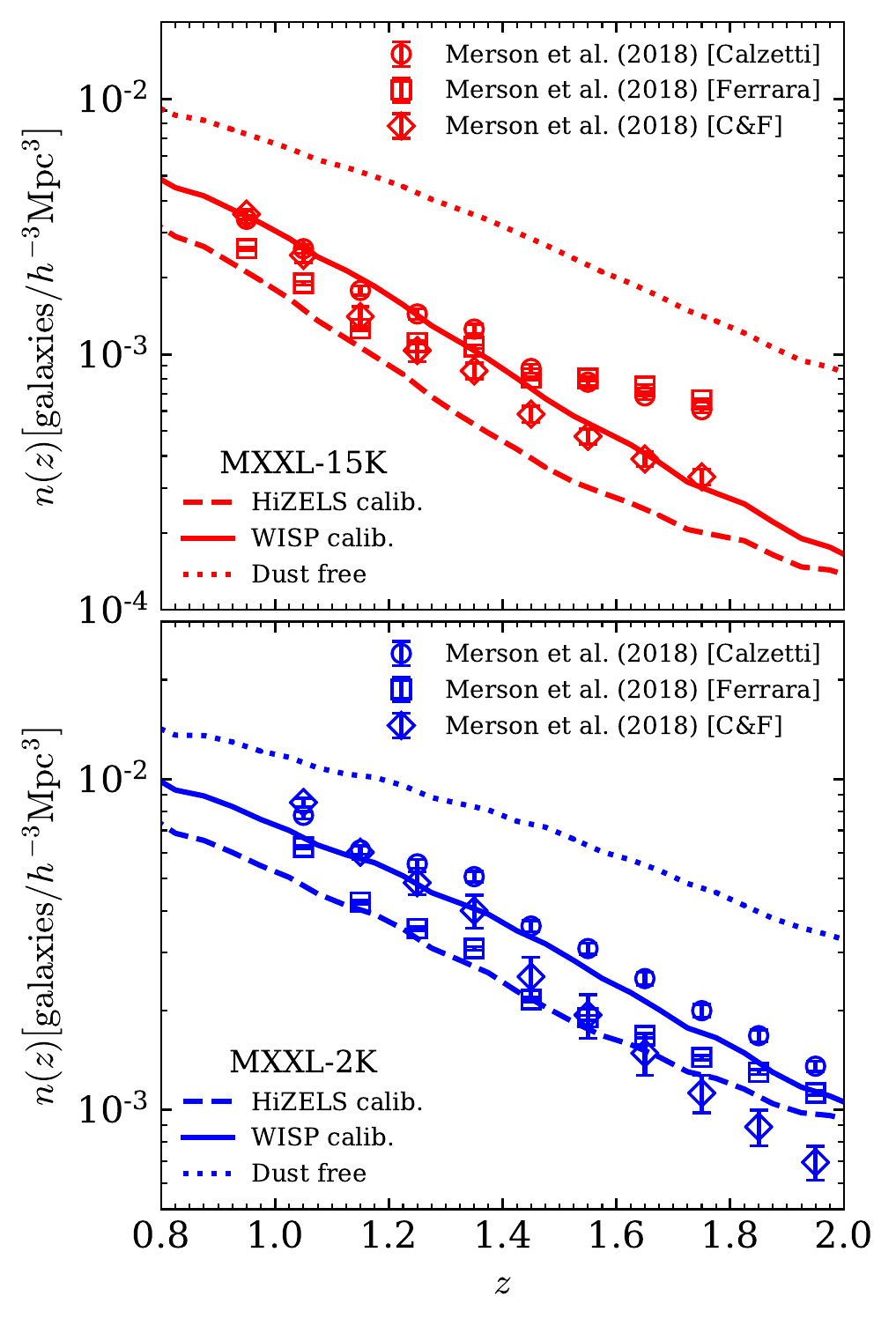}
  \caption{Number density per unit volume of $\halpha$-emitting galaxies in the MXXL-15K lightcone (upper panel) and MXXL-2K lightcone (lower panel). A flux limit of $2\times 10^{-16} \ergPerSecondPerCM{}$ has been applied to the dust-attenuated luminosities in the MXXL-15K lightcone and a limit of $1\times 10^{-16} \ergPerSecondPerCM{}$ to the dust-attenuated luminosities in the MXXL-2K lightcone. In each panel the solid line shows the number density obtained when adopting the WISP-calibrated dust attenuation, and the dashed line shows the number density obtained when adopting the HiZELS-calibrated dust attenuation. The dotted line shows the number density of the corresponding lightcone when no dust attenuation is applied. The various symbols show the predicted Euclid and WFIRST number densities from the \galacticus{} catalogues presented in \protect\citet{Merson18}, who considered three different dust attenuation methods: \protect\citet{Ferrara99}, \protect\citet{Calzetti00}, and \protect\citet[][C\&F]{Charlot00}. All flux selections assume a blended $\halpha+\nii$ flux.}
  \label{fig:lightcone_number_density}
\end{figure}

\begin{figure*}
  \centering
  \includegraphics[width=0.98\textwidth]{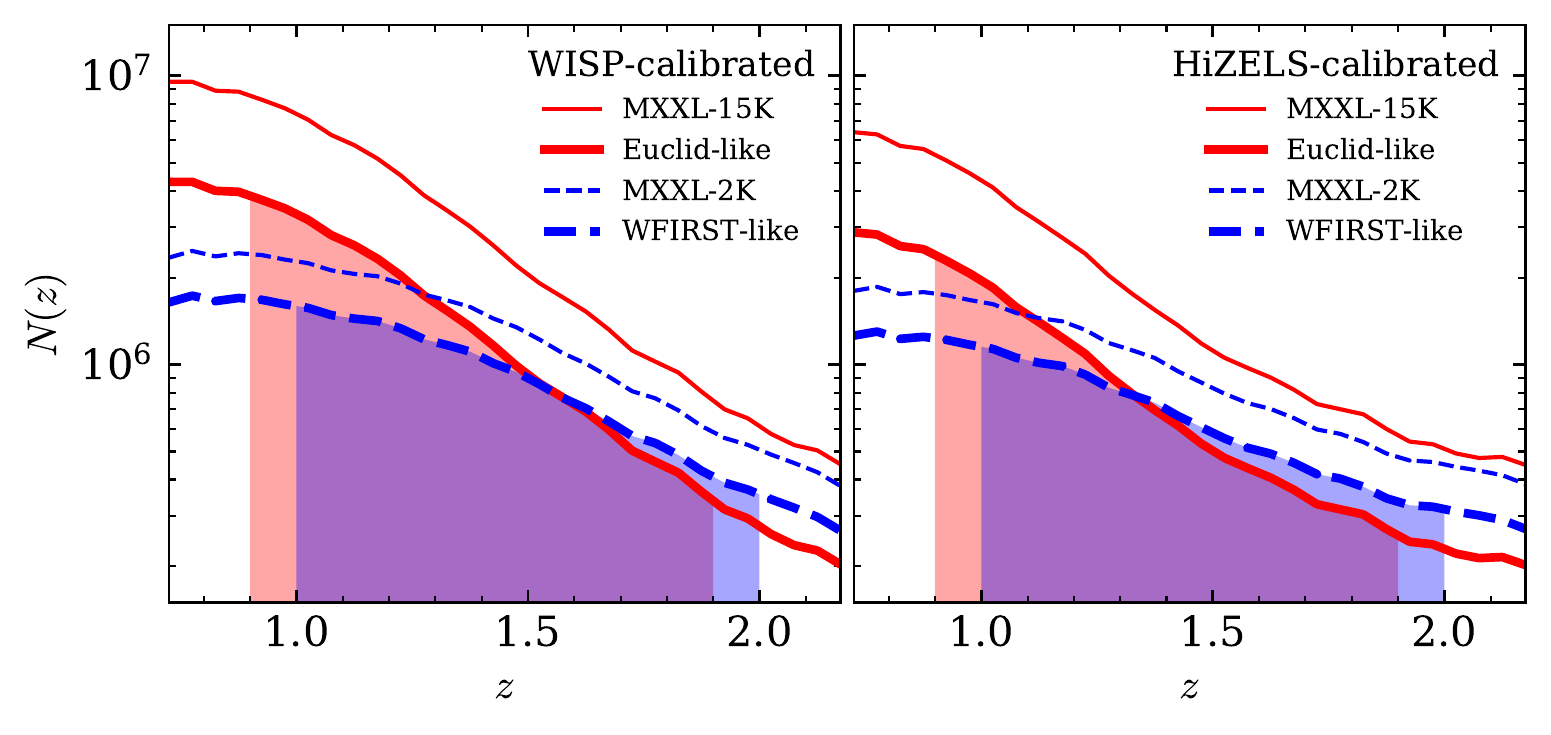}
  \caption{Redshift distributions for our Euclid-like and WFIRST-like surveys, assuming a WISP-calibrated dust attenuation (left-hand panel) and a HiZELS-calibrated dust attenuation (right-hand panel). Thin lines show the redshift distribution of the galaxies in the corresponding MXXL lightcone and thick lines show the redshift distribution for the Euclid-like and WFIRST-like surveys. Specifications of the lightcones and Euclid-like and WFIRST-like surveys are provided in Table~\ref{tab:surveys}. The shaded regions indicate the redshift range that is considered in our clustering analysis: $0.9\leqslant z<1.9$ for our Euclid-like survey and $1\leqslant z < 2$ for our WFIRST-like survey.}
  \label{fig:redshift_distributions}
\end{figure*}

Before progressing to examine the clustering of the galaxies in the lightcones, we examine the number densities of the galaxies as a check that our calibration yields number densities that are consistent with previous estimates. 

In Fig.~\ref{fig:lightcone_number_density} we show in the upper panel the number densities for the galaxies in the MXXL-15K lightcone, and in the lower panel the number densities for the galaxies in the MXXL-2K lightcone. We have applied a flux limit of $2\times 10^{-16} \ergPerSecondPerCM{}$ to the dust-attenuated luminosities in the MXXL-15K lightcone and a flux limit of $1\times 10^{-16} \ergPerSecondPerCM{}$ to the dust-attenuated luminosities in the MXXL-2K lightcone. In both panels the solid line shows the number densities when adopting the WISP-calibrated dust attenuations and the dashed line shows the number densities when adopting the HiZELS-calibrated dust attenuations. The dotted line corresponds to applying no dust attenuation. The symbols correspond to the number densities presented in \citet{Merson18} obtained by applying three different dust attenuation methods to a \galacticus{} lightcone. The dust attenuation methods considered in  \citet{Merson18} were: (i) interpolation over the library of dust curves from \citet{Ferrara99}, (ii) the \citet{Charlot00} model, where the dust attenuation follows a power law with wavelength, and (iii) the \citet{Calzetti00} dust screen law.

Considering first the MXXL-15K lightcone we see that when adopting a WISP-calibrated dust attenuation, the galaxy number density is consistent with the results from \citet{Merson18}. Adopting a HiZELS-calibrated dust attenuation leads to a number density that is slightly below the \citet{Merson18} results, but typically remains within approximately a factor of two. The fact that the number densities for the WISP-calibrated attenuations are in better agreement than the HiZELS-calibrated attenuations is not surprising as the \galacticus{} lightcone in \citet{Merson18} was calibrated to reproduce the WISP number counts from \citet{Mehta15}. Turning to the MXXL-2K lightcone we see that the number densities for the WISP-calibrated attenuations and the HiZELS-calibrated attenuations are both consistent with the number densities from \citet{Merson18}, with the difference between the two being consistent in size with the scatter between the number densities predicted for the different dust methods used in \citet{Merson18}. The redshift distributions for the MXXL-15K and MXXL-2K lightcones are provided in Appendix~\ref{sec:lightcone_redshift_distributions}.

Overall, when we apply the WISP-calibrated and HiZELS-calibrated dust attenuations to the MXXL-15K and MXXL-2K lightcones, the number densities that we obtain are entirely consistent with previous number density estimates.

\section{Linear bias forecasts}
\label{sec:linear_bias}

Having calibrated our lightcone catalogues we now proceed to make forecasts for the linear bias of $\halpha$-emitters as a function of redshift for a Euclid-like survey and a WFIRST-like survey. To calculate the linear bias we will first need to compute the correlation functions for the galaxies in our catalogues.

\subsection{Catalogue preparation}
\label{sec:catalogue_preparation}

The first step is to build catalogues for our Euclid-like and WFIRST-like surveys. We firstly apply a flux limit to the dust attenuated luminosities from the appropriate lightcone. For our Euclid-like surveys we apply a flux limit of $2\times 10^{-16} \ergPerSecondPerCM{}$ to select galaxies from the MXXL-15K lightcones and for our WFIRST-like surveys we apply a flux limit of $1\times 10^{-16} \ergPerSecondPerCM{}$ to select galaxies from the MXXL-2K lightcones. 

Secondly, we introduce incompleteness into the surveys by applying the appropriate completeness expected for each redshift survey: 45 per cent for Euclid \citep{Laureijs11} and 70 per cent for WFIRST \citep{Spergel15}. Both surveys have completeness significantly below 100 per cent due to slitless effects. This incompleteness will arise in the Euclid and WFIRST galaxy redshift surveys due to the inability to measure redshifts for faint galaxies with particularly noisy spectra, or for galaxies in high density regions where slitless spectra will overlap and could be confused.

Incompleteness is introduced by using random number generation to discard the appropriate fraction of galaxies. Note that we simply draw random numbers from a uniform distribution and make no attempt to introduce a density-dependent or luminosity-dependent incompleteness. For our Euclid-like surveys we apply a completeness limit of 0.45, consistent with the completeness limit estimated in \citet{Laureijs11}, and reject a random 55 per cent of the galaxies. For our WFIRST-like surveys we apply a completeness limit of 0.7, consistent with the completeness limit provided in \citet{Spergel15}, and reject a random 30 per cent of the galaxies. The specifications for our Euclid-like and WFIRST-like surveys are listed in Table~\ref{tab:surveys} alongside the specifications of the MXXL-15K and MXXL-2K lightcones. In Fig.~\ref{fig:redshift_distributions} we show the redshift distributions for our Euclid-like and WFIRST-like surveys, as well as for the MXXL-15K and the MXXL-2K lightcones.

In our clustering analysis we consider two versions of a Euclid-like survey and WFIRST-like survey: one calibrated to reproduce the WISP cumulative number counts, and one calibrated to reproduce the HiZELS luminosity functions. Additionally, for each version we repeat the analysis twice. In the first instance we take the observed redshifts of the galaxies (i.e. with peculiar velocities included), which we shall refer to as \emph{redshift-space}, and in the second instance we take the cosmological redshifts of the galaxies (i.e. with peculiar velocities removed), which we will refer to as \emph{real-space}. Note that galaxies are not observed in real-space, but we include this analysis to show the impact of redshift-space distortions.

\subsection{Clustering analysis}
\label{sec:clustering_analysis}

For our clustering analysis we split each of our Euclid-like and WFIRST-like surveys into 5 equal redshift bins of width $\Delta z=0.2$ and compute the two-point galaxy correlation function in each bin. For the Euclid-like surveys these bins cover the redshift range $0.9\leqslant z < 1.9$, and for the WFIRST-like surveys these bins cover the redshift range $1\leqslant z < 2$. 

In order to compute the galaxy correlation functions we generate individual catalogues of randoms for each redshift bin. In each instance we generate a sufficiently large number of randoms (at least ten times the number of galaxies in the redshift bin), with the number chosen arbitrarily such that adding any additional randoms does not improve our recovery of the galaxy correlation function. The number of randoms required increases with redshift as the number density of galaxies decreases. For the lower redshifts the number of randoms was also limited by available computing resources. The right ascension and declination of the randoms were found by sampling uniform distributions bounded by the right ascension limits and the cosine of the declination limits of the particular lightcone. To assign redshifts to the randoms we do not sample a uniform distribution as this would not reproduce the correct shape of the galaxy redshift distribution across the bin. Instead we assign redshifts by first replicating the redshifts of the galaxies, until we match the required number of randoms, and then randomly shuffling the galaxy redshifts.

We compute the angle-averaged galaxy two-point correlation function, $\xi(r)$, using the open-source \textsc{TreeCorr} package \citep{Jarvis04}, and the \citet{Landy93} estimator,
\begin{equation}
\xi_{\rm LS}(r) = \frac{DD(r)-2DR(r)+RR(r)}{RR(r)},
\label{eq:LandySzalay}
\end{equation}
where $DD(r)$ are the weighted galaxy-galaxy pair counts, $DR(r)$ are the weighted galaxy-random pair counts and $RR(r)$ are the weighted random-random pair counts as a function of separation $r$ in units of $\hMpc$. For each redshift bin the correlation function is calculated 5 times, each time using a different random seed when applying the survey incompleteness (see $\S$~\ref{sec:catalogue_preparation}) and using a different catalogue of randomly generated positions. The correlation function is calculated between $50\hMpc$ and $150\hMpc$ using 48 linearly spaced bins.

In Table~\ref{tab:bias_results_euclid} we report for our Euclid-like surveys the average number of galaxies in each redshift bin after introducing incompleteness and the effective redshift of these galaxies. We also report the average ratio of the number of galaxies to the number of random positions. The number of randoms was increased with increasing redshift to account for the decreasing number densities of the galaxy samples. Memory limitations of available computing resources forced us to limit the number of randoms, but in each case we ensured that a sufficient number was used to adequately recover the clustering signal of the galaxies. We report equivalent numbers for our WFIRST-like surveys in Table~\ref{tab:bias_results_wfirst}. 
 
\begin{table*}
\centering
\caption{Catalogue specifications and linear bias fits for our Euclid-like catalogues, measured in redshift-space (upper table) and in real-space (lower table). The properties shown are: the effective redshift of the galaxies in the redshift slice, $z_{\rm eff}$; the mean number of galaxies used to compute the correlation function, $\bar{N}_{\rm gal}$; the mean value for the ratio of randoms to galaxies, $\bar{N}_{\rm ran}/\bar{N}_{\rm gal}$; the mean and standard deviation of the halo mass, $M_h$, for that galaxy sample; and the linear bias fit. The columns show the results for each redshift slice considered. Each table shows the equivalent results when adopting a WISP-calibrated lightcone or a HiZELS-calibrated lightcone.}
\begin{tabular}{|c|c|c|c|c|c|c|c|}
\hline
Calibration&Property&$0.90\,\leqslant\,z\,<\,1.1$&$1.1\,\leqslant\,z\,<\,1.3$&$1.3\,\leqslant\,z\,<\,1.5$&$1.5\,\leqslant\,z\,<\,1.7$&$1.7\,\leqslant\,z\,<\,1.9$\\
\hline\hline
\multicolumn{7}{|c|}{\textbf{Euclid-like (redshift-space)}}\\
WISP&$z_{\rm eff}$&0.991&1.19&1.38&1.59&1.79\\
&$\bar{N}_{\rm gal}$&13175397&8688180&5045749&2912673&1748220\\
&$\bar{N}_{\rm ran}/\bar{N}_{\rm gal}$&10&10&14&25&30\\
&$\log_{10}\left (M_h/h^{-1}{\rm M_{\odot}}\right )$&$11.8\,\pm\,0.4$&$11.8\,\pm\,0.3$&$11.8\,\pm\,0.3$&$11.9\,\pm\,0.3$&$11.9\,\pm\,0.3$\\
&$b_{\rm lin}\,\pm\,\delta b_{\rm lin}$&$1.40\,\pm\,0.03$&$1.51\,\pm\,0.05$&$1.70\,\pm\,0.04$&$1.82\,\pm\,0.02$&$1.96\,\pm\,0.09$\\
&&&&&&\\
HiZELS&$z_{\rm eff}$&0.988&1.19&1.39&1.59&1.79\\
&$\bar{N}_{\rm gal}$&7785058&4642807&2635798&1687841&1215424\\
&$\bar{N}_{\rm ran}/\bar{N}_{\rm gal}$&10&10&15&25&29\\
&$\log_{10}\left (M_h/h^{-1}{\rm M_{\odot}}\right )$&$11.8\,\pm\,0.3$&$11.9\,\pm\,0.3$&$11.9\,\pm\,0.3$&$11.9\,\pm\,0.3$&$11.9\,\pm\,0.3$\\
&$b_{\rm lin}\,\pm\,\delta b_{\rm lin}$&$1.42\,\pm\,0.03$&$1.55\,\pm\,0.05$&$1.69\,\pm\,0.04$&$1.88\,\pm\,0.05$&$1.9\,\pm\,0.1$\\

\hline
\multicolumn{7}{|c|}{\textbf{Euclid-like (real-space)}}\\
WISP&$z_{\rm eff}$&0.991&1.19&1.38&1.59&1.79\\
&$\bar{N}_{\rm gal}$&13171311&8682758&5050012&2910250&1748428\\
&$\bar{N}_{\rm ran}/\bar{N}_{\rm gal}$&10&10&15&30&35\\
&$\log_{10}\left (M_h/h^{-1}{\rm M_{\odot}}\right )$&$11.8\,\pm\,0.4$&$11.8\,\pm\,0.3$&$11.8\,\pm\,0.3$&$11.9\,\pm\,0.3$&$11.9\,\pm\,0.3$\\
&$b_{\rm lin}\,\pm\,\delta b_{\rm lin}$&$1.05\,\pm\,0.01$&$1.17\,\pm\,0.04$&$1.30\,\pm\,0.03$&$1.44\,\pm\,0.05$&$1.6\,\pm\,0.1$\\
&&&&&&\\
HiZELS&$z_{\rm eff}$&0.988&1.19&1.39&1.59&1.79\\
&$\bar{N}_{\rm gal}$&7782514&4639483&2638085&1686470&1215545\\
&$\bar{N}_{\rm ran}/\bar{N}_{\rm gal}$&10&10&14&30&35\\
&$\log_{10}\left (M_h/h^{-1}{\rm M_{\odot}}\right )$&$11.8\,\pm\,0.3$&$11.9\,\pm\,0.3$&$11.9\,\pm\,0.3$&$11.9\,\pm\,0.3$&$11.9\,\pm\,0.3$\\
&$b_{\rm lin}\,\pm\,\delta b_{\rm lin}$&$1.055\,\pm\,0.009$&$1.21\,\pm\,0.04$&$1.32\,\pm\,0.03$&$1.49\,\pm\,0.08$&$1.6\,\pm\,0.1$\\

\hline
\end{tabular}
\label{tab:bias_results_euclid}
\end{table*}

\begin{table*}
\centering
\caption{Catalogue specifications and linear bias fits for our WFIRST-like catalogues, measured in redshift-space (upper table) and in real-space (lower table). The meanings of the various properties are the same as in Table~\ref{tab:bias_results_euclid}. The columns show the results for each redshift slice considered. Each table shows the equivalent results when adopting a WISP-calibrated lightcone or a HiZELS-calibrated lightcone.}
\begin{tabular}{|c|c|c|c|c|c|c|c|}
\hline
Calibration&Property&$1.0\,\leqslant\,z\,<\,1.2$&$1.2\,\leqslant\,z\,<\,1.4$&$1.4\,\leqslant\,z\,<\,1.6$&$1.6\,\leqslant\,z\,<\,1.8$&$1.8\,\leqslant\,z\,<\,2.0$\\
\hline\hline
\multicolumn{7}{|c|}{\textbf{WFIRST-like (redshift-space)}}\\
WISP&$z_{\rm eff}$&1.10&1.29&1.49&1.69&1.89\\
&$\bar{N}_{\rm gal}$&5915244&4839323&3576093&2442191&1673105\\
&$\bar{N}_{\rm ran}/\bar{N}_{\rm gal}$&20&20&25&30&40\\
&$\log_{10}\left (M_h/h^{-1}{\rm M_{\odot}}\right )$&$11.7\,\pm\,0.4$&$11.7\,\pm\,0.4$&$11.7\,\pm\,0.4$&$11.8\,\pm\,0.4$&$11.8\,\pm\,0.4$\\
&$b_{\rm lin}\,\pm\,\delta b_{\rm lin}$&$1.46\,\pm\,0.02$&$1.63\,\pm\,0.01$&$1.61\,\pm\,0.04$&$2.0\,\pm\,0.1$&$2.15\,\pm\,0.04$\\
&&&&&&\\
HiZELS&$z_{\rm eff}$&1.09&1.29&1.49&1.69&1.89\\
&$\bar{N}_{\rm gal}$&4193549&3278049&2341326&1768644&1369560\\
&$\bar{N}_{\rm ran}/\bar{N}_{\rm gal}$&20&20&25&30&40\\
&$\log_{10}\left (M_h/h^{-1}{\rm M_{\odot}}\right )$&$11.7\,\pm\,0.4$&$11.8\,\pm\,0.4$&$11.8\,\pm\,0.3$&$11.8\,\pm\,0.3$&$11.8\,\pm\,0.4$\\
&$b_{\rm lin}\,\pm\,\delta b_{\rm lin}$&$1.47\,\pm\,0.02$&$1.64\,\pm\,0.01$&$1.62\,\pm\,0.04$&$2.1\,\pm\,0.1$&$2.14\,\pm\,0.06$\\

\hline
\multicolumn{7}{|c|}{\textbf{WFIRST-like (real-space)}}\\
WISP&$z_{\rm eff}$&1.10&1.29&1.49&1.69&1.89\\
&$\bar{N}_{\rm gal}$&5918784&4832447&3578623&2449899&1673483\\
&$\bar{N}_{\rm ran}/\bar{N}_{\rm gal}$&20&20&25&30&40\\
&$\log_{10}\left (M_h/h^{-1}{\rm M_{\odot}}\right )$&$11.7\,\pm\,0.4$&$11.7\,\pm\,0.4$&$11.7\,\pm\,0.4$&$11.8\,\pm\,0.4$&$11.8\,\pm\,0.4$\\
&$b_{\rm lin}\,\pm\,\delta b_{\rm lin}$&$1.13\,\pm\,0.02$&$1.27\,\pm\,0.04$&$1.22\,\pm\,0.06$&$1.57\,\pm\,0.06$&$1.70\,\pm\,0.05$\\
&&&&&&\\
HiZELS&$z_{\rm eff}$&1.09&1.29&1.49&1.69&1.89\\
&$\bar{N}_{\rm gal}$&4196198&3273552&2342872&1773791&1370214\\
&$\bar{N}_{\rm ran}/\bar{N}_{\rm gal}$&20&20&25&30&40\\
&$\log_{10}\left (M_h/h^{-1}{\rm M_{\odot}}\right )$&$11.7\,\pm\,0.4$&$11.8\,\pm\,0.4$&$11.8\,\pm\,0.3$&$11.8\,\pm\,0.3$&$11.8\,\pm\,0.4$\\
&$b_{\rm lin}\,\pm\,\delta b_{\rm lin}$&$1.15\,\pm\,0.02$&$1.28\,\pm\,0.02$&$1.3\,\pm\,0.1$&$1.56\,\pm\,0.07$&$1.72\,\pm\,0.04$\\

\hline
\end{tabular}
\label{tab:bias_results_wfirst}
\end{table*}

\begin{figure*}
  \centering
  \includegraphics[width=0.99\textwidth]{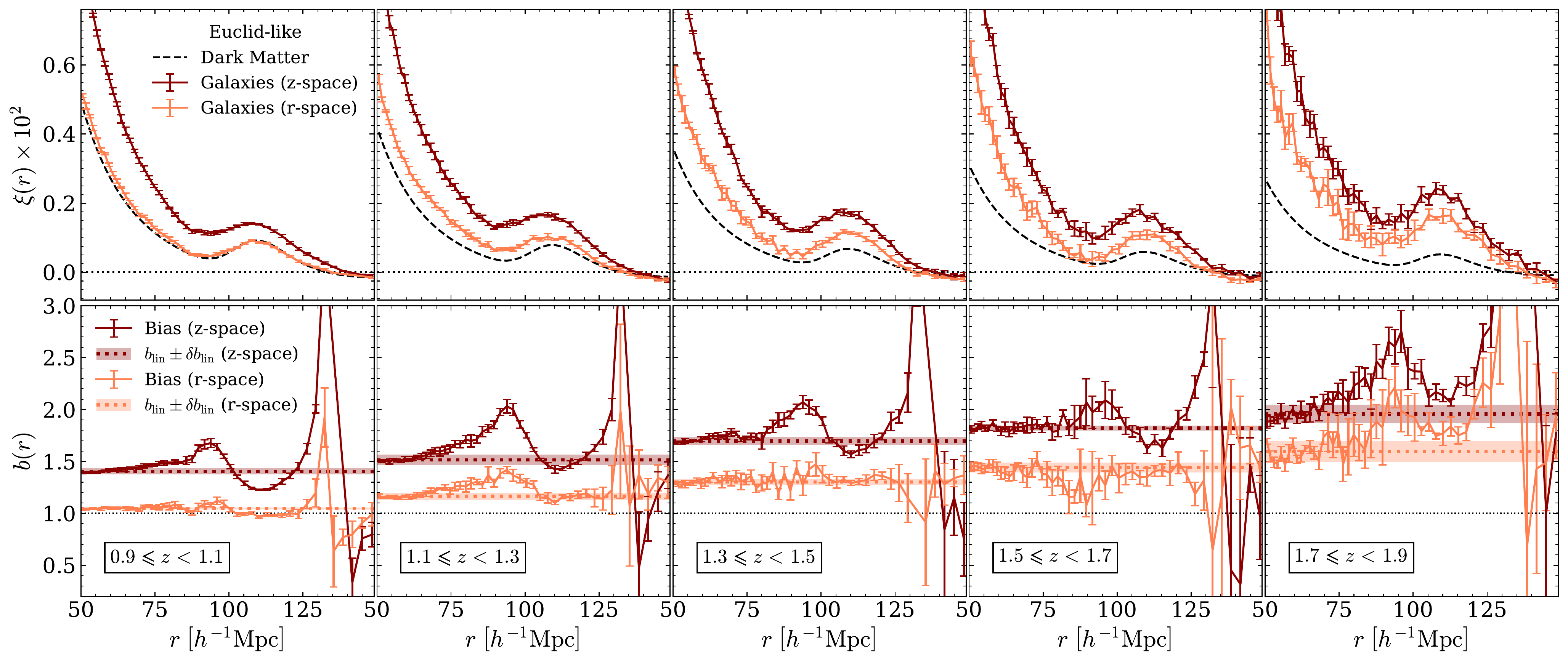}
  \includegraphics[width=0.99\textwidth]{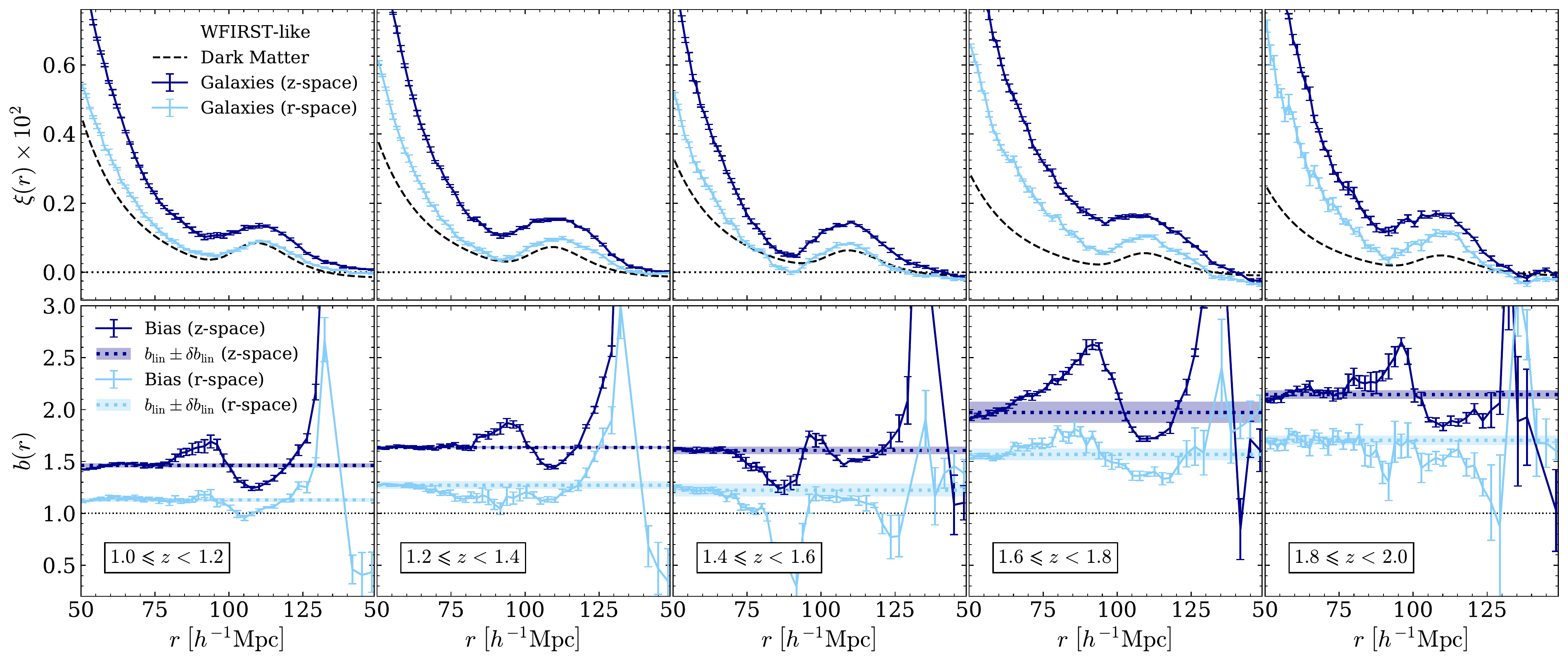}
  \caption{Clustering results and bias fits for a Euclid-like survey with dust-attenuated flux limit $2\times 10^{-16}\ergPerSecondPerCM{}$ (upper grid) and a WFIRST-like survey with dust-attenuated flux limit $1\times 10^{-16}\ergPerSecondPerCM{}$ (lower grid), constructed from lightcones whose luminosities were attenuated using WISP-calibrated dust attenuation values. In each grid the darker lines show the results in redshift-space, the fainter lines correspond to real-space (i.e. assuming the cosmological redshifts of the galaxies with no peculiar velocity component), and the black dashed lines show the dark matter correlation function at the effective redshift of the bin. In the lower panels of each grid, the horizontal dotted lines and shaded regions correspond to the linear bias fits assuming the appropriate correlation function (see text for details). The redshift range used for selection is shown in the bottom left-hand corner of the lower panels.}
  \label{fig:lightconeBias_wisp}
\end{figure*}

\begin{figure*}
  \centering
  \includegraphics[width=0.99\textwidth]{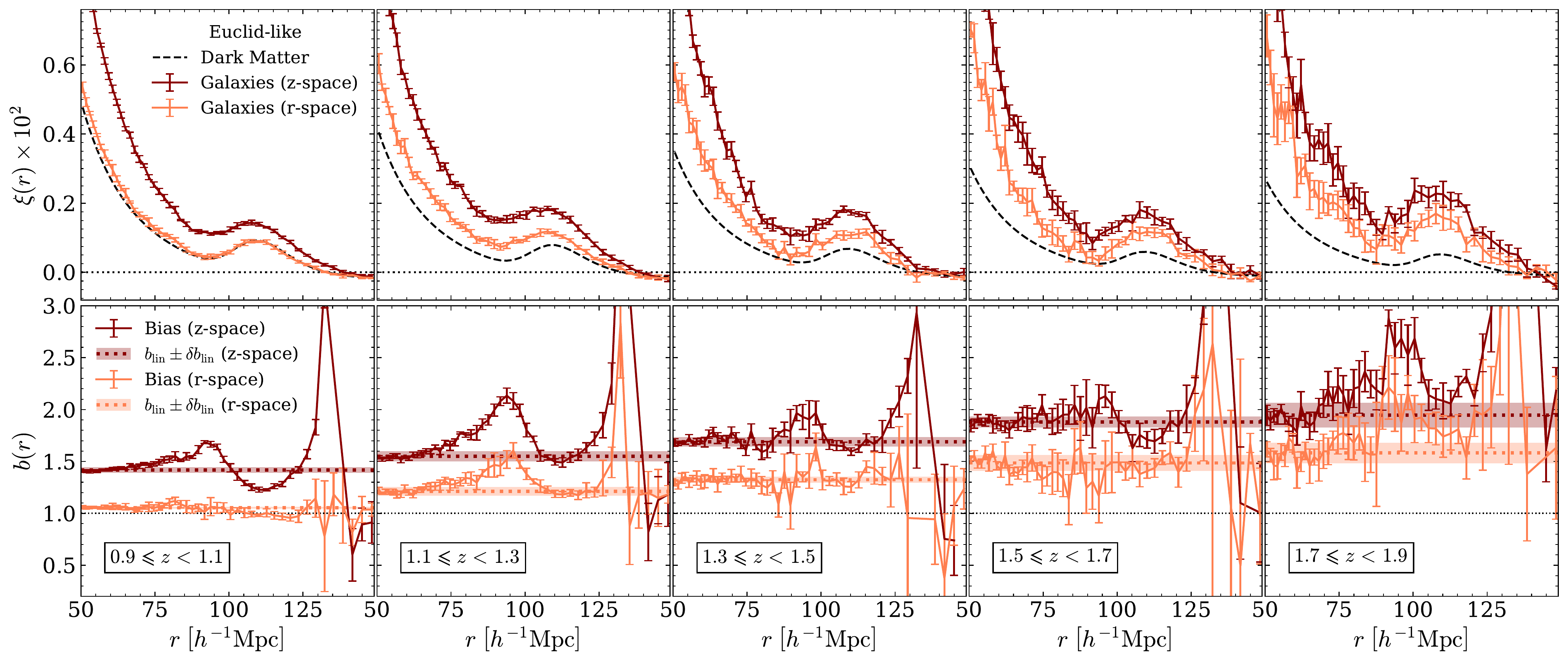}
  \includegraphics[width=0.99\textwidth]{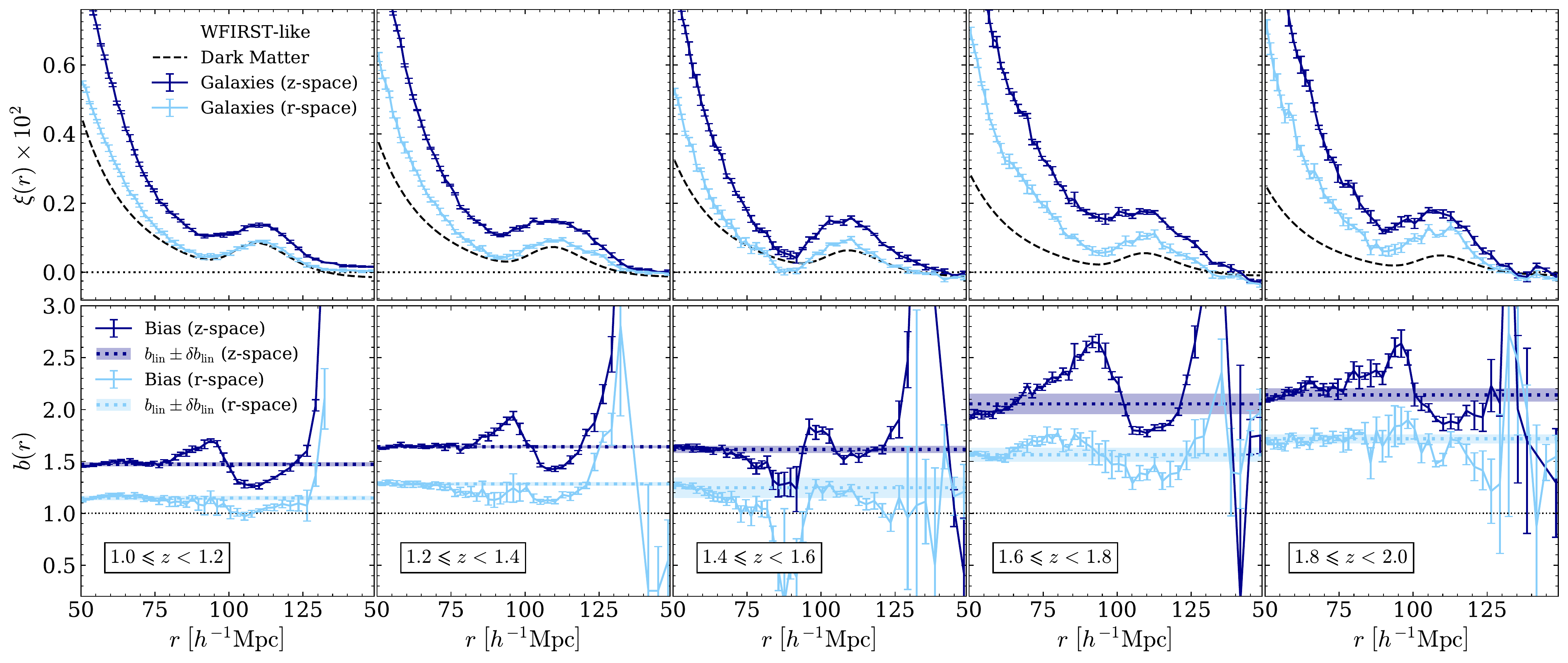}  
  \caption{Clustering results and bias fits for a Euclid-like survey with dust-attenuated flux limit $2\times 10^{-16}\ergPerSecondPerCM{}$ (upper grid) and a WFIRST-like survey with dust-attenuated flux limit $1\times 10^{-16}\ergPerSecondPerCM{}$ (lower grid), constructed from lightcones whose luminosities were attenuated using HiZELS-calibrated dust attenuation values. All the lines have the same meanings as in Fig.~\ref{fig:lightconeBias_wisp}.}
  \label{fig:lightconeBias_hizels}
\end{figure*}

The results of our clustering analysis are shown in Fig.~\ref{fig:lightconeBias_wisp}, which assumes WISP-calibrated dust attenuation, and Fig.~\ref{fig:lightconeBias_hizels}, which assumes HiZELS-calibrated dust attenuation. In each case the upper grid shows the results for the Euclid-like survey and the lower grid shows the results for the WFIRST-like survey. The top row of each grid shows the correlation function in each redshift bin, with the darker lines showing the redshift-space result and the fainter lines showing the real-space result. The lines show the average result over 5 repeats (each time changing the random seed) and the error bars show the standard deviation. 

We can see that in each instance we are able to recover the BAO peak in every redshift bin, albeit with increased noise in the highest redshift bin for the Euclid-like surveys. However, in each instance the recovery is significant over the statistical noise, particularly in real-space. For $r \lesssim 140\hMpc$ the correlation functions measured in redshift-space have a higher amplitude than the equivalent measurement in real-space, consistent with the expectations of \citet{Kaiser87}, which are based on linear perturbation theory. The impact of redshift-space distortions (RSDs) leads to some smearing out of the BAO peak.

\subsection{Linear bias measurements and fitting}
\label{sec:linear_bias_measurements}

\subsubsection{Linear bias measurements}
\label{sec:bias_measurements}

Given our galaxy clustering results, we can now use Eq.~\ref{eq:bias} to compute the linear bias in each redshift bin for our Euclid-like and WFIRST-like surveys. Each galaxy correlation function is divided by the non-linear dark matter correlation function computed at the same effective redshift. The dark matter correlation functions, which we compute using the \textsc{CLASS} and \textsc{HaloFit} functionality in the \texttt{NBodyKit} python package and assuming an MXXL cosmology (see Fig.~\ref{fig:theoryPk}), are shown with black dashed lines in the panels in the upper row of each grid in Fig.~\ref{fig:lightconeBias_wisp} and Fig.~\ref{fig:lightconeBias_hizels}. 

The lower row of panels in each grid in Fig.~\ref{fig:lightconeBias_wisp} and Fig.~\ref{fig:lightconeBias_hizels} show the measured linear bias as a function of separation. As with the correlation functions, the darker lines correspond to the bias estimated using the redshift-space galaxy correlation functions and the fainter lines correspond to the bias estimated using the real-space galaxy correlation functions. The lines show the bias estimate in each bin of separation averaged over the 5 repeat clustering calculations and the error bars show the standard deviation. 

Examining first the bias measurements in real-space, we see that for both the Euclid-like and WFIRST-like surveys the measured bias is consistent with a constant value for scales $r\lesssim 75 \hMpc$. For scales larger than this we begin to see deviations away from a constant value, particularly for the WFIRST-like survey. In the $1.4 \leqslant z \leqslant 1.6$ redshift bin for the WFIRST-like analysis, we see a significant negative deviation in the measured bias on scales $75\hMpc\lesssim r\lesssim 100\hMpc$. The cause of this decrease is not clear and is most likely due to cosmic variance. At the largest scales, $r\gtrsim 125\hMpc$, the bias measurements in each redshift bin grow rapidly as the galaxy and matter correlation functions tend toward zero and ultimately become negative due to the integral constraint. If we move to redshift-space we see a similar behaviour, although the deviations away from a constant on scales  $75\hMpc\lesssim r \lesssim 125\hMpc$ are now more pronounced than in real-space. In several instances these scale-dependent deviations display a sinusoidal-like shape and in every instance occur around the BAO scale. Similar sinusoidal features can be seen in \citet{Hada19} when the authors examine the difference between the correlation function for a mock galaxy field in redshift-space and the correlation function for the initial linear density field (see also the plots showing the scale dependence of the bias of star forming galaxies in \citealt{Angulo14}). It is possible that these deviations arise due to a distortion of the galaxy correlation function relative to the correlation function expected from linear theory. This distortion, which is manifested as a smoothing of the BAO peak and slight shift of the peak away from the linear theory position, is understood to arise from a combination of non-linear collapse, mode coupling, redshift-space distortions, as well as the bias itself \citep{Jeong06, Eisenstein07a, Angulo08, Crocce08, Sanchez08, Smith08, Anselmi16}. It is possible to model these effects to correct the shape of the BAO in the linear theory predictions \citep[e.g.][]{Crocce08,Smith08}.

\subsubsection{Linear bias fitting}
\label{sec:bias_fitting}

In each redshift bin we attempt to fit the real-space and redshift-space measurements of the bias with a constant, scale-independent value, $b_{\rm lin}$. To do this we use $\chi^2$ minimisation to fit a zeroth-order polynomial to the measured bias values $b(r<r_{\rm cut})$. Given the impact of the large-scale deviations around the BAO scale, care must be taken when selecting the scale at which to fit a constant bias value, unless the BAO distortions are corrected for. We adopt $r_{\rm cut}= 75\hMpc$ for each redshift bin, as below this scale our bias measurements are broadly consistent with a constant value. The results of these fits are shown in Fig.~\ref{fig:lightconeBias_wisp} and Fig.~\ref{fig:lightconeBias_hizels} as horizontal dotted lines with shaded regions showing an associated uncertainty, $\delta b_{\rm lin}$. The associated uncertainties were originally estimated as the $1\sigma$ uncertainties on the linear bias fits obtained from the $\chi^2$ minimisation procedure, which  were on the order of 0.1 per cent. These uncertainties appear quite conservative given our measurements of $b(r)$. Instead, we estimate the uncertainties on the linear bias fits using the root-mean-square (RMS) of the difference between the measured values and the fitted value, $b_{\rm lin}$; 
\begin{equation}
    \delta b_{\rm lin} = \sqrt{\frac{1}{N}\sum_{r\leqslant r_{\rm cut}}\left (b(r)-b_{\rm lin}\right )^2},
    \label{eq:fit_uncertainty}
\end{equation}
where $N$ is the number of measured bias values with $r\leqslant r_{\rm cut}$. These RMS uncertainties, $\delta b_{\rm lin}$, which are on the order of 1 per cent, are reported alongside the linear bias values, $b_{\rm lin}$, in Table~\ref{tab:bias_results_euclid} and Table~\ref{tab:bias_results_wfirst}.

Comparing the fitted linear bias values in Table~\ref{tab:bias_results_euclid} and Table~\ref{tab:bias_results_wfirst} we can see that the bias values for the WISP-calibrated instance and the HiZELS-calibrated instance are in good agreement and consistent within error. The choice of calibration, whether to match the WISP number counts or the HiZELS luminosity function, does not therefore appear to significantly impact the measurement of the bias. Secondly we see that the fitted linear bias values for the Euclid-like and WFIRST-like surveys are also broadly consistent with one another. We will return to this point in our discussion in $\S$\ref{sec:discussion}.

\subsection{Linear bias evolution with redshift}
\label{sec:redshift_evolution}

\begin{figure*}
  \centering
   \includegraphics[width=0.99\textwidth]{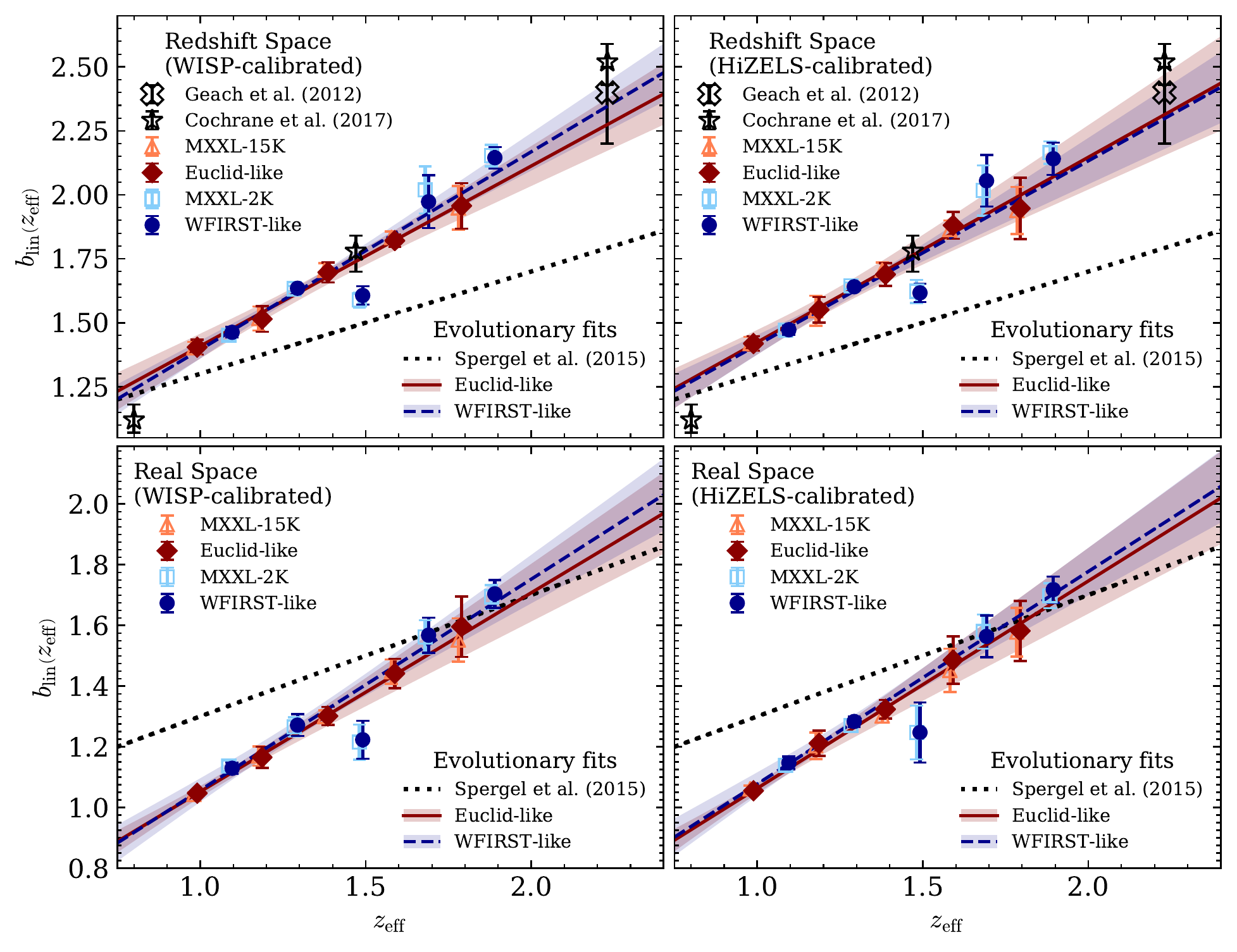}
   \caption{Linear bias as a function of effective redshift for both redshift-space (upper panels) and real-space (lower panels) when adopting a WISP-calibrated lightcone (left-hand column) or a HiZELS-calibrated lightcone (right-hand column). Red colours indicate results for our Euclid-like survey and blue colours indicate results for our WFIRST-like survey. The various red and blue symbols show the fitted linear bias values from the clustering analyses in each of the individual redshift bins. Filled symbols show the results for our Euclid-like and WFIRST-like surveys, whilst empty symbols show equivalent results from a clustering analysis of the MXXL-15 and MXXL-2K lightcones, without the introduction of any incompleteness effects (see Appexndix~\ref{sec:lightcone_bias_fits}). Linear relations were fit to the bias values using $\chi^2$ minimisation. The red solid lines show the linear relations for the Euclid-like surveys that minimise the $\chi^2$ statistic and the red shaded regions correspond to the range of fits that have a gradient and intercept within the $1\sigma$ uncertainty range. The blue dashed lines and blue shaded regions show the equivalent results for the WFIRST-like surveys. The black cross shows the bias result from \protect\citet{Geach12}, the black stars show the bias results from \protect\citet{Cochrane17} and the black dotted line shows the linear relation assumed by \protect\citet{Spergel15}.}
  \label{fig:redshiftVsBias}
\end{figure*}

\begin{table*}
\centering
\caption{Gradient and intercept parameter results obtained from using $\chi^2$ minimisation to fit linear relations for the redshift evolution of the linear bias for our Euclid-like and WFIRST-like surveys. The upper half of the table shows the parameter fits to our bias measurements in redshift-space, whilist the lower half of the table shows the fits to the bias measurements in real-space. The parameter values stated correspond to the values that minimise the $\chi^2$ statistic, and the uncertainties correspond to the bounds of the $1\sigma$ contours as shown in Fig.~\ref{fig:redshiftVsBiasParameters}. Parameters are reported for both the WISP-calibrated and HiZELS-calibrated versions of the surveys.}
\begin{tabular}{|c|c|c|c|c|}
\hline
Calibration&\multicolumn{2}{|c|}{Euclid-like}&\multicolumn{2}{|c|}{WFIRST-like}\\
&Gradient&Intercept&Gradient&Intercept\\
\hline\hline
\multicolumn{5}{|c|}{\textbf{Redshift-space}}\\
WISP&$0.70\,\pm\,0.11$&$0.70\,\pm\,0.15$&$0.77\,\pm\,0.10$&$0.62\,\pm\,0.13$\\
HiZELS&$0.72\,\pm\,0.15$&$0.70^{+0.19}_{-0.18}$&$0.72\,\pm\,0.12$&$0.69\,\pm\,0.16$\\

\hline
\multicolumn{5}{|c|}{\textbf{Real-space}}\\
WISP&$0.66\,\pm\,0.10$&$0.40\,\pm\,0.11$&$0.70\,\pm\,0.10$&$0.36\,\pm\,0.13$\\
HiZELS&$0.68\,\pm\,0.11$&$0.38\,\pm\,0.12$&$0.70^{+0.10}_{-0.11}$&$0.38\,\pm\,0.14$\\

\hline
\end{tabular}
\label{tab:bias_fits_euclid_wfirst}
\end{table*}

\begin{figure*}
  \centering
   \includegraphics[width=0.99\textwidth]{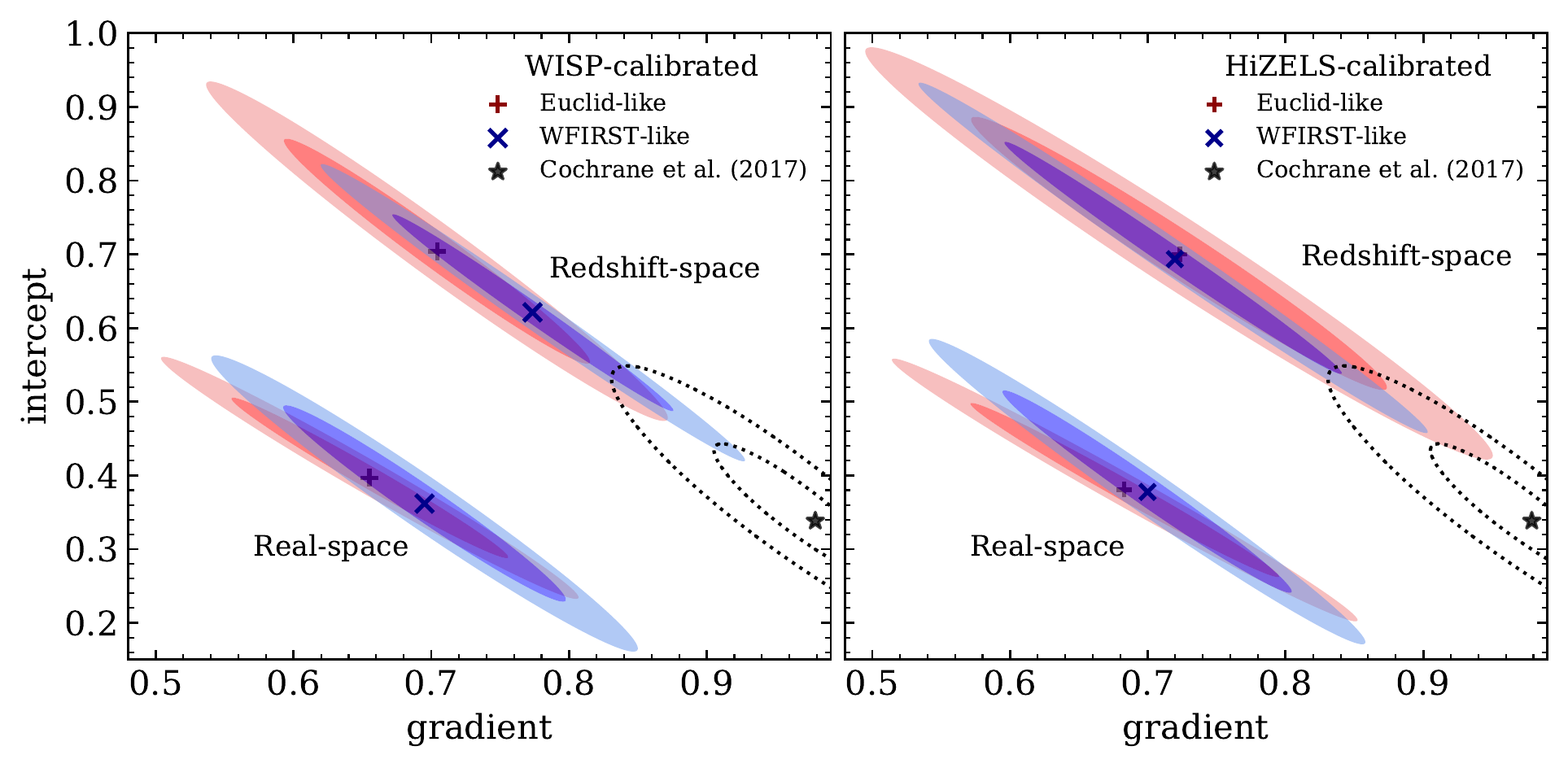}
   \caption{Gradient and intercept parameters for the linear relations describing the redshift evolution of the linear bias for our Euclid-like and WFIRST-like surveys. Euclid-like results are coloured red and WFIRST-like results are coloured blue. Symbols indicate the best-fit parameters, corresponding to the minimum $\chi^2$ fit. Darker shaded ellipses correspond to the $1\sigma$ uncertainties and lighter shaded ellipses correspond to the  $2\sigma$ uncertainties. The upper pairs of shaded ellipses correspond to the redshift-space results and the lower pairs of shaded ellipses correspond to the real-space results. The left-hand panel shows the results for the WISP-calibrated instance and the right-hand panel shows the results for the HiZELS-calibrated instance. The star and dotted ellipses show the parameter fit and $1\sigma$ and $2\sigma$ uncertainties obtained when using $\chi^2$ minimisation to fit a linear relation to the bias results from \citet{Cochrane17}.}
  \label{fig:redshiftVsBiasParameters}
\end{figure*}

We now examine how the linear bias of $\halpha$-emitting galaxies evolves with redshift. We can see from Fig.~\ref{fig:lightconeBias_wisp} and Fig.~\ref{fig:lightconeBias_hizels} that our fitted values for the linear bias increase with increasing redshift, as we would expect. In Fig.~\ref{fig:redshiftVsBias} we plot the fits for the linear bias from the individual redshift bins as a function of effective redshift. The filled red diamonds show the fitted bias values for the redshift bins of the Euclid-like survey, whilst the filled blue circles show the fitted bias values for the redshift bins of the WFIRST-like survey. We can now see clearly the consistency between the bias values predicted for the Euclid-like surveys and those predicted for the WFIRST-like surveys. 

The faint symbols in Fig.~\ref{fig:redshiftVsBias} show the fitted linear bias values obtained if we perform the clustering analysis with the MXXL-15K and MXXL-2K lightcones, i.e. assuming that both surveys are 100 per cent complete. This analysis is presented in Appendix~\ref{sec:lightcone_bias_fits}. The level of agreement between the linear bias values from the MXXL-15K and MXXL-2K lightcones and the Euclid-like and WFIRST-like surveys suggests that the introduction of incompleteness does not significantly affect our recovery of the bias. 

\subsubsection{Evolutionary fits}
\label{sec:evolutionaryFits}

It is clear from Fig.~\ref{fig:redshiftVsBias} that we can model the evolution of the linear bias, $b_{\rm lin}$, with redshift using a linear relation $b_{\rm lin}\left (z_{\rm eff}\right )=m\,z_{\rm eff}+c$, where $z_{\rm eff}$ is the effective redshift of the galaxy sample, $m$ is the gradient and $c$ is the intercept. We use $\chi^2$ minimisation to fit linear relations to the linear bias values for the Euclid-like and WFIRST-like surveys, in both redshift-space and real-space and for both the WISP-calibrated values and the HiZELS-calibrated values. The parameter fits, including the $1\sigma$ and $2\sigma$ contours, are shown in Fig.~\ref{fig:redshiftVsBiasParameters} and are reported in Table~\ref{tab:bias_fits_euclid_wfirst}. These fits are also shown in Fig.~\ref{fig:redshiftVsBias} where the solid and dashed lines show the linear fits whose parameters minimise the $\chi^2$ statistic. The red solid lines correspond to the fits for the Euclid-like survey and the blue dashed lines correspond to the fits for the WFIRST-like survey. The shaded regions in Fig.~\ref{fig:redshiftVsBias} indicate the range of linear fits that fall within the corresponding $1\sigma$ contour in Fig.~\ref{fig:redshiftVsBiasParameters}. 

Comparing the gradients and intercepts fits in Fig.~\ref{fig:redshiftVsBiasParameters} we see that for both the WISP-calibrated and HiZELS-calibrated instances the linear fits for the Euclid-like and WFIRST-like surveys are consistent with one another within the $1\sigma$ uncertainties. For the HiZELS-calibrated instances the Euclid-like and WFIRST-like fits have gradients and intercepts that are in excellent agreement. For the WISP-calibrated instance the WFIRST-like surveys have a slightly steeper gradient and smaller intercept than the Euclid-like surveys, though the fits are still consistent at the $1\sigma$ level. In each instance the $1\sigma$ contours for the WFIRST-like survey are typically narrower than the $1\sigma$ contours for the Euclid-like survey. The impact of this can be seen by comparing the width of the shaded regions in each panel of Fig.~\ref{fig:redshiftVsBias}. This likely arises due to the individual linear bias values in the Euclid-like surveys having larger uncertainties, particularly at high redshift, compared to the linear bias values for the WFIRST-like surveys. We note however, that the individual linear bias values for the WFIRST-like surveys typically have a larger scatter. For the WFIRST-like surveys there is a visible outlier at $z_{\rm eff}\simeq 1.5$ that is most likely due to cosmic variance, though we note from the WFIRST-like clustering analysis that the measured bias shows a significant negative deviation on scales $75\hMpc\lesssim r\lesssim 100\hMpc$ in the $1.4 \leqslant z \leqslant 1.6$ bin. Whilst this deviation, which we discussed in $\S$~\ref{sec:linear_bias_measurements}, did not impact our fit for the linear bias on scales $r<75\hMpc$, it may signify the presence of cosmic variance. We argue that this outlier is not having a significant impact on our fits in Fig.~\ref{fig:redshiftVsBias}. Overall we see that in redshift-space the Euclid-like and WFIRST-like fits are both consistent within $1\sigma$ with a linear relation $b(z)=0.7z + 0.7$, in both the WISP-calibrated and the HiZELS-calibrated instances.

\subsubsection{Comparison to observations}
\label{sec:comparisonToObs}

To compare with linear bias estimates in the literature we show in the upper panels of Fig.~\ref{fig:redshiftVsBias} the linear bias estimates from \citet{Geach12} and \citet{Cochrane17}, both of which were obtained from clustering analyses of HiZELS galaxies. The \citet{Cochrane17} results that we plot correspond to the effective bias results from their `full' samples, with no additional luminosity selection applied. If we extrapolate our linear fits for our Euclid-like and WFIRST-like surveys out to $z=2.23$, we estimate that $b_{\rm lin}\left (z_{\rm eff}=2.23\right )\sim 2.3$, which is in excellent agreement with the \citet{Geach12} result of $b_{\rm lin}=2.4^{+0.1}_{-0.2}$. Indeed, we see in Fig.~\ref{fig:redshiftVsBias} that the \citet{Geach12} result falls within the shaded regions corresponding to the $1\sigma$ uncertainty range for our evolutionary fits.

We see that our linear bias relations are also in excellent agreement with the \citet{Cochrane17} result at $z=1.47$, though our forecasts suggest a shallower evolutionary relation with redshift, leading to our forecasts predicting a higher bias at $z=0.8$ and a lower bias at $z=2.23$ compared to the \citet{Cochrane17} results. This can be seen clearly if we use $\chi^2$ minisation to fit a linear relation to the three \citet{Cochrane17} bias measurements. The best-fitting parameters for the \citet{Cochrane17} measurements are indicated by the star in Fig.~\ref{fig:redshiftVsBiasParameters}, with the dotted lines indicating the $1\sigma$ and $2\sigma$ uncertainty contours. The relation that we obtain is $b(z)=\left (0.98^{+0.06}_{-0.07}\right )z + \left (0.34^{+0.11}_{-0.09}\right )$, where the stated uncertainties correspond to the bounds of the $1\sigma$ ellipse. We can see from  Fig.~\ref{fig:redshiftVsBiasParameters} that the fit to the \citet{Cochrane17} measurements is consistent with our fits at the $2\sigma$ level.

Overall our forecasts are in good agreement with the observational bias estimates from HiZELS. 

The dotted lines in Fig.~\ref{fig:redshiftVsBias} correspond to the linear bias relation $b_{\rm lin}(z) = 0.4z + 0.9$ that was assumed by \citet{Spergel15}. We find that the \citet{Spergel15} relation is consistent with our relations in real-space, particularly at $z\sim2$. However, a difference in the slopes leads to our forecasts predicting a smaller bias for $z\lesssim2$. Comparing the \citet{Spergel15} relation to our relations in redshift-space we see that our relations predict a larger bias for all redshifts $z>0.9$. This is understandable given that the \citet{Spergel15} relation was based upon the clustering analysis of \citet{Orsi10}, which was carried out in real-space. 

\section{Discussion}
\label{sec:discussion}

\begin{figure*}
  \centering
  \includegraphics[width=0.99\textwidth]{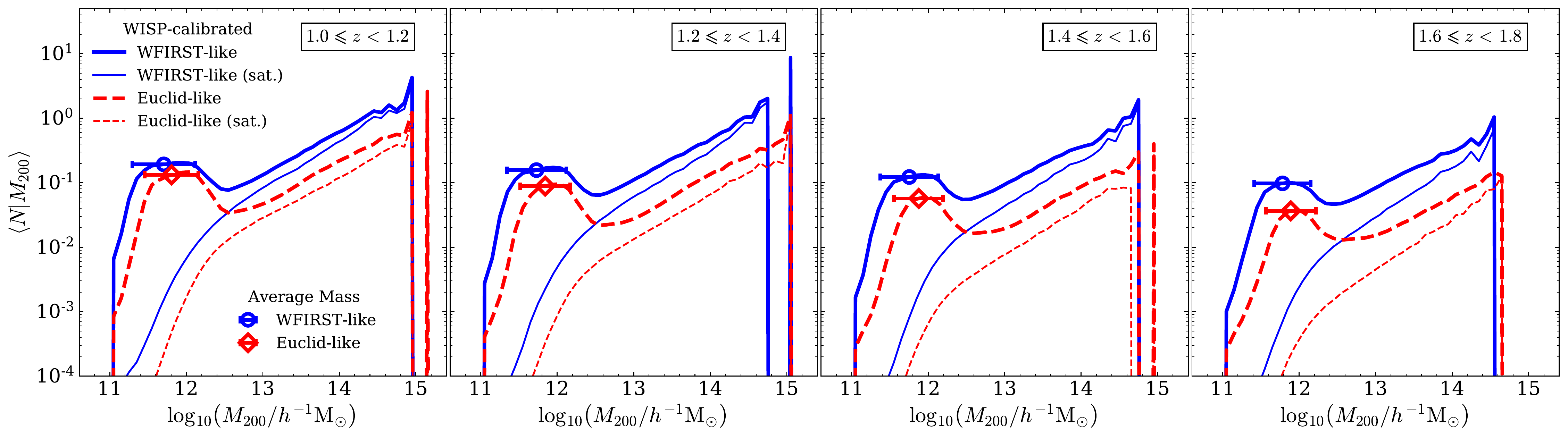}
  \includegraphics[width=0.99\textwidth]{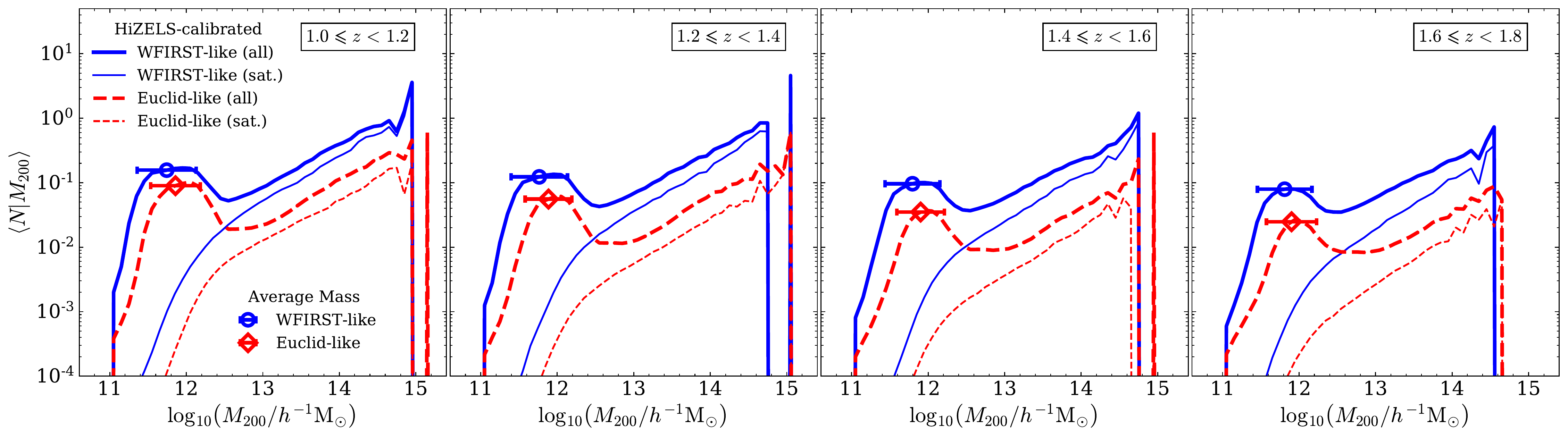}
  \caption{Halo occupation distributions (HODs) for our Euclid-like and WFIRST-like surveys. The upper row of panels show the HODs for the WISP-calibrated instance and the lower row of panels show the HODs for the HiZELS-calibrated instance. For the purposes of comparison, we compute the HODs in the four redshift bins from our WFIRST-like analysis that overlap with the redshift range for our Euclid-like analysis. The blue solid lines show the HODs for the WFIRST-like survey and the red dashed lines show the HODs for the Euclid-like survey. The thick lines show the HODs for all galaxies and the thin lines show the HODs for satellite galaxies only. The symbols show the mean and standard deviation of the halo masses hosting the galaxies in our Euclid-like and WFIRST-like samples.}
  \label{fig:euclid_wfirst_hods}
\end{figure*}

As noted previously, by comparing the linear bias fits for our Euclid-like and WFIRST-like surveys, shown in Fig.~\ref{fig:lightconeBias_wisp} and Fig.~\ref{fig:lightconeBias_hizels}, we find that the two surveys yield linear bias values that are broadly consistent and in reasonable agreement with each other. Additionally, Fig.~\ref{fig:redshiftVsBias} and Fig.~\ref{fig:redshiftVsBiasParameters} show that the redshift evolution of the bias is  consistent between the two surveys. At first sight this appears surprising, as one might expect the Euclid-like survey, which has a brighter flux-limit, to have a higher bias. However, consideration of the dark matter haloes in which the $\halpha$-emitting galaxies reside may help explain this result.

In Fig.~\ref{fig:euclid_wfirst_hods} we show the HODs for our Euclid-like and WFIRST-like surveys. For the purposes of comparison, we compute the HODs in four redshift bins ($1.0\leqslant z<1.2$, $1.2\leqslant z<1.4$, $1.4\leqslant z<1.6$ and $1.6\leqslant z<1.8$) that span the redshift range in common to both the Euclid-like and WFIRST-like surveys. The solid blue lines show the HODs for the WFIRST-like galaxies and the red dashed lines show the HODs for the Euclid-like galaxies that fall within the specified redshift bin. Thick lines correspond to the HODs for all galaxies, whilst thin lines correspond to the HODs for satellite galaxies only. The open symbols indicate the mean and standard deviation of the halo masses hosting the Euclid-like and WFIRST-like galaxies. Note that the mean and standard deviations of the halo masses hosting the galaxies used in our clustering analysis are also provided in Table~\ref{tab:bias_results_euclid} and Table~\ref{tab:bias_results_wfirst}. The mean halo masses for the Euclid-like survey and the WFIRST-like survey are consistent within error, with a value $M_{200}\sim 10^{11.8}\hMsol$. This halo mass is in excellent agreement with the results from \citet{Geach12}, who concluded from their clustering analysis of HiZELS galaxies at $z=2.23$ that $\halpha$-emitting galaxies are typically hosted by dark matter haloes with mass $\log_{10}\left (M_h\right /h^{-1} M_{\odot})=11.7\pm0.1$. Also, based upon their empirical modelling of the stellar mass-halo mass relation, \citet{Behroozi13} determined that haloes of mass $10^{11.8}\hMsol$ are the most efficient at forming stars for all redshifts $z<8$, which also is in very good agreement with our typical halo mass values.

We can see from Fig.~\ref{fig:euclid_wfirst_hods} that the Euclid-like and WFIRST-like surveys have HODs that are broadly similar in shape, with a peak at $M_{200}\lesssim 10^{12}\hMsol$ and a power-law extending to higher masses. However, the HODs differ in their normalisation. In each redshift bin we see that for any given halo mass the WFIRST-like survey has on average a larger number of galaxies per halo. This is understandable given the deeper flux limit for the WFIRST mission. The mean halo masses, shown by the open symbols, suggest that the WFIRST-like galaxies are also typically found in lower mass haloes than the Euclid-like galaxies. This can also be seen by comparing the peaks in the HODs at $M_{200}\lesssim 10^{12}\hMsol$, where the peak in the Euclid-like HOD is narrower and skewed towards slightly larger halo mass. We note however that the difference between the mean halo masses for the two surveys is smaller than the widths of the halo mass distributions and so within error the two surveys have an equivalent mean halo mass.

Since the WFIRST-like galaxies are typically placed in lower mass haloes we might again expect the WFIRST-like survey to have a smaller linear bias value than the Euclid-like survey. Halo bias models \citep[e.g.][]{Tinker10} suggest that at $M_{200}\lesssim 10^{12}\hMsol$ the halo bias is a weak function of halo mass and so the small difference in the mean halo mass between the Euclid-like and WFIRST-like surveys will have only a negligible impact on our linear bias values. However, subtle differences in the shapes of the Euclid-like and WFIRST-like HODs can help explain why the two surveys have consistent linear bias values. Considering the power-law slopes of the HODs, we see that the WFIRST-like HOD has a slightly steeper power-law slope than the Euclid-like HOD. Examining the HODs for satellite galaxies only, shown in Fig.~\ref{fig:euclid_wfirst_hods} by the thin lines, we see that this is caused by the satellite HOD for the WFIRST-like survey having a steeper power-law slope than the satellite HOD for the Euclid-like survey. As such, as we move towards larger halo mass the typical number of satellite galaxies found in any given halo will grow more rapidly for the WFIRST-like survey than for the Euclid-like survey. This can be seen in Fig.~\ref{fig:hizelsHalphaHOD_luminositySelection} and Fig.~\ref{fig:halphaHOD_fits} where, as we push down to fainter luminosities, the power-law slope of the HOD becomes steeper. Therefore, the number of faint satellites observed in high mass haloes will increase more rapidly as we push down to fainter flux limits. Since massive haloes are known to be more highly biased, the growing contribution of satellites in high mass haloes will lead to a boost in the linear bias of the galaxy sample. The fainter WFIRST-like survey would therefore receive a relatively larger boost in the measured linear bias compared to the Euclid-like survey. As a result, this could lead to the WFIRST-like survey having a linear bias that is more consistent with, or potentially greater than, the linear bias of the Euclid-like survey.

We also see that our choice of calibration does not appear to have a large impact on our measurements of the bias, or linear bias fits. We can see in Table~\ref{tab:bias_results_euclid} or Table~\ref{tab:bias_results_wfirst} that the linear bias fits using a WISP-calibrated lightcone and the fits using a HiZELS-calibrated lightcone are consistent within error with one another, despite the WISP-calibrated lightcones and HiZELS-calibrated lightcones showing a clear difference in the number density of galaxies (see Fig.~\ref{fig:lightcone_number_density}). The difference in the number density of galaxies between the two calibrations is a result of the different dust attenuation values leading to a small difference in the number of galaxies passing the flux selection. The weaker dust attenuation in the WISP-calibrated instance means that additional faint galaxies will be selected, that would not otherwise be selected in the HiZELS-calibrated instance. This is equivalent to having a applied a slightly fainter flux limit in the WISP-calibrated instance. Given the luminosity-dependent power-law slope of the HOD, the WISP-calibrated instance will thus include a larger contribution of faint satellites in highly biased, massive haloes. As a result, the WISP-calibrated instance will see a slight boost in the measured linear bias relative to the HiZELS-calibrated instance, such that the biases for the two instances are consistent despite the difference in the galaxy number densities.

\section{Conclusions \& Summary}
\label{sec:conclusions}

In this work we forecast the linear bias, as a function of redshift, for $\halpha$-emitting galaxies in a Euclid-like survey and also in a WFIRST-like survey.

To simulate our Euclid-like and WFIRST-like surveys, we use the methodology of \citet{Smith17}, whereby dark matter haloes in a lightcone are populated with galaxies by sampling from a set of luminosity-dependent halo occupation distributions (HODs). The lightcone of dark matter haloes that we use is the halo lightcone created by \citet{Smith17} from the Millennium XXL simulation \citep[MXXL,][]{Angulo12}, which has a cosmology consistent with the first year results of the Wilkinson Microwave Anisotropy Probe \citep{Spergel03}. This lightcone covers the entire sky out to a redshift of $z=2.2$ and has a halo mass resolution of $M_{200}\geqslant 10^{11}\hMsol$, where $M_{200}$ is the mass enclosed within a sphere whose mean density is 200 times the mean density of the Universe. 

Unlike broad-band photometric samples, the HOD of $\halpha$-emitters is less well understood and to-date has not been extensively parametrised. As such, instead of using a parametric form for the HOD, we generate a library of HODs as a function of limiting $\halpha+\nii$ luminosity using a physically-motivated galaxy formation model. The model that we use is the open source \galacticus{} semi-analytical model \citep{Benson10}, which predicts emission line luminosities based upon outputs from the \cloudy{} photo-ionisation code \citep{Ferland13}. Unfortunately the mass resolution of the MXXL halo merger trees  prevents us from running \galacticus{} on the simulation directly. However, we apply \galacticus{} to the Millennium Simulation \citep{Springel05}, which has an identical cosmology to the MXXL. We acknowledge that this cosmology is not consistent with current cosmological results from \citet{Planck18e}, but argue that the uncertainties are most likely dominated by our ignorance of the process of galaxy formation.

Examining the $\halpha$-emitter HODs predicted by \galacticus{}, we see that for faint $\halpha+\nii$ luminosity limits the shape of the HOD resembles a smoothed step-like function, whilst towards brighter $\halpha+\nii$ luminosity limits the amplitude of the HOD decreases, with the HOD displaying a broad peak at masses $M_{200}\sim 10^{12}\hMsol$ (Fig.~\ref{fig:hizelsHalphaHOD_luminositySelection}). This shape, which is consistent with the shapes of HODs of emission line galaxies and star-forming galaxies seen elsewhere in the literature, can be understood as resulting from feedback processes quenching star-formation in more massive haloes. Galaxies are placed into the dark matter haloes of the MXXL halo lightcone by interpolating over the library of HODs and using random sampling to draw a population of galaxies. Central galaxies are placed at the centre-of-mass of their host halo and move at the halo velocity. Satellite galaxies are placed randomly within the halo following a \citet{Navarro97} profile, with velocities drawn randomly from an isotropic Maxwell-Boltzmann distribution.

We build two lightcone galaxy catalogues: one covering $15\,000\,{\rm deg}^2$ with a $\halpha+\nii$ flux limit of $2\times 10^{-16}\ergPerSecondPerCM$, from which we build our Euclid-like survey, and one covering $2\,000\,{\rm deg}^2$ with a $\halpha+\nii$ flux limit of $1\times 10^{-16}\ergPerSecondPerCM$, from which we build our WFIRST-like survey. To calibrate the lightcones so that they have the correct number density and total number of galaxies, we apply dust attenuation to the $\halpha+\nii$ luminosities such that the lightcone luminosity functions are consistent with the luminosity functions from HiZELS \citep{Sobral13} and the lightcone cumulative number counts are consistent with the cumulative number counts from the WISP survey \citep{Mehta15}. The values for the attenuation, $A_{\halpha}$, are determined through a $\chi^2$ analysis. We find that the lightcones are able to independently reproduce the WISP number counts (Fig.~\ref{fig:lightconeCounts}) and the HiZELS luminosity functions (Fig.~\ref{fig:lightconeLF}), but that different values for $A_{\halpha}$ are required to reproduce these datasets (Fig.~\ref{fig:dustAttenuation}). The cause of this difference is uncertain, but may be due to cosmic variance between the WISP and HiZELS surveys, differences in the selection functions of the surveys or the WISP number counts probing a restricted luminosity range compared to the HiZELS luminosity functions. We therefore proceed with our bias forecasts considering both a WISP-calibrated version and a HiZELS-calibrated version of each lightcone.

To make bias forecasts we first build our Euclid-like and WFIRST-like surveys by using random sampling to apply incompleteness to the appropriate lightcone. We assume a completeness of 45 per cent for our Euclid-like survey and 70 per cent for our WFIRST-like survey. The \textsc{TreeCorr} software \citep{Jarvis04} is then used to compute the angle averaged galaxy correlation function in five redshift bins of $\Delta z=0.2$ between $0.9\leqslant z < 1.9$ for the Euclid-like survey and between $1 \leqslant z < 2$ for the WFIRST-like survey. For each redshift bin, the galaxy correlation function is computed for both WISP-calibrated and HiZELS-calibrated versions of the surveys. The correlation function is computed five times for each bin, each time using a different random seed to apply survey incompleteness. The entire set of calculations is repeated twice, with galaxies selected in redshift-space in the first instance and in real-space in the second instance. By computing the correlation function between $50\hMpc$ and $150\hMpc$ we are able to recover the BAO peak for each redshift bin (upper panels of Fig.~\ref{fig:lightconeBias_wisp} and Fig.~\ref{fig:lightconeBias_hizels}). Comparing the correlation functions in real-space and redshift-space we see that the correlation function in redshift-space has a higher amplitude, consistent with theory of redshift-space distortions from \citet{Kaiser87}. 

To compute the linear bias we divide the galaxy correlation function in each redshift bin by the non-linear dark matter correlation function, which we compute at the effective redshift of the galaxies in that bin using the \textsc{Nbodykit} \citep{Hand18} python package. We find that for scales $r\lesssim 75 \hMpc$ the measured linear bias in each redshift bin is consistent with a fixed, scale-independent value. However, at larger scales the measured bias displays scale-dependent deviations away from a constant. These sinusoidal-like deviations, which are more pronounced in redshift-space, occur at scales around the BAO scale, which suggests that they could be being caused by distortions of the shape and position of the BAO in the galaxy correlation function relative to the shape and postion predicted by linear theory. Such distortions can arise due to a combination of non-linear effects, mode coupling, redshift-space distortions, as well as even the bias itself. 

Taking the measured linear bias for separations $r<75\hMpc$, we use $\chi^2$ minimisation to fit a constant, scale-independent value for the linear bias in each redshift bin. Note that the presence of the deviations, due to distortion of the BAO, means that care must be taken when considering the scales at which to fit the linear bias. We estimate the uncertainties on the fitted linear bias values by computing the root-mean-square (RMS) of the difference between the measured and the fitted values (lower panels of Fig.~\ref{fig:lightconeBias_wisp} and Fig.~\ref{fig:lightconeBias_hizels}). The fitted linear bias values increase with increasing redshift, as we would expect. 

Using $\chi^2$ minimisation we fit linear relations to the linear bias as a function of effective redshift (Fig.~\ref{fig:redshiftVsBias}). The best-fitting linear relations for the Euclid-like and WFIRST-like surveys, (shown in Fig.~\ref{fig:redshiftVsBiasParameters} and listed in Table~\ref{tab:bias_fits_euclid_wfirst}), are consistent with one another at the $1\sigma$ level. For the HiZELS-calibrated instance the best-fitting parameters for the Euclid-like and WFIRST-like surveys are in excellent agreement, though for the WISP-calibrated instance the Euclid-like relation has a shallower gradient. In redshift-space, the best-fitting parameters for the Euclid-like and WFIRST-like surveys are both consistent with the relation $b(z)=0.7z + 0.7$. By extrapolating these linear relations we find that they are in excellent agreement with the bias result from \citet{Geach12}. Our linear bias results are consistent with the results from \citet{Cochrane17}, particularly at $z=1.47$, though our results suggest a shallower slope for the evolution of bias with redshift. Using $\chi^2$ to fit the \citet{Cochrane17} bias measurements, we find that our best-fitting parameters are consistent with the \citet{Cochrane17} fit at a $2\sigma$ level. Comparing to the linear bias relation assumed by \citet{Spergel15}, we find that our real-space relations are consistent with the \citet{Spergel15} relation, but in redshift-space our relations predict larger linear bias for all redshifts $z>0.9$. 

A surprising result is that our linear bias forecasts for the Euclid-like and WFIRST-like surveys are consistent with one another. This can be seen clearly when the linear bias values are plotted as a function of effective redshift (Fig.~\ref{fig:redshiftVsBias}). We would expect the Euclid-like survey to yield higher bias values due to the brighter flux limit. However, examination of the HODs for the Euclid-like and WFIRST-like surveys shows that the HOD for the fainter WFIRST-like survey has a steeper power-law slope at high masses, leading to a more rapidly growing number of satellite galaxies than in the HOD for the Euclid-like survey (Fig.~\ref{fig:euclid_wfirst_hods}). The greater contribution of faint satellite galaxies in high mass haloes can act to boost the bias for the WFIRST-like survey such that it is consistent with or even greater than the bias for the Euclid-like survey. Additionally, by comparing the fitted linear bias values for the WISP-calibrated and HiZELS-calibrated versions of the surveys, we find that the choice of calibration has negligible impact on the linear bias values.

\section*{Acknowledgements}
We thank the anonymous referee for their many helpful comments and suggestions that greatly enhanced the discussion of this work. We additionally thank Anahita Alvari, Iary Davidzon, Andreas Faisst, Zhongxu Zhai, and the members of the JPL Darksector group for various insightful conversations that helped improve this work. AM acknowledges sponsorship of a NASA Postdoctoral Program Fellowship. AM was supported by JPL, which is run under contract by California Institute of Technology for NASA. This work was supported by NASA ROSES grant 12-EUCLID12-0004 and by NASA grant 15-WFIRST15-0008 “Cosmology with the High Latitude Survey” WFIRST Science Investigation Team (SIT). This work used the DiRAC@Durham facility managed by the Institute for Computational Cosmology on behalf of the STFC DiRAC HPC Facility (www.dirac.ac.uk). The equipment was funded by BEIS capital funding via STFC capital grants ST/K00042X/1, ST/P002293/1, ST/R002371/1 and ST/S002502/1, Durham University and STFC operations grant ST/R000832/1. DiRAC is part of the National e-Infrastructure. Copyright 2019. All rights reserved.


\bibliographystyle{mn2e_trunc8}
\bibliography{aim,smith_thesis}

\begin{thebibliography}{115}
\expandafter\ifx\csname natexlab\endcsname\relax\def\natexlab#1{#1}\fi

\bibitem[{{Albrecht} {et~al.}(2006){Albrecht}, {Bernstein}, {Cahn}, {Freedman},
  {Hewitt}, {Hu}, {Huth}, {Kamionkowski}, {Kolb}, {Knox}, {Mather}, {Staggs},
  \& {Suntzeff}}]{Albrecht06}
{Albrecht} A., {Bernstein} G., {Cahn} R., {Freedman} W.~L., {Hewitt} J., {Hu}
  W., {Huth} J., {Kamionkowski} M. {et~al}, 2006, arXiv:0609591

\bibitem[{{Angulo} {et~al.}(2008){Angulo}, {Baugh}, {Frenk}, \&
  {Lacey}}]{Angulo08}
{Angulo} R.~E., {Baugh} C.~M., {Frenk} C.~S., {Lacey} C.~G., 2008, \mnras, 383,
  755

\bibitem[{{Angulo} {et~al.}(2012){Angulo}, {Springel}, {White}, {Jenkins},
  {Baugh}, \& {Frenk}}]{Angulo12}
{Angulo} R.~E., {Springel} V., {White} S.~D.~M., {Jenkins} A., {Baugh} C.~M.,
  {Frenk} C.~S., 2012, \mnras, 426, 2046

\bibitem[{{Angulo} {et~al.}(2014){Angulo}, {White}, {Springel}, \&
  {Henriques}}]{Angulo14}
{Angulo} R.~E., {White} S.~D.~M., {Springel} V., {Henriques} B., 2014, \mnras,
  442, 2131

\bibitem[{{Anselmi} {et~al.}(2016){Anselmi}, {Starkman}, \&
  {Sheth}}]{Anselmi16}
{Anselmi} S., {Starkman} G.~D., {Sheth} R.~K., 2016, \mnras, 455, 2474

\bibitem[{{Arnouts} {et~al.}(2005){Arnouts}, {Schiminovich}, {Ilbert},
  {Tresse}, {Milliard}, {Treyer}, {Bardelli}, {Budavari}, {Wyder}, {Zucca}, {Le
  F{\`e}vre}, {Martin}, {Vettolani}, {Adami}, {Arnaboldi}, {Barlow}, {Bianchi},
  {Bolzonella}, {Bottini}, {Byun}, {Cappi}, {Charlot}, {Contini}, {Donas},
  {Forster}, {Foucaud}, {Franzetti}, {Friedman}, {Garilli}, {Gavignaud},
  {Guzzo}, {Heckman}, {Hoopes}, {Iovino}, {Jelinsky}, {Le Brun}, {Lee},
  {Maccagni}, {Madore}, {Malina}, {Marano}, {Marinoni}, {McCracken}, {Mazure},
  {Meneux}, {Merighi}, {Morrissey}, {Neff}, {Paltani}, {Pell{\`o}}, {Picat},
  {Pollo}, {Pozzetti}, {Radovich}, {Rich}, {Scaramella}, {Scodeggio},
  {Seibert}, {Siegmund}, {Small}, {Szalay}, {Welsh}, {Xu}, {Zamorani}, \&
  {Zanichelli}}]{Arnouts05}
{Arnouts} S., {Schiminovich} D., {Ilbert} O., {Tresse} L., {Milliard} B.,
  {Treyer} M., {Bardelli} S., {Budavari} T. {et~al}, 2005, \apjl, 619, L43

\bibitem[{{Atek} {et~al.}(2010){Atek}, {Malkan}, {McCarthy}, {Teplitz},
  {Scarlata}, {Siana}, {Henry}, {Colbert}, {Ross}, {Bridge}, {Bunker},
  {Dressler}, {Fosbury}, {Martin}, \& {Shim}}]{Atek10}
{Atek} H., {Malkan} M., {McCarthy} P., {Teplitz} H.~I., {Scarlata} C., {Siana}
  B., {Henry} A., {Colbert} J.~W. {et~al}, 2010, \apj, 723, 104

\bibitem[{{Atek} {et~al.}(2011){Atek}, {Siana}, {Scarlata}, {Malkan},
  {McCarthy}, {Teplitz}, {Henry}, {Colbert}, {Bridge}, {Bunker}, {Dressler},
  {Fosbury}, {Hathi}, {Martin}, {Ross}, \& {Shim}}]{Atek11}
{Atek} H., {Siana} B., {Scarlata} C., {Malkan} M., {McCarthy} P., {Teplitz} H.,
  {Henry} A., {Colbert} J. {et~al}, 2011, \apj, 743, 121

\bibitem[{{Bardeen} {et~al.}(1986){Bardeen}, {Bond}, {Kaiser}, \&
  {Szalay}}]{Bardeen86}
{Bardeen} J.~M., {Bond} J.~R., {Kaiser} N., {Szalay} A.~S., 1986, \apj, 304, 15

\bibitem[{{Basilakos} {et~al.}(2012){Basilakos}, {Dent}, {Dutta},
  {Perivolaropoulos}, \& {Plionis}}]{Basilakos12}
{Basilakos} S., {Dent} J.~B., {Dutta} S., {Perivolaropoulos} L., {Plionis} M.,
  2012, \prd, 85, 123501

\bibitem[{{Behroozi} {et~al.}(2013){Behroozi}, {Wechsler}, \&
  {Conroy}}]{Behroozi13}
{Behroozi} P.~S., {Wechsler} R.~H., {Conroy} C., 2013, \apj, 770, 57

\bibitem[{{Benson}(2010)}]{Benson10}
{Benson} A.~J., 2010, \physrep, 495, 33

\bibitem[{{Benson}(2012)}]{Benson12}
---, 2012, \na, 17, 175

\bibitem[{{Benson}(2014)}]{Benson14}
---, 2014, \mnras, 444, 2599

\bibitem[{{Benson} {et~al.}(2000{\natexlab{a}}){Benson}, {Baugh}, {Cole},
  {Frenk}, \& {Lacey}}]{Benson00b}
{Benson} A.~J., {Baugh} C.~M., {Cole} S., {Frenk} C.~S., {Lacey} C.~G.,
  2000{\natexlab{a}}, \mnras, 316, 107

\bibitem[{{Benson} {et~al.}(2000{\natexlab{b}}){Benson}, {Cole}, {Frenk},
  {Baugh}, \& {Lacey}}]{Benson00a}
{Benson} A.~J., {Cole} S., {Frenk} C.~S., {Baugh} C.~M., {Lacey} C.~G.,
  2000{\natexlab{b}}, \mnras, 311, 793

\bibitem[{{Berlind} \& {Weinberg}(2002)}]{Berlind02}
{Berlind} A.~A., {Weinberg} D.~H., 2002, \apj, 575, 587

\bibitem[{{Blake} \& {Glazebrook}(2003)}]{Blake03}
{Blake} C., {Glazebrook} K., 2003, \apj, 594, 665

\bibitem[{{Blanton} {et~al.}(2000){Blanton}, {Cen}, {Ostriker}, {Strauss}, \&
  {Tegmark}}]{Blanton00}
{Blanton} M., {Cen} R., {Ostriker} J.~P., {Strauss} M.~A., {Tegmark} M., 2000,
  \apj, 531, 1

\bibitem[{{Brammer} {et~al.}(2011){Brammer}, {Whitaker}, {van Dokkum},
  {Marchesini}, {Franx}, {Kriek}, {Labb{\'e}}, {Lee}, {Muzzin}, {Quadri},
  {Rudnick}, \& {Williams}}]{Brammer11}
{Brammer} G.~B., {Whitaker} K.~E., {van Dokkum} P.~G., {Marchesini} D., {Franx}
  M., {Kriek} M., {Labb{\'e}} I., {Lee} K.-S. {et~al}, 2011, \apj, 739, 24

\bibitem[{{Calzetti} {et~al.}(2000){Calzetti}, {Armus}, {Bohlin}, {Kinney},
  {Koornneef}, \& {Storchi-Bergmann}}]{Calzetti00}
{Calzetti} D., {Armus} L., {Bohlin} R.~C., {Kinney} A.~L., {Koornneef} J.,
  {Storchi-Bergmann} T., 2000, \apj, 533, 682

\bibitem[{{Chabrier}(2003)}]{Chabrier03}
{Chabrier} G., 2003, \apjl, 586, L133

\bibitem[{{Charlot} \& {Fall}(2000)}]{Charlot00}
{Charlot} S., {Fall} S.~M., 2000, \apj, 539, 718

\bibitem[{{Clerkin} {et~al.}(2015){Clerkin}, {Kirk}, {Lahav}, {Abdalla}, \&
  {Gazta{\~n}aga}}]{Clerkin15}
{Clerkin} L., {Kirk} D., {Lahav} O., {Abdalla} F.~B., {Gazta{\~n}aga} E., 2015,
  \mnras, 448, 1389

\bibitem[{{Cochrane} \& {Best}(2018)}]{Cochrane18b}
{Cochrane} R.~K., {Best} P.~N., 2018, \mnras, 480, 864

\bibitem[{{Cochrane} {et~al.}(2018){Cochrane}, {Best}, {Sobral}, {Smail},
  {Geach}, {Stott}, \& {Wake}}]{Cochrane18a}
{Cochrane} R.~K., {Best} P.~N., {Sobral} D., {Smail} I., {Geach} J.~E., {Stott}
  J.~P., {Wake} D.~A., 2018, \mnras, 475, 3730

\bibitem[{{Cochrane} {et~al.}(2017){Cochrane}, {Best}, {Sobral}, {Smail},
  {Wake}, {Stott}, \& {Geach}}]{Cochrane17}
{Cochrane} R.~K., {Best} P.~N., {Sobral} D., {Smail} I., {Wake} D.~A., {Stott}
  J.~P., {Geach} J.~E., 2017, \mnras, 469, 2913

\bibitem[{{Conroy} {et~al.}(2010){Conroy}, {Schiminovich}, \&
  {Blanton}}]{Conroy10}
{Conroy} C., {Schiminovich} D., {Blanton} M.~R., 2010, \apj, 718, 184

\bibitem[{{Contreras} {et~al.}(2013){Contreras}, {Baugh}, {Norberg}, \&
  {Padilla}}]{Contreras13}
{Contreras} S., {Baugh} C.~M., {Norberg} P., {Padilla} N., 2013, \mnras, 432,
  2717

\bibitem[{{Crocce} \& {Scoccimarro}(2008)}]{Crocce08}
{Crocce} M., {Scoccimarro} R., 2008, \prd, 77, 023533

\bibitem[{{Davis} \& {Geller}(1976)}]{Davis76}
{Davis} M., {Geller} M.~J., 1976, \apj, 208, 13

\bibitem[{{DESI Collaboration} {et~al.}(2016){DESI Collaboration}, {Aghamousa},
  {Aguilar}, {Ahlen}, {Alam}, {Allen}, {Allende Prieto}, {Annis}, {Bailey},
  {Balland}, \& et~al.}]{DESI16a}
{DESI Collaboration}, {Aghamousa} A., {Aguilar} J., {Ahlen} S., {Alam} S.,
  {Allen} L.~E., {Allende Prieto} C., {Annis} J. {et~al}, 2016, arXiv:161100036

\bibitem[{{Desjacques} {et~al.}(2018){Desjacques}, {Jeong}, \&
  {Schmidt}}]{Desjacques18}
{Desjacques} V., {Jeong} D., {Schmidt} F., 2018, \physrep, 733, 1

\bibitem[{{Dom{\'{\i}}nguez} {et~al.}(2013){Dom{\'{\i}}nguez}, {Siana},
  {Henry}, {Scarlata}, {Bedregal}, {Malkan}, {Atek}, {Ross}, {Colbert},
  {Teplitz}, {Rafelski}, {McCarthy}, {Bunker}, {Hathi}, {Dressler}, {Martin},
  \& {Masters}}]{Dominguez13}
{Dom{\'{\i}}nguez} A., {Siana} B., {Henry} A.~L., {Scarlata} C., {Bedregal}
  A.~G., {Malkan} M., {Atek} H., {Ross} N.~R. {et~al}, 2013, \apj, 763, 145

\bibitem[{{Dressler}(1980)}]{Dressler80}
{Dressler} A., 1980, \apj, 236, 351

\bibitem[{{Dressler} {et~al.}(2012){Dressler}, {Spergel}, {Mountain},
  {Postman}, {Elliott}, {Bendek}, {Bennett}, {Dalcanton}, {Gaudi}, {Gehrels},
  {Guyon}, {Hirata}, {Kalirai}, {Kasdin}, {Kruk}, {Macintosh}, {Malhotra},
  {Penny}, {Perlmutter}, {Rieke}, {Riess}, {Rhoads}, {Shaklan}, {Somerville},
  {Stern}, {Thompson}, \& {Weinberg}}]{Dressler12}
{Dressler} A., {Spergel} D., {Mountain} M., {Postman} M., {Elliott} E.,
  {Bendek} E., {Bennett} D., {Dalcanton} J. {et~al}, 2012, arXiv:12107809

\bibitem[{{Durkalec} {et~al.}(2018){Durkalec}, {Le F{\`e}vre}, {Pollo},
  {Zamorani}, {Lemaux}, {Garilli}, {Bardelli}, {Hathi}, {Koekemoer}, {Pforr},
  \& {Zucca}}]{Durkalec18}
{Durkalec} A., {Le F{\`e}vre} O., {Pollo} A., {Zamorani} G., {Lemaux} B.~C.,
  {Garilli} B., {Bardelli} S., {Hathi} N. {et~al}, 2018, \aap, 612, A42

\bibitem[{{Eisenstein} {et~al.}(2007){Eisenstein}, {Seo}, \&
  {White}}]{Eisenstein07a}
{Eisenstein} D.~J., {Seo} H.-J., {White} M., 2007, \apj, 664, 660

\bibitem[{{Faisst} {et~al.}(2018){Faisst}, {Masters}, {Wang}, {Merson},
  {Capak}, {Malhotra}, \& {Rhoads}}]{Faisst18}
{Faisst} A.~L., {Masters} D., {Wang} Y., {Merson} A., {Capak} P., {Malhotra}
  S., {Rhoads} J.~E., 2018, \apj, 855, 132

\bibitem[{{Favole} {et~al.}(2017){Favole}, {Rodr{\'{\i}}guez-Torres},
  {Comparat}, {Prada}, {Guo}, {Klypin}, \& {Montero-Dorta}}]{Favole17}
{Favole} G., {Rodr{\'{\i}}guez-Torres} S.~A., {Comparat} J., {Prada} F., {Guo}
  H., {Klypin} A., {Montero-Dorta} A.~D., 2017, \mnras, 472, 550

\bibitem[{{Ferland} {et~al.}(2013){Ferland}, {Porter}, {van Hoof}, {Williams},
  {Abel}, {Lykins}, {Shaw}, {Henney}, \& {Stancil}}]{Ferland13}
{Ferland} G.~J., {Porter} R.~L., {van Hoof} P.~A.~M., {Williams} R.~J.~R.,
  {Abel} N.~P., {Lykins} M.~L., {Shaw} G., {Henney} W.~J. {et~al}, 2013,
  \rmxaa, 49, 137

\bibitem[{{Ferrara} {et~al.}(1999){Ferrara}, {Bianchi}, {Cimatti}, \&
  {Giovanardi}}]{Ferrara99}
{Ferrara} A., {Bianchi} S., {Cimatti} A., {Giovanardi} C., 1999, \apjs, 123,
  437

\bibitem[{{Fry}(1996)}]{Fry96}
{Fry} J.~N., 1996, \apjl, 461, L65

\bibitem[{{Garn} \& {Best}(2010)}]{Garn10b}
{Garn} T., {Best} P.~N., 2010, \mnras, 409, 421

\bibitem[{{Garn} {et~al.}(2010){Garn}, {Sobral}, {Best}, {Geach}, {Smail},
  {Cirasuolo}, {Dalton}, {Dunlop}, {McLure}, \& {Farrah}}]{Garn10a}
{Garn} T., {Sobral} D., {Best} P.~N., {Geach} J.~E., {Smail} I., {Cirasuolo}
  M., {Dalton} G.~B., {Dunlop} J.~S. {et~al}, 2010, \mnras, 402, 2017

\bibitem[{{Gazta{\~n}aga} {et~al.}(2012){Gazta{\~n}aga}, {Eriksen}, {Crocce},
  {Castander}, {Fosalba}, {Marti}, {Miquel}, \& {Cabr{\'e}}}]{Gaztanaga12}
{Gazta{\~n}aga} E., {Eriksen} M., {Crocce} M., {Castander} F.~J., {Fosalba} P.,
  {Marti} P., {Miquel} R., {Cabr{\'e}} A., 2012, \mnras, 422, 2904

\bibitem[{{Geach} {et~al.}(2008){Geach}, {Smail}, {Best}, {Kurk}, {Casali},
  {Ivison}, \& {Coppin}}]{Geach08}
{Geach} J.~E., {Smail} I., {Best} P.~N., {Kurk} J., {Casali} M., {Ivison}
  R.~J., {Coppin} K., 2008, \mnras, 388, 1473

\bibitem[{{Geach} {et~al.}(2012){Geach}, {Sobral}, {Hickox}, {Wake}, {Smail},
  {Best}, {Baugh}, \& {Stott}}]{Geach12}
{Geach} J.~E., {Sobral} D., {Hickox} R.~C., {Wake} D.~A., {Smail} I., {Best}
  P.~N., {Baugh} C.~M., {Stott} J.~P., 2012, \mnras, 426, 679

\bibitem[{{Gonzalez-Perez} {et~al.}(2018){Gonzalez-Perez}, {Comparat},
  {Norberg}, {Baugh}, {Contreras}, {Lacey}, {McCullagh}, {Orsi}, {Helly}, \&
  {Humphries}}]{Gonzalez-Perez18}
{Gonzalez-Perez} V., {Comparat} J., {Norberg} P., {Baugh} C.~M., {Contreras}
  S., {Lacey} C., {McCullagh} N., {Orsi} A. {et~al}, 2018, \mnras, 474, 4024

\bibitem[{{Green} {et~al.}(2012){Green}, {Schechter}, {Baltay}, {Bean},
  {Bennett}, {Brown}, {Conselice}, {Donahue}, {Fan}, {Gaudi}, {Hirata},
  {Kalirai}, {Lauer}, {Nichol}, {Padmanabhan}, {Perlmutter}, {Rauscher},
  {Rhodes}, {Roellig}, {Stern}, {Sumi}, {Tanner}, {Wang}, {Weinberg}, {Wright},
  {Gehrels}, {Sambruna}, {Traub}, {Anderson}, {Cook}, {Garnavich},
  {Hillenbrand}, {Ivezic}, {Kerins}, {Lunine}, {McDonald}, {Penny}, {Phillips},
  {Rieke}, {Riess}, {van der Marel}, {Barry}, {Cheng}, {Content}, {Cutri},
  {Goullioud}, {Grady}, {Helou}, {Jackson}, {Kruk}, {Melton}, {Peddie},
  {Rioux}, \& {Seiffert}}]{Green12}
{Green} J., {Schechter} P., {Baltay} C., {Bean} R., {Bennett} D., {Brown} R.,
  {Conselice} C., {Donahue} M. {et~al}, 2012, arXiv:12084012

\bibitem[{{Guo} {et~al.}(2013){Guo}, {Zehavi}, {Zheng}, {Weinberg}, {Berlind},
  {Blanton}, {Chen}, {Eisenstein}, {Ho}, {Kazin}, {Manera}, {Maraston},
  {McBride}, {Nuza}, {Padmanabhan}, {Parejko}, {Percival}, {Ross}, {Ross},
  {Samushia}, {S{\'a}nchez}, {Schlegel}, {Schneider}, {Skibba}, {Swanson},
  {Tinker}, {Tojeiro}, {Wake}, {White}, {Bahcall}, {Bizyaev}, {Brewington},
  {Bundy}, {da Costa}, {Ebelke}, {Malanushenko}, {Malanushenko}, {Oravetz},
  {Rossi}, {Simmons}, {Snedden}, {Streblyanska}, \& {Thomas}}]{Guo13}
{Guo} H., {Zehavi} I., {Zheng} Z., {Weinberg} D.~H., {Berlind} A.~A., {Blanton}
  M., {Chen} Y., {Eisenstein} D.~J. {et~al}, 2013, \apj, 767, 122

\bibitem[{{Guzzo} {et~al.}(2008){Guzzo}, {Pierleoni}, {Meneux}, {Branchini},
  {Le F{\`e}vre}, {Marinoni}, {Garilli}, {Blaizot}, {De Lucia}, {Pollo},
  {McCracken}, {Bottini}, {Le Brun}, {Maccagni}, {Picat}, {Scaramella},
  {Scodeggio}, {Tresse}, {Vettolani}, {Zanichelli}, {Adami}, {Arnouts},
  {Bardelli}, {Bolzonella}, {Bongiorno}, {Cappi}, {Charlot}, {Ciliegi},
  {Contini}, {Cucciati}, {de la Torre}, {Dolag}, {Foucaud}, {Franzetti},
  {Gavignaud}, {Ilbert}, {Iovino}, {Lamareille}, {Marano}, {Mazure}, {Memeo},
  {Merighi}, {Moscardini}, {Paltani}, {Pell{\`o}}, {Perez-Montero}, {Pozzetti},
  {Radovich}, {Vergani}, {Zamorani}, \& {Zucca}}]{Guzzo08}
{Guzzo} L., {Pierleoni} M., {Meneux} B., {Branchini} E., {Le F{\`e}vre} O.,
  {Marinoni} C., {Garilli} B., {Blaizot} J. {et~al}, 2008, \nat, 451, 541

\bibitem[{{Guzzo} {et~al.}(1997){Guzzo}, {Strauss}, {Fisher}, {Giovanelli}, \&
  {Haynes}}]{Guzzo97}
{Guzzo} L., {Strauss} M.~A., {Fisher} K.~B., {Giovanelli} R., {Haynes} M.~P.,
  1997, \apj, 489, 37

\bibitem[{{Hada} \& {Eisenstein}(2019)}]{Hada19}
{Hada} R., {Eisenstein} D.~J., 2019, \mnras, 482, 5685

\bibitem[{{Hand} {et~al.}(2018){Hand}, {Feng}, {Beutler}, {Li}, {Modi},
  {Seljak}, \& {Slepian}}]{Hand18}
{Hand} N., {Feng} Y., {Beutler} F., {Li} Y., {Modi} C., {Seljak} U., {Slepian}
  Z., 2018, \aj, 156, 160

\bibitem[{{Hatfield} {et~al.}(2016){Hatfield}, {Lindsay}, {Jarvis},
  {H{\"a}u{\ss}ler}, {Vaccari}, \& {Verma}}]{Hatfield16}
{Hatfield} P.~W., {Lindsay} S.~N., {Jarvis} M.~J., {H{\"a}u{\ss}ler} B.,
  {Vaccari} M., {Verma} A., 2016, \mnras, 459, 2618

\bibitem[{{Hui} \& {Parfrey}(2008)}]{Hui08}
{Hui} L., {Parfrey} K.~P., 2008, \prd, 77, 043527

\bibitem[{{Jarvis} {et~al.}(2004){Jarvis}, {Bernstein}, \& {Jain}}]{Jarvis04}
{Jarvis} M., {Bernstein} G., {Jain} B., 2004, \mnras, 352, 338

\bibitem[{{Jeong} \& {Komatsu}(2006)}]{Jeong06}
{Jeong} D., {Komatsu} E., 2006, \apj, 651, 619

\bibitem[{{Kaiser}(1984)}]{Kaiser84}
{Kaiser} N., 1984, \apjl, 284, L9

\bibitem[{{Kaiser}(1987)}]{Kaiser87}
---, 1987, \mnras, 227, 1

\bibitem[{{Kauffmann} {et~al.}(1997){Kauffmann}, {Nusser}, \&
  {Steinmetz}}]{Kauffmann97}
{Kauffmann} G., {Nusser} A., {Steinmetz} M., 1997, \mnras, 286, 795

\bibitem[{{Kim} {et~al.}(2015){Kim}, {Im}, {Lee}, {Edge}, {Wake}, {Merson}, \&
  {Jeon}}]{Kim15}
{Kim} J.-W., {Im} M., {Lee} S.-K., {Edge} A.~C., {Wake} D.~A., {Merson} A.~I.,
  {Jeon} Y., 2015, \apj, 806, 189

\bibitem[{{Knebe} {et~al.}(2015){Knebe}, {Pearce}, {Thomas}, {Benson},
  {Blaizot}, {Bower}, {Carretero}, {Castander}, {Cattaneo}, {Cora}, {Croton},
  {Cui}, {Cunnama}, {De Lucia}, {Devriendt}, {Elahi}, {Font}, {Fontanot},
  {Garcia-Bellido}, {Gargiulo}, {Gonzalez-Perez}, {Helly}, {Henriques},
  {Hirschmann}, {Lee}, {Mamon}, {Monaco}, {Onions}, {Padilla}, {Power},
  {Pujol}, {Skibba}, {Somerville}, {Srisawat}, {Vega-Mart{\'{\i}}nez}, \&
  {Yi}}]{Knebe15}
{Knebe} A., {Pearce} F.~R., {Thomas} P.~A., {Benson} A., {Blaizot} J., {Bower}
  R., {Carretero} J., {Castander} F.~J. {et~al}, 2015, \mnras, 451, 4029

\bibitem[{{Landy} \& {Szalay}(1993)}]{Landy93}
{Landy} S.~D., {Szalay} A.~S., 1993, \apj, 412, 64

\bibitem[{{Laureijs} {et~al.}(2011){Laureijs}, {Amiaux}, {Arduini},
  {Augu{\`e}res}, {Brinchmann}, {Cole}, {Cropper}, {Dabin}, {Duvet}, {Ealet},
  \& et~al.}]{Laureijs11}
{Laureijs} R., {Amiaux} J., {Arduini} S., {Augu{\`e}res} J.~., {Brinchmann} J.,
  {Cole} R., {Cropper} M., {Dabin} C. {et~al}, 2011, arXiv:11103193

\bibitem[{{Law-Smith} \& {Eisenstein}(2017)}]{Law-Smith17}
{Law-Smith} J., {Eisenstein} D.~J., 2017, \apj, 836, 87

\bibitem[{{Lemson} \& {Virgo Consortium}(2006)}]{Lemson06}
{Lemson} G., {Virgo Consortium} t., 2006, arXiv:0608019

\bibitem[{{Li} \& {White}(2009)}]{Li09}
{Li} C., {White} S.~D.~M., 2009, \mnras, 398, 2177

\bibitem[{{Ly} {et~al.}(2007){Ly}, {Malkan}, {Kashikawa}, {Shimasaku}, {Doi},
  {Nagao}, {Iye}, {Kodama}, {Morokuma}, \& {Motohara}}]{Ly07}
{Ly} C., {Malkan} M.~A., {Kashikawa} N., {Shimasaku} K., {Doi} M., {Nagao} T.,
  {Iye} M., {Kodama} T. {et~al}, 2007, \apj, 657, 738

\bibitem[{{Madau} \& {Dickinson}(2014)}]{Madau14}
{Madau} P., {Dickinson} M., 2014, \araa, 52, 415

\bibitem[{{Madau} {et~al.}(1998){Madau}, {Pozzetti}, \& {Dickinson}}]{Madau98}
{Madau} P., {Pozzetti} L., {Dickinson} M., 1998, \apj, 498, 106

\bibitem[{{Mann} {et~al.}(1998){Mann}, {Peacock}, \& {Heavens}}]{Mann98}
{Mann} R.~G., {Peacock} J.~A., {Heavens} A.~F., 1998, \mnras, 293, 209

\bibitem[{{Martens} {et~al.}(2019){Martens}, {Fang}, {Troxel}, {DeRose},
  {Hirata}, {Wechsler}, \& {Wang}}]{Martens19}
{Martens} D., {Fang} X., {Troxel} M.~A., {DeRose} J., {Hirata} C.~M.,
  {Wechsler} R.~H., {Wang} Y., 2019, \mnras, 485, 211

\bibitem[{{McCracken} {et~al.}(2015){McCracken}, {Wolk}, {Colombi},
  {Kilbinger}, {Ilbert}, {Peirani}, {Coupon}, {Dunlop}, {Milvang-Jensen},
  {Caputi}, {Aussel}, {B{\'e}thermin}, \& {Le F{\`e}vre}}]{McCracken15}
{McCracken} H.~J., {Wolk} M., {Colombi} S., {Kilbinger} M., {Ilbert} O.,
  {Peirani} S., {Coupon} J., {Dunlop} J. {et~al}, 2015, \mnras, 449, 901

\bibitem[{{Mehta} {et~al.}(2015){Mehta}, {Scarlata}, {Colbert}, {Dai},
  {Dressler}, {Henry}, {Malkan}, {Rafelski}, {Siana}, {Teplitz}, {Bagley},
  {Beck}, {Ross}, {Rutkowski}, \& {Wang}}]{Mehta15}
{Mehta} V., {Scarlata} C., {Colbert} J.~W., {Dai} Y.~S., {Dressler} A., {Henry}
  A., {Malkan} M., {Rafelski} M. {et~al}, 2015, \apj, 811, 141

\bibitem[{{Merson} {et~al.}(2018){Merson}, {Wang}, {Benson}, {Faisst},
  {Masters}, {Kiessling}, \& {Rhodes}}]{Merson18}
{Merson} A., {Wang} Y., {Benson} A., {Faisst} A., {Masters} D., {Kiessling} A.,
  {Rhodes} J., 2018, \mnras, 474, 177

\bibitem[{{Merson} {et~al.}(2013){Merson}, {Baugh}, {Helly}, {Gonzalez-Perez},
  {Cole}, {Bielby}, {Norberg}, {Frenk}, {Benson}, {Bower}, {Lacey}, \&
  {Lagos}}]{Merson13}
{Merson} A.~I., {Baugh} C.~M., {Helly} J.~C., {Gonzalez-Perez} V., {Cole} S.,
  {Bielby} R., {Norberg} P., {Frenk} C.~S. {et~al}, 2013, \mnras, 429, 556

\bibitem[{{Mirbabayi} {et~al.}(2015){Mirbabayi}, {Schmidt}, \&
  {Zaldarriaga}}]{Mirbabayi15}
{Mirbabayi} M., {Schmidt} F., {Zaldarriaga} M., 2015, \jcap, 7, 030

\bibitem[{{Navarro} {et~al.}(1997){Navarro}, {Frenk}, \& {White}}]{Navarro97}
{Navarro} J.~F., {Frenk} C.~S., {White} S.~D.~M., 1997, \apj, 490, 493

\bibitem[{{Norberg} {et~al.}(2002){Norberg}, {Baugh}, {Hawkins}, {Maddox},
  {Madgwick}, {Lahav}, {Cole}, {Frenk}, {Baldry}, {Bland-Hawthorn}, {Bridges},
  {Cannon}, {Colless}, {Collins}, {Couch}, {Dalton}, {De Propris}, {Driver},
  {Efstathiou}, {Ellis}, {Glazebrook}, {Jackson}, {Lewis}, {Lumsden},
  {Peacock}, {Peterson}, {Sutherland}, \& {Taylor}}]{Norberg02}
{Norberg} P., {Baugh} C.~M., {Hawkins} E., {Maddox} S., {Madgwick} D., {Lahav}
  O., {Cole} S., {Frenk} C.~S. {et~al}, 2002, \mnras, 332, 827

\bibitem[{{Norberg} {et~al.}(2001){Norberg}, {Baugh}, {Hawkins}, {Maddox},
  {Peacock}, {Cole}, {Frenk}, {Bland-Hawthorn}, {Bridges}, {Cannon}, {Colless},
  {Collins}, {Couch}, {Dalton}, {De Propris}, {Driver}, {Efstathiou}, {Ellis},
  {Glazebrook}, {Jackson}, {Lahav}, {Lewis}, {Lumsden}, {Madgwick}, {Peterson},
  {Sutherland}, \& {Taylor}}]{Norberg01}
{Norberg} P., {Baugh} C.~M., {Hawkins} E., {Maddox} S., {Peacock} J.~A., {Cole}
  S., {Frenk} C.~S., {Bland-Hawthorn} J. {et~al}, 2001, \mnras, 328, 64

\bibitem[{{Orsi} {et~al.}(2010){Orsi}, {Baugh}, {Lacey}, {Cimatti}, {Wang}, \&
  {Zamorani}}]{Orsi10}
{Orsi} A., {Baugh} C.~M., {Lacey} C.~G., {Cimatti} A., {Wang} Y., {Zamorani}
  G., 2010, \mnras, 405, 1006

\bibitem[{{Peacock} \& {Smith}(2000)}]{Peacock00}
{Peacock} J.~A., {Smith} R.~E., 2000, \mnras, 318, 1144

\bibitem[{{Planck Collaboration} {et~al.}(2018){Planck Collaboration},
  {Aghanim}, {Akrami}, {Ashdown}, {Aumont}, {Baccigalupi}, {Ballardini},
  {Banday}, {Barreiro}, {Bartolo}, {Basak}, {Battye}, {Benabed}, {Bernard},
  {Bersanelli}, {Bielewicz}, {Bock}, {Bond}, {Borrill}, {Bouchet}, {Boulanger},
  {Bucher}, {Burigana}, {Butler}, {Calabrese}, {Cardoso}, {Carron},
  {Challinor}, {Chiang}, {Chluba}, {Colombo}, {Combet}, {Contreras}, {Crill},
  {Cuttaia}, {de Bernardis}, {de Zotti}, {Delabrouille}, {Delouis}, {Di
  Valentino}, {Diego}, {Dor{\'e}}, {Douspis}, {Ducout}, {Dupac}, {Dusini},
  {Efstathiou}, {Elsner}, {En{\ss}lin}, {Eriksen}, {Fantaye}, {Farhang},
  {Fergusson}, {Fernandez-Cobos}, {Finelli}, {Forastieri}, {Frailis},
  {Franceschi}, {Frolov}, {Galeotta}, {Galli}, {Ganga}, {G{\'e}nova-Santos},
  {Gerbino}, {Ghosh}, {Gonz{\'a}lez-Nuevo}, {G{\'o}rski}, {Gratton},
  {Gruppuso}, {Gudmundsson}, {Hamann}, {Handley}, {Herranz}, {Hivon}, {Huang},
  {Jaffe}, {Jones}, {Karakci}, {Keih{\"a}nen}, {Keskitalo}, {Kiiveri}, {Kim},
  {Kisner}, {Knox}, {Krachmalnicoff}, {Kunz}, {Kurki-Suonio}, {Lagache},
  {Lamarre}, {Lasenby}, {Lattanzi}, {Lawrence}, {Le Jeune}, {Lemos},
  {Lesgourgues}, {Levrier}, {Lewis}, {Liguori}, {Lilje}, {Lilley}, {Lindholm},
  {L{\'o}pez-Caniego}, {Lubin}, {Ma}, {Mac{\'{\i}}as-P{\'e}rez}, {Maggio},
  {Maino}, {Mandolesi}, {Mangilli}, {Marcos-Caballero}, {Maris}, {Martin},
  {Martinelli}, {Mart{\'{\i}}nez-Gonz{\'a}lez}, {Matarrese}, {Mauri}, {McEwen},
  {Meinhold}, {Melchiorri}, {Mennella}, {Migliaccio}, {Millea}, {Mitra},
  {Miville-Desch{\^e}nes}, {Molinari}, {Montier}, {Morgante}, {Moss}, {Natoli},
  {N{\o}rgaard-Nielsen}, {Pagano}, {Paoletti}, {Partridge}, {Patanchon},
  {Peiris}, {Perrotta}, {Pettorino}, {Piacentini}, {Polastri}, {Polenta},
  {Puget}, {Rachen}, {Reinecke}, {Remazeilles}, {Renzi}, {Rocha}, {Rosset},
  {Roudier}, {Rubi{\~n}o-Mart{\'{\i}}n}, {Ruiz-Granados}, {Salvati}, {Sandri},
  {Savelainen}, {Scott}, {Shellard}, {Sirignano}, {Sirri}, {Spencer},
  {Sunyaev}, {Suur-Uski}, {Tauber}, {Tavagnacco}, {Tenti}, {Toffolatti},
  {Tomasi}, {Trombetti}, {Valenziano}, {Valiviita}, {Van Tent}, {Vibert},
  {Vielva}, {Villa}, {Vittorio}, {Wandelt}, {Wehus}, {White}, {White},
  {Zacchei}, \& {Zonca}}]{Planck18e}
{Planck Collaboration}, {Aghanim} N., {Akrami} Y., {Ashdown} M., {Aumont} J.,
  {Baccigalupi} C., {Ballardini} M., {Banday} A.~J. {et~al}, 2018,
  arXiv:180706209

\bibitem[{{Pozzetti} {et~al.}(2016){Pozzetti}, {Hirata}, {Geach}, {Cimatti},
  {Baugh}, {Cucciati}, {Merson}, {Norberg}, \& {Shi}}]{Pozzetti16}
{Pozzetti} L., {Hirata} C.~M., {Geach} J.~E., {Cimatti} A., {Baugh} C.,
  {Cucciati} O., {Merson} A., {Norberg} P. {et~al}, 2016, \aap, 590, A3

\bibitem[{{Salvador} {et~al.}(2019){Salvador}, {S{\'a}nchez}, {Pagul},
  {Garc{\'{\i}}a-Bellido}, {Sanchez}, {Pujol}, {Frieman}, {Gaztanaga}, {Ross},
  {Sevilla-Noarbe}, {Abbott}, {Allam}, {Annis}, {Avila}, {Bertin}, {Brooks},
  {Burke}, {Carnero Rosell}, {Carrasco Kind}, {Carretero}, {Castander},
  {Cunha}, {De Vicente}, {Diehl}, {Doel}, {Evrard}, {Fosalba}, {Gruen},
  {Gruendl}, {Gschwend}, {Gutierrez}, {Hartley}, {Hollowood}, {James}, {Kuehn},
  {Kuropatkin}, {Lahav}, {Lima}, {March}, {Marshall}, {Menanteau}, {Miquel},
  {Romer}, {Roodman}, {Scarpine}, {Schindler}, {Smith}, {Soares-Santos},
  {Sobreira}, {Suchyta}, {Swanson}, {Tarle}, {Thomas}, {Vikram}, \&
  {Walker}}]{Salvador19}
{Salvador} A.~I., {S{\'a}nchez} F.~J., {Pagul} A., {Garc{\'{\i}}a-Bellido} J.,
  {Sanchez} E., {Pujol} A., {Frieman} J., {Gaztanaga} E. {et~al}, 2019, \mnras,
  482, 1435

\bibitem[{{S{\'a}nchez} {et~al.}(2008){S{\'a}nchez}, {Baugh}, \&
  {Angulo}}]{Sanchez08}
{S{\'a}nchez} A.~G., {Baugh} C.~M., {Angulo} R.~E., 2008, \mnras, 390, 1470

\bibitem[{{Schechter}(1976)}]{Schechter76}
{Schechter} P., 1976, \apj, 203, 297

\bibitem[{{Seljak}(2000)}]{Seljak00}
{Seljak} U., 2000, \mnras, 318, 203

\bibitem[{{Seo} \& {Eisenstein}(2003)}]{Seo03}
{Seo} H.-J., {Eisenstein} D.~J., 2003, \apj, 598, 720

\bibitem[{{Shim} {et~al.}(2009){Shim}, {Colbert}, {Teplitz}, {Henry}, {Malkan},
  {McCarthy}, \& {Yan}}]{Shim09}
{Shim} H., {Colbert} J., {Teplitz} H., {Henry} A., {Malkan} M., {McCarthy} P.,
  {Yan} L., 2009, \apj, 696, 785

\bibitem[{{Skibba} {et~al.}(2006){Skibba}, {Sheth}, {Connolly}, \&
  {Scranton}}]{Skibba06}
{Skibba} R., {Sheth} R.~K., {Connolly} A.~J., {Scranton} R., 2006, \mnras, 369,
  68

\bibitem[{{Skibba} {et~al.}(2014){Skibba}, {Smith}, {Coil}, {Moustakas},
  {Aird}, {Blanton}, {Bray}, {Cool}, {Eisenstein}, {Mendez}, {Wong}, \&
  {Zhu}}]{Skibba14}
{Skibba} R.~A., {Smith} M.~S.~M., {Coil} A.~L., {Moustakas} J., {Aird} J.,
  {Blanton} M.~R., {Bray} A.~D., {Cool} R.~J. {et~al}, 2014, \apj, 784, 128

\bibitem[{Smith(2018)}]{SmithThesis}
Smith A., 2018, PhD thesis, Durham University

\bibitem[{{Smith} {et~al.}(2017){Smith}, {Cole}, {Baugh}, {Zheng}, {Angulo},
  {Norberg}, \& {Zehavi}}]{Smith17}
{Smith} A., {Cole} S., {Baugh} C., {Zheng} Z., {Angulo} R., {Norberg} P.,
  {Zehavi} I., 2017, \mnras, 470, 4646

\bibitem[{{Smith} {et~al.}(2008){Smith}, {Scoccimarro}, \& {Sheth}}]{Smith08}
{Smith} R.~E., {Scoccimarro} R., {Sheth} R.~K., 2008, \prd, 77, 043525

\bibitem[{{Sobral} {et~al.}(2010){Sobral}, {Best}, {Geach}, {Smail},
  {Cirasuolo}, {Garn}, {Dalton}, \& {Kurk}}]{Sobral10}
{Sobral} D., {Best} P.~N., {Geach} J.~E., {Smail} I., {Cirasuolo} M., {Garn}
  T., {Dalton} G.~B., {Kurk} J., 2010, \mnras, 404, 1551

\bibitem[{{Sobral} {et~al.}(2009){Sobral}, {Best}, {Geach}, {Smail}, {Kurk},
  {Cirasuolo}, {Casali}, {Ivison}, {Coppin}, \& {Dalton}}]{Sobral09}
{Sobral} D., {Best} P.~N., {Geach} J.~E., {Smail} I., {Kurk} J., {Cirasuolo}
  M., {Casali} M., {Ivison} R.~J. {et~al}, 2009, \mnras, 398, 75

\bibitem[{{Sobral} {et~al.}(2012){Sobral}, {Best}, {Matsuda}, {Smail}, {Geach},
  \& {Cirasuolo}}]{Sobral12}
{Sobral} D., {Best} P.~N., {Matsuda} Y., {Smail} I., {Geach} J.~E., {Cirasuolo}
  M., 2012, \mnras, 420, 1926

\bibitem[{{Sobral} {et~al.}(2013){Sobral}, {Smail}, {Best}, {Geach}, {Matsuda},
  {Stott}, {Cirasuolo}, \& {Kurk}}]{Sobral13}
{Sobral} D., {Smail} I., {Best} P.~N., {Geach} J.~E., {Matsuda} Y., {Stott}
  J.~P., {Cirasuolo} M., {Kurk} J., 2013, \mnras, 428, 1128

\bibitem[{{Song} \& {Percival}(2009)}]{Song09}
{Song} Y.-S., {Percival} W.~J., 2009, \jcap, 10, 004

\bibitem[{{Spergel} {et~al.}(2015){Spergel}, {Gehrels}, {Baltay}, {Bennett},
  {Breckinridge}, {Donahue}, {Dressler}, {Gaudi}, {Greene}, {Guyon}, {Hirata},
  {Kalirai}, {Kasdin}, {Macintosh}, {Moos}, {Perlmutter}, {Postman},
  {Rauscher}, {Rhodes}, {Wang}, {Weinberg}, {Benford}, {Hudson}, {Jeong},
  {Mellier}, {Traub}, {Yamada}, {Capak}, {Colbert}, {Masters}, {Penny},
  {Savransky}, {Stern}, {Zimmerman}, {Barry}, {Bartusek}, {Carpenter}, {Cheng},
  {Content}, {Dekens}, {Demers}, {Grady}, {Jackson}, {Kuan}, {Kruk}, {Melton},
  {Nemati}, {Parvin}, {Poberezhskiy}, {Peddie}, {Ruffa}, {Wallace}, {Whipple},
  {Wollack}, \& {Zhao}}]{Spergel15}
{Spergel} D., {Gehrels} N., {Baltay} C., {Bennett} D., {Breckinridge} J.,
  {Donahue} M., {Dressler} A., {Gaudi} B.~S. {et~al}, 2015, arXiv:150303757

\bibitem[{{Spergel} {et~al.}(2003){Spergel}, {Verde}, {Peiris}, {Komatsu},
  {Nolta}, {Bennett}, {Halpern}, {Hinshaw}, {Jarosik}, {Kogut}, {Limon},
  {Meyer}, {Page}, {Tucker}, {Weiland}, {Wollack}, \& {Wright}}]{Spergel03}
{Spergel} D.~N., {Verde} L., {Peiris} H.~V., {Komatsu} E., {Nolta} M.~R.,
  {Bennett} C.~L., {Halpern} M., {Hinshaw} G. {et~al}, 2003, \apjs, 148, 175

\bibitem[{{Springel} {et~al.}(2005){Springel}, {White}, {Jenkins}, {Frenk},
  {Yoshida}, {Gao}, {Navarro}, {Thacker}, {Croton}, {Helly}, {Peacock}, {Cole},
  {Thomas}, {Couchman}, {Evrard}, {Colberg}, \& {Pearce}}]{Springel05}
{Springel} V., {White} S.~D.~M., {Jenkins} A., {Frenk} C.~S., {Yoshida} N.,
  {Gao} L., {Navarro} J., {Thacker} R. {et~al}, 2005, \nat, 435, 629

\bibitem[{{Springel} {et~al.}(2001){Springel}, {White}, {Tormen}, \&
  {Kauffmann}}]{Springel01}
{Springel} V., {White} S.~D.~M., {Tormen} G., {Kauffmann} G., 2001, \mnras,
  328, 726

\bibitem[{{Tegmark} \& {Peebles}(1998)}]{Tegmark98}
{Tegmark} M., {Peebles} P.~J.~E., 1998, \apjl, 500, L79

\bibitem[{{Tinker} {et~al.}(2010){Tinker}, {Robertson}, {Kravtsov}, {Klypin},
  {Warren}, {Yepes}, \& {Gottl{\"o}ber}}]{Tinker10}
{Tinker} J.~L., {Robertson} B.~E., {Kravtsov} A.~V., {Klypin} A., {Warren}
  M.~S., {Yepes} G., {Gottl{\"o}ber} S., 2010, \apj, 724, 878

\bibitem[{{Valentino} {et~al.}(2017){Valentino}, {Daddi}, {Silverman},
  {Puglisi}, {Kashino}, {Renzini}, {Cimatti}, {Pozzetti}, {Rodighiero},
  {Pannella}, {Gobat}, \& {Zamorani}}]{Valentino17}
{Valentino} F., {Daddi} E., {Silverman} J.~D., {Puglisi} A., {Kashino} D.,
  {Renzini} A., {Cimatti} A., {Pozzetti} L. {et~al}, 2017, \mnras, 472, 4878

\bibitem[{{Verde} {et~al.}(2002){Verde}, {Heavens}, {Percival}, {Matarrese},
  {Baugh}, {Bland-Hawthorn}, {Bridges}, {Cannon}, {Cole}, {Colless}, {Collins},
  {Couch}, {Dalton}, {De Propris}, {Driver}, {Efstathiou}, {Ellis}, {Frenk},
  {Glazebrook}, {Jackson}, {Lahav}, {Lewis}, {Lumsden}, {Maddox}, {Madgwick},
  {Norberg}, {Peacock}, {Peterson}, {Sutherland}, \& {Taylor}}]{Verde02}
{Verde} L., {Heavens} A.~F., {Percival} W.~J., {Matarrese} S., {Baugh} C.~M.,
  {Bland-Hawthorn} J., {Bridges} T., {Cannon} R. {et~al}, 2002, \mnras, 335,
  432

\bibitem[{{Wang}(2008{\natexlab{a}})}]{Wang08a}
{Wang} Y., 2008{\natexlab{a}}, \jcap, 5, 021

\bibitem[{{Wang}(2008{\natexlab{b}})}]{Wang08b}
---, 2008{\natexlab{b}}, \prd, 77, 123525

\bibitem[{{Zehavi} {et~al.}(2002){Zehavi}, {Blanton}, {Frieman}, {Weinberg},
  {Mo}, {Strauss}, {Anderson}, {Annis}, {Bahcall}, {Bernardi}, {Briggs},
  {Brinkmann}, {Burles}, {Carey}, {Castander}, {Connolly}, {Csabai},
  {Dalcanton}, {Dodelson}, {Doi}, {Eisenstein}, {Evans}, {Finkbeiner},
  {Friedman}, {Fukugita}, {Gunn}, {Hennessy}, {Hindsley}, {Ivezi{\'c}}, {Kent},
  {Knapp}, {Kron}, {Kunszt}, {Lamb}, {Leger}, {Long}, {Loveday}, {Lupton},
  {McKay}, {Meiksin}, {Merrelli}, {Munn}, {Narayanan}, {Newcomb}, {Nichol},
  {Owen}, {Peoples}, {Pope}, {Rockosi}, {Schlegel}, {Schneider}, {Scoccimarro},
  {Sheth}, {Siegmund}, {Smee}, {Snir}, {Stebbins}, {Stoughton}, {SubbaRao},
  {Szalay}, {Szapudi}, {Tegmark}, {Tucker}, {Uomoto}, {Vanden Berk}, {Vogeley},
  {Waddell}, {Yanny}, \& {York}}]{Zehavi02}
{Zehavi} I., {Blanton} M.~R., {Frieman} J.~A., {Weinberg} D.~H., {Mo} H.~J.,
  {Strauss} M.~A., {Anderson} S.~F., {Annis} J. {et~al}, 2002, \apj, 571, 172

\bibitem[{{Zehavi} {et~al.}(2011){Zehavi}, {Zheng}, {Weinberg}, {Blanton},
  {Bahcall}, {Berlind}, {Brinkmann}, {Frieman}, {Gunn}, {Lupton}, {Nichol},
  {Percival}, {Schneider}, {Skibba}, {Strauss}, {Tegmark}, \&
  {York}}]{Zehavi11}
{Zehavi} I., {Zheng} Z., {Weinberg} D.~H., {Blanton} M.~R., {Bahcall} N.~A.,
  {Berlind} A.~A., {Brinkmann} J., {Frieman} J.~A. {et~al}, 2011, \apj, 736, 59

\bibitem[{{Zhai} {et~al.}(2017){Zhai}, {Tinker}, {Hahn}, {Seo}, {Blanton},
  {Tojeiro}, {Camacho}, {Lima}, {Carnero Rosell}, {Sobreira}, {da Costa},
  {Bautista}, {Brownstein}, {Comparat}, {Dawson}, {Newman}, {Prakash},
  {Roman-Lopes}, \& {Schneider}}]{Zhai17}
{Zhai} Z., {Tinker} J.~L., {Hahn} C., {Seo} H.-J., {Blanton} M.~R., {Tojeiro}
  R., {Camacho} H.~O., {Lima} M. {et~al}, 2017, \apj, 848, 76

\end{thebibliography}

\appendix

\section{Component fitting for the halo occupation distributions}
\label{sec:hod_component_fits}

As described in $\S$~\ref{sec:generating_hods}, we generate a library of luminosity-dependent HODs obtained by running the \galacticus{} galaxy formation model on snapshots from the Millennium Simulation. In order to randomly sample the HODS for halo masses beyond the tabulated range, we choose to fit functional forms to the separate central galaxy and satellite galaxy components. 

\subsection{Fitting the central galaxy component}
\label{sec:hod_central_fits}

Given the occupation numbers of central galaxies, $N_{\rm cen}$, we find that for the faintest luminosity limits $N_{\rm cen}\sim1$ for $M_{200}>10^{11}\hMsol$. As discussed in  $\S$~\ref{sec:generating_hods}, for brighter luminosity limits $N_{\rm cen}$ becomes peaked around  $10^{12}\hMsol$, with a slight increase towards higher mass. We find that the increase in $N_{\rm cen}$ for halo masses above $10^{14}\hMsol$ can be modelled using the linear relation,
\begin{equation}
  \log_{10}\left (-\log_{10}\left (N_{\rm cen}\right )\right ) \propto m_{\rm fit} \log_{10}\left ( M_{200}\right )
\end{equation}
where $m_{\rm fit}$ is the fitted slope, which will take a different value for each luminosity limit. Unfortunately, for the brightest luminosity limits the tabulated HODs are quite noisy and there is the potential for the HODs for the different luminosity limits to cross, which must be avoided if we are to draw random samples from the HODs. In order to address these two issues, we take the values for $m_{\rm fit}$ and fit a second linear relation such that we have a description for the slope as a function of blended luminosity threshold, $m_{\rm fit}\left (\log_{10}\left (L_{\halpha+\nii}\right )\right )$. This relation can be then be extrapolated to fit the slopes for luminosities for which the tabulated HOD is noisy and also ensures that the central galaxy occupation distributions for the luminosity limited samples never cross. 

\begin{figure*}
  \centering
  \includegraphics[width=0.99\textwidth]{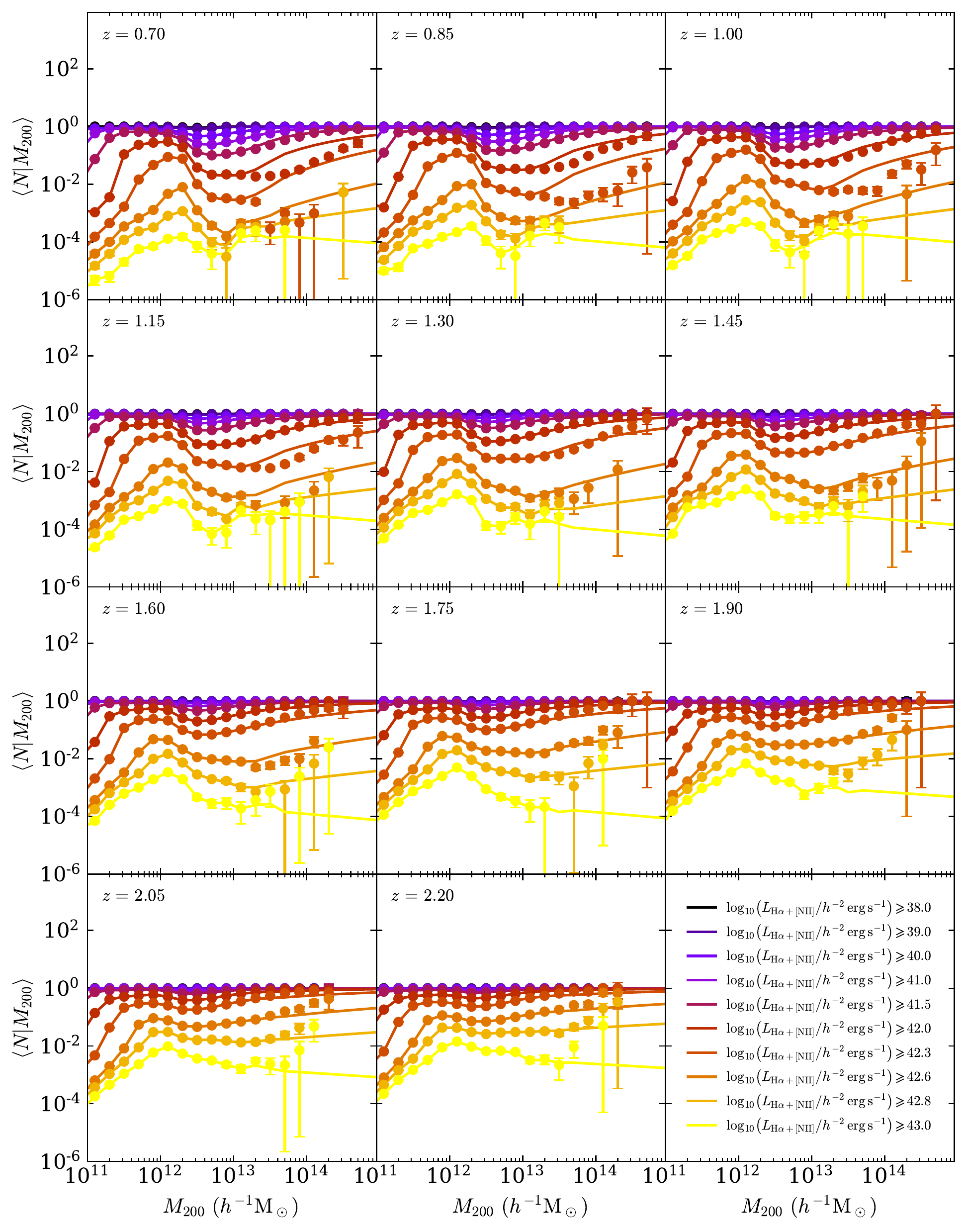}
  \caption{Central galaxy occupation distributions for 10 blended luminosity-selected samples from \galacticus{}, plotted at selected redshifts between $z=0.7$ and $z=2.2$. The solid lines show the fits to the occupation numbers as discussed in Appendix~\ref{sec:hod_central_fits}. The various colours of the lines indicate the blended $\halpha+\nii$ luminosity limit used to select the galaxies, as indicated in the lower right-hand panel. The halo mass assumed is the mass within an over-density with average density corresponding to 200 times the mean density of the Universe.}
  \label{fig:centralHOD_fits}
\end{figure*}

The fits to the tabulated occupation numbers of central galaxies are shown in Fig.~\ref{fig:centralHOD_fits}. Whilst these linear relations generally provide reasonably good fits to the tabulated numbers, there are some small discrepancies, particularly when examining the brightest luminosity limits at lower redshift and higher halo mass. However, we note from Fig.~\ref{fig:hizelsHalphaHOD_luminositySelection} that there are very few bright $\halpha$-emitting galaxies in the most massive haloes at these redshifts and so these small discrepancies in the fitting do not have a significant impact on the number densities of central galaxies in our lightcone catalogues.

\subsection{Fitting the satellite galaxy component}
\label{sec:hod_satellite_fits}

For the occupation numbers of satellite galaxies, we find that the satellite component of the tabulated HODs are well fit by a double power law. For the faintest luminosity limits a power law fit for masses above $10^{13}\hMsol$ is used to extrapolate the occupation number to higher halo mass. 

\begin{figure*}
  \centering
  \includegraphics[width=0.99\textwidth]{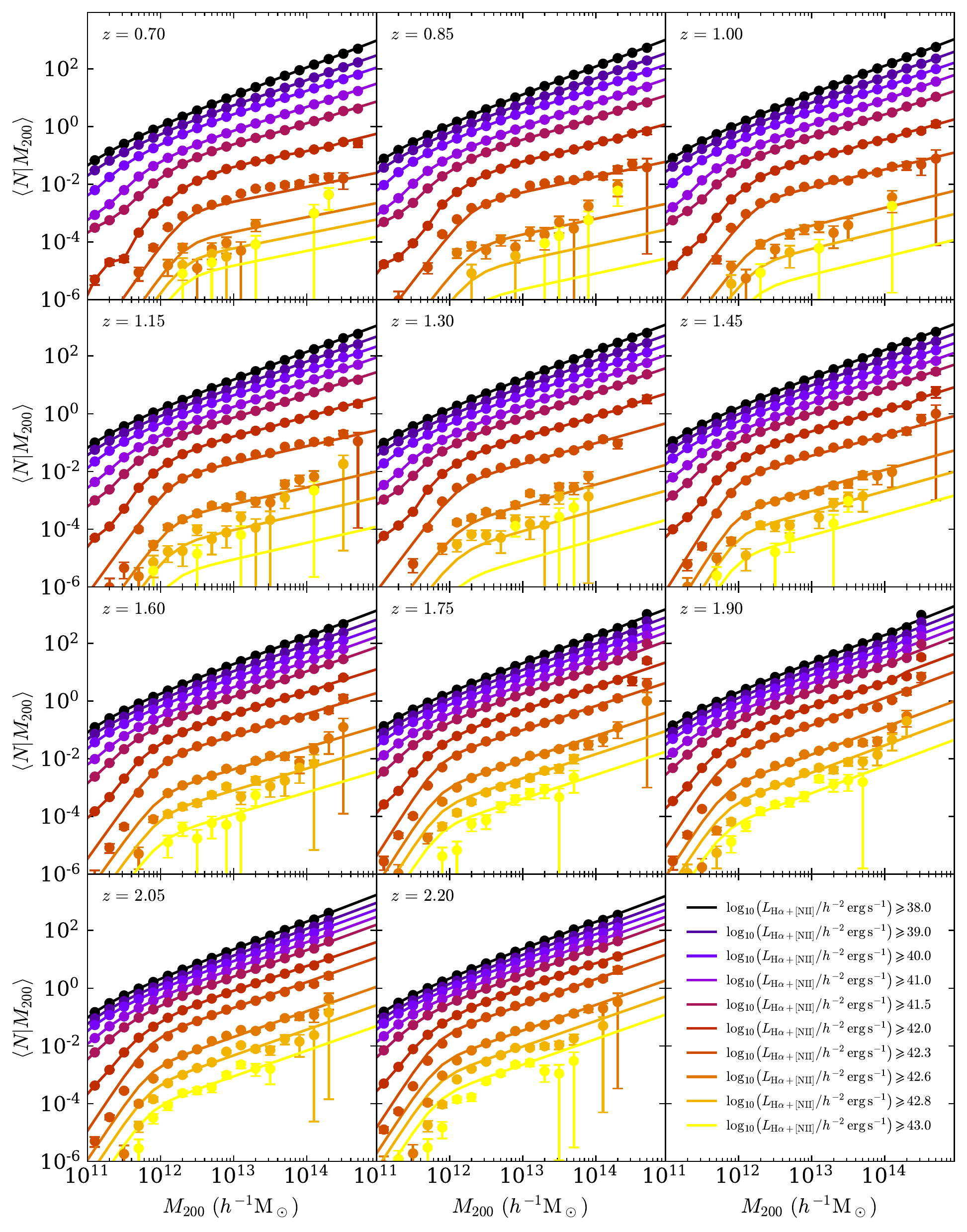}
  \caption{Satellite galaxy occupation distributions for 10 blended luminosity-selected samples from \galacticus{}, plotted at selected redshifts between $z=0.7$ and $z=2.2$. The solid lines show the fits to the occupation numbers as discussed in Appendix~\ref{sec:hod_satellite_fits}. The various colours of the lines indicate the blended $\halpha+\nii$ luminosity limit used to select the galaxies, as indicated in the lower right-hand panel. The halo mass assumed is the mass within an over-density with average density corresponding to 200 times the mean density of the Universe.}
  \label{fig:satelliteHOD_fits}
\end{figure*}

The fits to the tabulated satellite galaxy occupation numbers are shown in Fig.~\ref{fig:satelliteHOD_fits}. These fits are able to provide a reasonable match to the tabulated satellite galaxy occupation numbers, even for the brightest luminosity limits.

Just as with the central galaxies, at lower redshifts the satellite occupation numbers for the brightest luminosity limits are quite noisy and power law fits for these bright luminosity limits lead to crossing of the HODs, which we wish to minimise. Therefore for these brightest limits we simply apply a vertical offset to the double power law fitted to occupation numbers for the $\log_{10}\left (L_{\rm \halpha+\nii}/\ergPerSecondPerCM\right ) > 42.14$ limit. These fits, which can be seen in the upper row of panels in Fig.~\ref{fig:satelliteHOD_fits}, lead to an underestimate of the number of satellite galaxies, though since the fraction of satellite galaxies in these bright luminosity samples is relatively small, these under-estimates do not have a significant impact on the number density of satellite galaxies in our lightcone catalogues.

\section{Lightcone redshift distributions}
\label{sec:lightcone_redshift_distributions}

\begin{table}
\centering
\caption{Sky-averaged redshift distributions, ${\rm d}N/{\rm d}z$, for galaxies in the WISP-calibrated and HiZELS-calibrated versions of the MXXL-15 and MXXL-2K lightcones. The MXXL-15K lightcone has an $\halpha+\nii$ blended flux limit of $2\times 10^{-16}\ergPerSecondPerCM$ and the MXXL-2K lightcone has an $\halpha+\nii$ blended flux limit of $1\times 10^{-16}\ergPerSecondPerCM$. The redshift column corresponds to the bin centre and the distributions correspond to the number of galaxies per unit redshift per square degree on the sky. Galaxy counts do not include any incompleteness due to instrument-dependent efficiency.}
\begin{tabular}{|c|c|c|c|c|}
\hline
Redshift&\multicolumn{2}{|c|}{MXXL-15K}&\multicolumn{2}{|c|}{MXXL-2K}\\
&WISP&HiZELS&WISP&HiZELS\\
\hline\hline
0.725&12338&8247&23471&18003\\
0.775&12334&8103&24751&18587\\
0.825&11484&7392&23701&17539\\
0.875&11394&7214&24299&17827\\
0.925&10688&6574&23952&17398\\
0.975&9983&5943&23100&16732\\
1.025&9115&5307&22454&16194\\
1.075&8068&4542&21200&15093\\
1.125&7435&4027&20606&14489\\
1.175&6681&3555&20225&14120\\
1.225&5849&3125&19074&13176\\
1.275&4993&2630&17489&11882\\
1.325&4419&2279&16699&11220\\
1.375&3888&1997&15842&10537\\
1.425&3345&1766&14444&9477\\
1.475&2845&1530&13468&8670\\
1.525&2473&1365&12213&7935\\
1.575&2212&1259&10957&7357\\
1.625&1974&1165&10082&7019\\
1.675&1707&1058&9072&6527\\
1.725&1446&943&8087&5961\\
1.775&1323&905&7642&5756\\
1.825&1213&870&6939&5392\\
1.875&1041&774&6122&4920\\
1.925&904&699&5566&4654\\
1.975&842&685&5270&4599\\
2.025&743&636&4879&4424\\
2.075&680&613&4560&4292\\
2.125&652&619&4240&4139\\
2.175&583&581&3797&3855\\

\hline
\end{tabular}
\label{tab:lightcone_redshift_distributions}
\end{table}

In Table~\ref{tab:lightcone_redshift_distributions} we provide the redshift distributions, ${\rm d}N/{\rm d}z$, for the MXXL-15K and the MXXL-2K lightcones. Galaxy counts are shown per unit redshift, per square degree. The MXXL-15K lightcone has an $\halpha+\nii$ blended flux limit of $2\times 10^{-16}\ergPerSecondPerCM$ and the MXXL-2K lightcone has an $\halpha+\nii$ blended flux limit of $1\times 10^{-16}\ergPerSecondPerCM$. The galaxy fluxes have had dust attenuation applied and we provide the galaxy counts for both a WISP-calibrated attenuation and a HiZELS-calibrated attenuation.

\section{Linear bias fits for MXXL-15K and MXXL-2K lightcones}
\label{sec:lightcone_bias_fits}

In this section we report the results of the clustering analysis for the MXXL-15K and MXXL-2K lightcones. This analysis is carried out in an identical manner to that for the Euclid-like and WFIRST-like surveys (see $\S$~\ref{sec:linear_bias}) but with the difference that we do not apply any incompleteness to the lightcone catalogues. In other words, this analysis corresponds to making linear bias forecasts for idealised Euclid-like and WFIRST-like surveys that have 100 per cent completeness. The MXXL-15K lightcone, which has an $\halpha+\nii$ blended flux limit of $2\times 10^{-16}\ergPerSecondPerCM$, corresponds to an idealised Euclid-like survey, whilst the MXXL-2K lightcone, which has an $\halpha+\nii$ blended flux limit of $1\times 10^{-16}\ergPerSecondPerCM$, corresponds to an idealised WFIRST-like survey.

Just as with the clustering analysis for the Euclid-like and WFIRST-like surveys, we divide the MXXL-15K and MXXL-2K lightcones into 5 equal redshift bins of width $\Delta z=0.2$ spanning $0.9 \leqslant z < 1.9$ for the MXXL-15K lightcone and $1 \leqslant z < 2$ for the MXXL-2K lightcone. The correlation function is computed 5 times for each redshift bin, each time using a different random seed to select the random catalogue. We again consider two versions of the lightcones: one calibrated to reproduce the WISP number counts and one calibrated to reproduce the HiZELS luminosity functions. The whole analysis is again repeated twice, firstly selecting the galaxies in redshift-space and secondly selecting the galaxies in real-space. In Table~\ref{tab:bias_results_15k} we show for each instance of the MXXL-15K analysis the effective redshift, the mean number of galaxies and the mean ratio for the number of galaxies to the number of randoms used for the clustering analysis. Table~\ref{tab:bias_results_2k} shows the equivalent numbers for the MXXL-2K analyses.

\begin{figure*}
  \centering
  \includegraphics[width=0.99\textwidth]{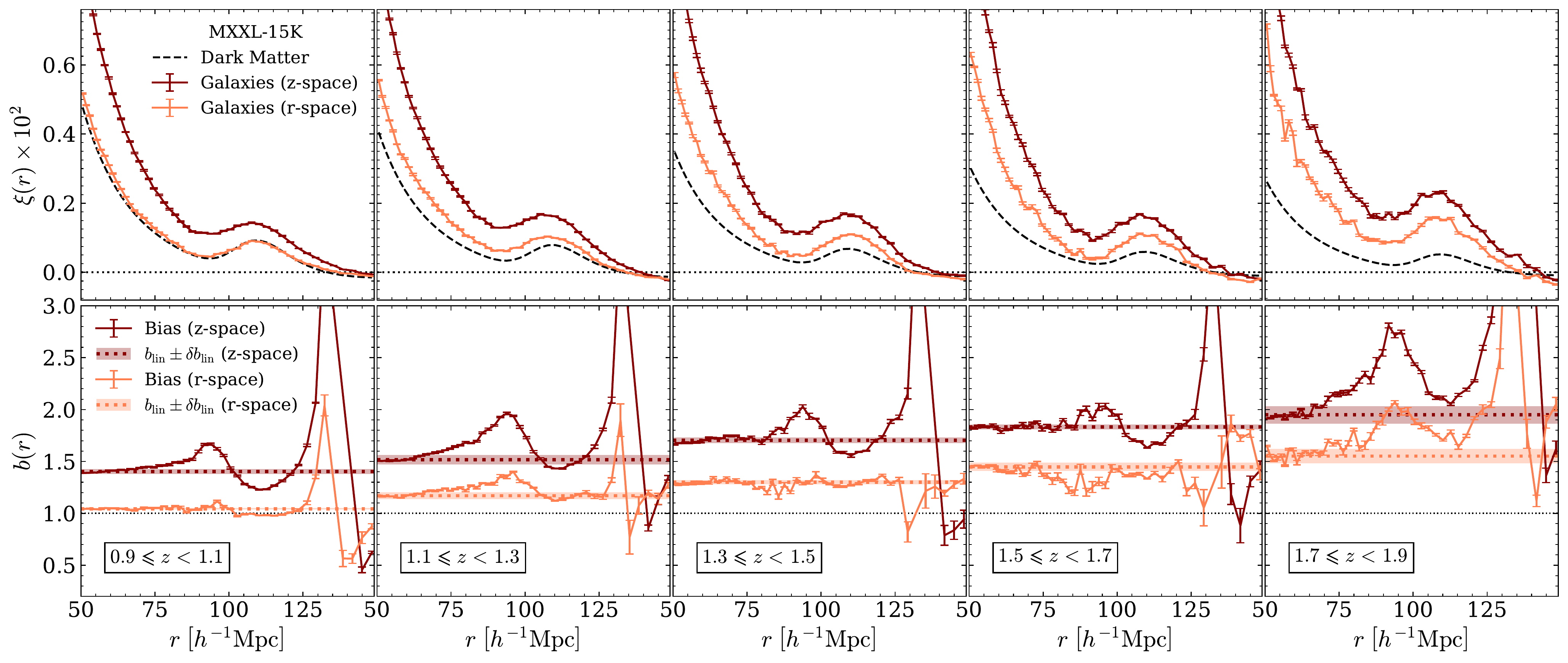}
  \includegraphics[width=0.99\textwidth]{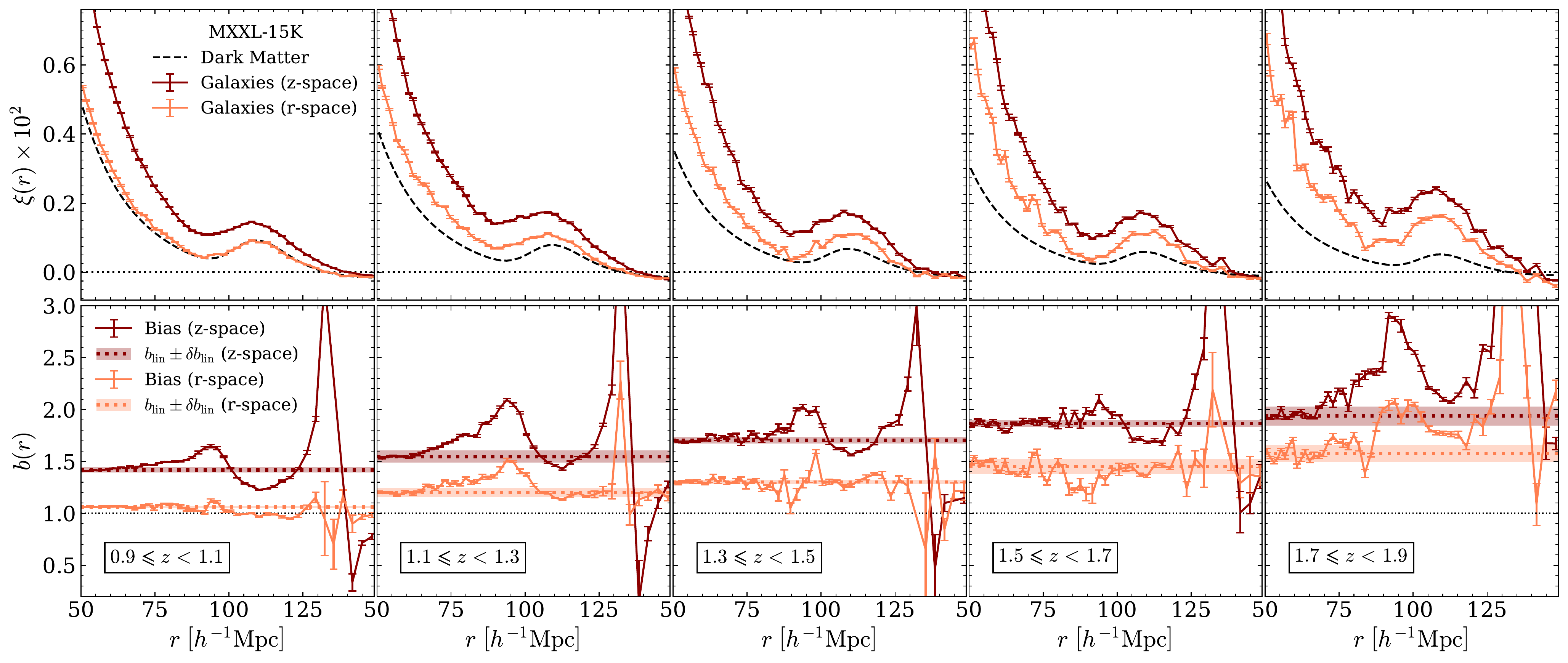}
  \caption{Clustering results and bias fits for the MXXL-15K lightcone. The upper grid of panels shows the results when adopting a WISP-calibrated version of the lightcone and the lower grid of panels shows the results when adopting a HiZELS-calibrated version of the lightcone. In each instance the darker lines show the results in redshift-space, the fainter lines correspond to real-space (i.e. assuming the cosmological redshifts of the galaxies with no peculiar velocity component), and the black dashed lines show the dark matter correlation function computed at the effective redshift of the bin. In the lower panels of each grid, the horizontal dotted lines and shaded regions correspond to the linear bias fits assuming the appropriate correlation function (see text for details). The redshift range used for selection is shown in the bottom left-hand corner of the lower panels.}
  \label{fig:lightconeBias_15k}
\end{figure*}

\begin{figure*}
  \centering
  \includegraphics[width=0.99\textwidth]{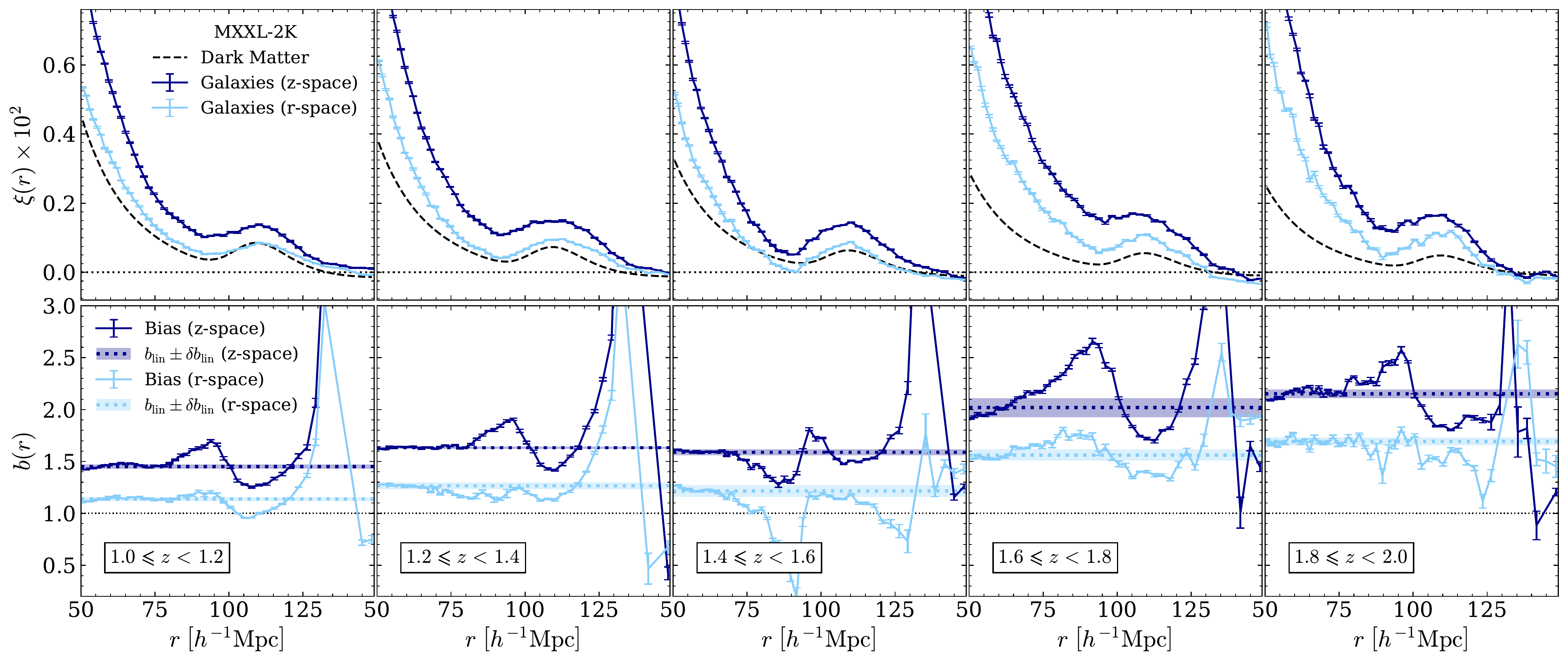}
  \includegraphics[width=0.99\textwidth]{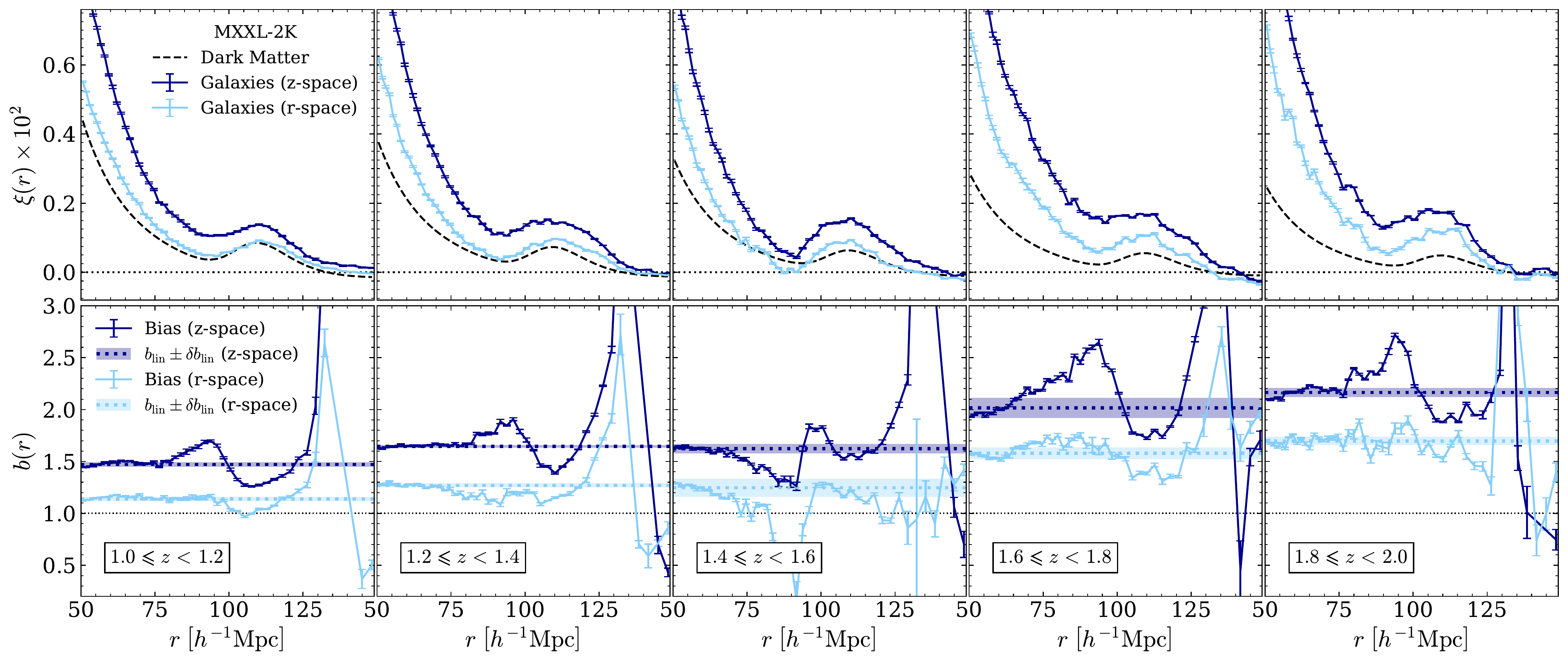}
  \caption{Clustering results and bias fits for the MXXL-2K lightcone. The upper grid of panels shows the results when adopting a WISP-calibrated version of the lightcone and the lower grid of panels shows the results when adopting a HiZELS-calibrated version of the lightcone. All the lines and shaded regions have the same meanings as in Fig.~\ref{fig:lightconeBias_15k}. The redshift range used for selection is shown in the bottom left-hand corner of the lower panels.}
  \label{fig:lightconeBias_2k}
\end{figure*}

The galaxy correlation functions for the MXXL-15K lightcone are shown in upper row of each grid in Fig.~\ref{fig:lightconeBias_15k}, where the dark red lines show the mean correlation functions in redshift-space (averaged over 5 repeat calculations), and the fainter red lines show the mean correlation functions in real-space (also averaged over 5 repeat calculations). The upper grid assumes a WISP-calibrated version of the lightcone and the lower grid assumes a HiZELS-calibrated version of the lightcone. In every instance the BAO peak is clearly identifiable and we see a larger clustering amplitude in redshift-space compared to real-space, consistent with the expectations from \citet{Kaiser87}. Comparing Fig.~\ref{fig:lightconeBias_15k} with the upper grids of Fig.~\ref{fig:lightconeBias_wisp} and Fig.~\ref{fig:lightconeBias_hizels} we can see that for the MXXL-15K the correlation functions are less noisy than those for the Euclid-like survey due to the lack of incompleteness. In Fig.~\ref{fig:lightconeBias_2k} we show the equivalent results for the MXXL-2K lightcone where we again are able to identify the BAO peak in each redshift bin, though with perhaps less significance in the highest redshift bins.

The lower row in each grid of panels in Fig.~\ref{fig:lightconeBias_15k} and Fig.~\ref{fig:lightconeBias_2k} show the measured bias, $b(r)$, computed from the galaxy and dark matter correlation functions. Comparing our bias measurements to the equivalent measurements for the Euclid-like and WFIRST-surveys (Fig.~\ref{fig:lightconeBias_wisp} and Fig.~\ref{fig:lightconeBias_hizels}) we see again see that the bias measurements are consistent with a constant for scales $r\lesssim 75\hMpc$ and that at larger scales we see pronounced scale-dependent deviations away from a constant value. As discussed in $\S$~\ref{sec:bias_measurements}, these deviations are most likely a result of distortions to the BAO caused by a combination of non-linear collapse, mode coupling effects and redshift-space distortions.

The horizontal dotted lines in the lower row in each grid in Fig.~\ref{fig:lightconeBias_15k} and Fig.~\ref{fig:lightconeBias_2k} show the fit for the linear bias on scales $r\lesssim 75 \hMpc$, obtained using $\chi^2$ minimisation to fit a zeroth order polynomial. The shaded regions show the RMS of the residuals (see $\S$\ref{sec:bias_fitting} for details). The linear bias fits for the MXXL-15K and MXXL-2K lightcones are reported in  Table~\ref{tab:bias_results_15k} and  Table~\ref{tab:bias_results_2k} respectively.  For the MXXL-15K and MXXL-2K bias results we again see that the linear bias is larger in redshift-space and increases with increasing redshift. Comparing the linear bias values in Table~\ref{tab:bias_results_15k} and Table~\ref{tab:bias_results_euclid} we see that for the corresponding redshift bins the linear bias values for the MXXL-15K lightcone and the Euclid-like survey are consistent with one another, suggesting that introduction of incompleteness has a negligible impact on our estimates for the linear bias. We also find that there is no difference in the mean halo mass. If we compare the linear bias values in Table~\ref{tab:bias_results_2k} and Table~\ref{tab:bias_results_wfirst} we see a similar result for the MXXL-2K lightcone and the WFIRST-like survey. 

\begin{table*}
\centering
\caption{Catalogue specifications and linear bias fits for the MXXL-15K lightcone, measured in redshift-space (upper table) and in real-space (lower table). In each instance, the properties shown are: the effective redshift of the slice, $z_{\rm eff}$; the mean number of galaxies used to compute the correlation function, $\bar{N}_{\rm gal}$; the mean value for the ratio of randoms to galaxies, $\bar{N}_{\rm ran}/\bar{N}_{\rm gal}$; the mean and standard deviation of the halo mass, $M_h$, for that galaxy sample; and the linear bias fit. The columns show the results for each redshift slice considered. Each table shows the bias fits when adopting a WISP-calibrated version of the lightcone and a HiZELS-calibrated version of the lightcone.}
\begin{tabular}{|c|c|c|c|c|c|c|c|}
\hline
Calibration&Property&$0.90\,\leqslant\,z\,<\,1.1$&$1.1\,\leqslant\,z\,<\,1.3$&$1.3\,\leqslant\,z\,<\,1.5$&$1.5\,\leqslant\,z\,<\,1.7$&$1.7\,\leqslant\,z\,<\,1.9$\\
\hline\hline
\multicolumn{7}{|c|}{\textbf{MXXL-15K (redshift-space)}}\\
WISP&$z_{\rm eff}$&0.991&1.19&1.38&1.59&1.79\\
&$\bar{N}_{\rm gal}$&29281758&19305864&11214261&6472414&3886909\\
&$\bar{N}_{\rm ran}/\bar{N}_{\rm gal}$&10&10&10&15&25\\
&$\log_{10}\left (M_h/h^{-1}{\rm M_{\odot}}\right )$&$11.8\,\pm\,0.4$&$11.8\,\pm\,0.3$&$11.8\,\pm\,0.3$&$11.9\,\pm\,0.3$&$11.9\,\pm\,0.3$\\
&$b_{\rm lin}\,\pm\,\delta b_{\rm lin}$&$1.40\,\pm\,0.02$&$1.52\,\pm\,0.05$&$1.70\,\pm\,0.03$&$1.83\,\pm\,0.02$&$1.95\,\pm\,0.09$\\
&&&&&&\\
HiZELS&$z_{\rm eff}$&0.988&1.19&1.39&1.59&1.79\\
&$\bar{N}_{\rm gal}$&7785058&4642807&2635798&1687841&1215424\\
&$\bar{N}_{\rm ran}/\bar{N}_{\rm gal}$&10&10&15&25&29\\
&$\log_{10}\left (M_h/h^{-1}{\rm M_{\odot}}\right )$&$11.8\,\pm\,0.3$&$11.9\,\pm\,0.3$&$11.9\,\pm\,0.3$&$11.9\,\pm\,0.3$&$11.9\,\pm\,0.3$\\
&$b_{\rm lin}\,\pm\,\delta b_{\rm lin}$&$1.42\,\pm\,0.03$&$1.55\,\pm\,0.05$&$1.69\,\pm\,0.04$&$1.88\,\pm\,0.05$&$1.9\,\pm\,0.1$\\

\hline
\multicolumn{7}{|c|}{\textbf{MXXL-15K (real-space)}}\\
WISP&$z_{\rm eff}$&0.991&1.19&1.38&1.59&1.79\\
&$\bar{N}_{\rm gal}$&29272710&19293708&11223651&6467080&3887359\\
&$\bar{N}_{\rm ran}/\bar{N}_{\rm gal}$&10&10&10&15&25\\
&$\log_{10}\left (M_h/h^{-1}{\rm M_{\odot}}\right )$&$11.8\,\pm\,0.4$&$11.8\,\pm\,0.3$&$11.8\,\pm\,0.3$&$11.9\,\pm\,0.3$&$11.9\,\pm\,0.3$\\
&$b_{\rm lin}\,\pm\,\delta b_{\rm lin}$&$1.043\,\pm\,0.007$&$1.17\,\pm\,0.03$&$1.30\,\pm\,0.02$&$1.45\,\pm\,0.04$&$1.55\,\pm\,0.07$\\
&&&&&&\\
HiZELS&$z_{\rm eff}$&0.988&1.19&1.39&1.59&1.79\\
&$\bar{N}_{\rm gal}$&17295737&10310660&5863143&3747817&2702461\\
&$\bar{N}_{\rm ran}/\bar{N}_{\rm gal}$&10&10&10&15&25\\
&$\log_{10}\left (M_h/h^{-1}{\rm M_{\odot}}\right )$&$11.8\,\pm\,0.3$&$11.9\,\pm\,0.3$&$11.9\,\pm\,0.3$&$11.9\,\pm\,0.3$&$11.9\,\pm\,0.3$\\
&$b_{\rm lin}\,\pm\,\delta b_{\rm lin}$&$1.06\,\pm\,0.01$&$1.20\,\pm\,0.04$&$1.30\,\pm\,0.02$&$1.45\,\pm\,0.07$&$1.58\,\pm\,0.08$\\

\hline
\end{tabular}
\label{tab:bias_results_15k}
\end{table*}

\begin{table*}
\centering
\caption{Catalogue specifications and linear bias fits for the MXXL-2K lightcone, measured in redshift-space (upper table) and in real-space (lower table). The meanings of the various properties are the same as in Table~\ref{tab:bias_results_15k}. The columns show the results for each redshift slice considered. Each table shows the bias fits when adopting a WISP-calibrated version of the lightcone and a HiZELS-calibrated version of the lightcone.}
\begin{tabular}{|c|c|c|c|c|c|c|c|}
\hline
Calibration&Property&$1.0\,\leqslant\,z\,<\,1.2$&$1.2\,\leqslant\,z\,<\,1.4$&$1.4\,\leqslant\,z\,<\,1.6$&$1.6\,\leqslant\,z\,<\,1.8$&$1.8\,\leqslant\,z\,<\,2.0$\\
\hline\hline
\multicolumn{7}{|c|}{\textbf{MXXL-2K (redshift-space)}}\\
WISP&$z_{\rm eff}$&1.10&1.29&1.49&1.69&1.89\\
&$\bar{N}_{\rm gal}$&8448622&6910632&5108407&3488459&2390012\\
&$\bar{N}_{\rm ran}/\bar{N}_{\rm gal}$&10&20&20&20&30\\
&$\log_{10}\left (M_h/h^{-1}{\rm M_{\odot}}\right )$&$11.7\,\pm\,0.4$&$11.7\,\pm\,0.4$&$11.7\,\pm\,0.4$&$11.8\,\pm\,0.4$&$11.8\,\pm\,0.4$\\
&$b_{\rm lin}\,\pm\,\delta b_{\rm lin}$&$1.45\,\pm\,0.02$&$1.63\,\pm\,0.01$&$1.59\,\pm\,0.03$&$2.02\,\pm\,0.09$&$2.15\,\pm\,0.04$\\
&&&&&&\\
HiZELS&$z_{\rm eff}$&1.09&1.29&1.49&1.69&1.89\\
&$\bar{N}_{\rm gal}$&5989807&4681804&3344242&2526554&1956709\\
&$\bar{N}_{\rm ran}/\bar{N}_{\rm gal}$&10&20&20&20&30\\
&$\log_{10}\left (M_h/h^{-1}{\rm M_{\odot}}\right )$&$11.7\,\pm\,0.4$&$11.8\,\pm\,0.4$&$11.8\,\pm\,0.3$&$11.8\,\pm\,0.3$&$11.8\,\pm\,0.4$\\
&$b_{\rm lin}\,\pm\,\delta b_{\rm lin}$&$1.47\,\pm\,0.02$&$1.64\,\pm\,0.01$&$1.62\,\pm\,0.05$&$2.0\,\pm\,0.1$&$2.17\,\pm\,0.04$\\

\hline
\multicolumn{7}{|c|}{\textbf{MXXL-2K (real-space)}}\\
WISP&$z_{\rm eff}$&1.10&1.29&1.49&1.69&1.89\\
&$\bar{N}_{\rm gal}$&8454205&6901558&5112198&3499443&2390544\\
&$\bar{N}_{\rm ran}/\bar{N}_{\rm gal}$&20&20&20&20&20\\
&$\log_{10}\left (M_h/h^{-1}{\rm M_{\odot}}\right )$&$11.7\,\pm\,0.4$&$11.7\,\pm\,0.4$&$11.7\,\pm\,0.4$&$11.8\,\pm\,0.4$&$11.8\,\pm\,0.4$\\
&$b_{\rm lin}\,\pm\,\delta b_{\rm lin}$&$1.14\,\pm\,0.02$&$1.27\,\pm\,0.03$&$1.22\,\pm\,0.06$&$1.56\,\pm\,0.06$&$1.69\,\pm\,0.04$\\
&&&&&&\\
HiZELS&$z_{\rm eff}$&1.09&1.29&1.49&1.69&1.89\\
&$\bar{N}_{\rm gal}$&5993966&4675862&3346543&2534087&1957573\\
&$\bar{N}_{\rm ran}/\bar{N}_{\rm gal}$&20&20&20&20&22\\
&$\log_{10}\left (M_h/h^{-1}{\rm M_{\odot}}\right )$&$11.7\,\pm\,0.4$&$11.8\,\pm\,0.4$&$11.8\,\pm\,0.3$&$11.8\,\pm\,0.3$&$11.8\,\pm\,0.4$\\
&$b_{\rm lin}\,\pm\,\delta b_{\rm lin}$&$1.14\,\pm\,0.02$&$1.27\,\pm\,0.02$&$1.25\,\pm\,0.09$&$1.58\,\pm\,0.06$&$1.70\,\pm\,0.04$\\

\hline
\end{tabular}
\label{tab:bias_results_2k}
\end{table*}

\end{document}